\documentclass[subeqn]{article}
  \usepackage[T1]{fontenc}
  \usepackage{amsmath,amssymb,mathrsfs}
  \usepackage{xspace}
  \usepackage[left=3.8cm, right=3.8cm, top=3cm, bottom=3cm]{geometry}
  \usepackage{enumitem}
\usepackage[toc,page]{appendix}

\usepackage[ansinew]{inputenc}
\usepackage{cases}
\usepackage{array}
\usepackage{mathenv}
\usepackage{hyperref}
\usepackage{graphicx,color}
\usepackage{eurosym}
\usepackage{amsmath,amsfonts,amssymb,amsthm}
\usepackage{wasysym}
\usepackage{mathrsfs}
\usepackage{oldstyle}
\usepackage[T1]{fontenc}
\usepackage{pb-diagram}

  \theoremstyle{theorem}
  \newtheorem{theorem}{Theorem}
    \newtheorem{proposition}{Proposition}
     \newtheorem{lemma}{Lemma}
 
  \theoremstyle{remark}
  
  \newtheorem{remark}{Remark}
   \newtheorem{notation}{Notation}

  \newenvironment{dem}{
\trivlist \item[\hskip \labelsep{\bf\ Proof:}]}{\hfill \makebox[2em]{\hfill{\footnotesize $\Box$}}
\vspace{2ex}}

\title{\textbf{Absence of binding in a mean-field approximation of quantum electrodynamics}}
\author{Sok J\'er\'emy\\
 Ceremade, UMR 7534, Universit\'e Paris-Dauphine,\\
  Place du Mar\'echal de Lattre de Tassigny,\\
  75775 Paris Cedex 16, France.\\ \\
}%

\begin{document}
%\pageblanche0
\maketitle
%\tableofcontents
%%%%%%%%%%%%%%%%%%%%%%%%%%%%%%
%\romanpagenumbers
%\input premieres.tex
%%%%%%%%%%%%%%%%%%%%%%%%%%%%%%
%%%%%%%%%%%%
\newcommand{\ket}[1]{\ensuremath{|#1\rangle}\xspace}
\newcommand{\bra}[1]{\ensuremath{\langle #1|}\xspace}
\newcommand{\psh}[2]{\ensuremath{\langle #1\,,\,#2\rangle}\xspace}
\newcommand{\upp}[1]{\ensuremath{^{(#1)}}\xspace}
\newcommand{\sqmod}[1]{\ensuremath{|#1|^2}\xspace}
%%%%%%%%%%%%%%%%%%%%%%%%%%%%%%
\newcommand{\ssum}{\ensuremath{\displaystyle\sum}}
\newcommand{\dint}{\ensuremath{\displaystyle\int}}
\newcommand{\diint}{\ensuremath{\displaystyle\int\!\!\!\!\int}}
\newcommand{\diiint}{\ensuremath{\displaystyle\int\!\!\!\!\int\!\!\!\!\int}}
\newcommand{\diiiint}{\ensuremath{\displaystyle\int\!\!\!\!\int\!\!\!\!\int\!\!\!\!\int}}

\newcommand{\malpha}{\ensuremath{\boldsymbol{\alpha}}}
\newcommand{\talpha}{\ensuremath{\widetilde{\alpha}}}

\newcommand{\pneg}{\ensuremath{\chi_{(\infty,0)}}}
\newcommand{\g}{\ensuremath{\gamma}}
\newcommand{\G}{\ensuremath{\Gamma}}
\newcommand{\hla}{\ensuremath{\mathfrak{h}_\Lambda}}

\newcommand{\inv}{\ensuremath{\frac{1}{|\cdot|}}}

\newcommand{\qter}[1]{\ensuremath{Q_{(#1)}}}
\newcommand{\rter}[1]{\ensuremath{\rho_{(#1)}}}

\newcommand{\nq}[1]{\ensuremath{\lvert\lvert#1\rvert\rvert_{\mathcal{Q}}}}
\newcommand{\nx}[1]{\ensuremath{\lvert\lvert#1\rvert\rvert_{\mathcal{X}}}}
\newcommand{\nqstar}[1]{\ensuremath{{}^*\lvert\lvert#1\rvert\rvert_{\mathcal{Q}}}}
\newcommand{\nc}[1]{\ensuremath{\lVert#1\rVert_{\mathfrak{C}}}}
\newcommand{\ns}[2]{\ensuremath{\lVert#2\rVert_{\mathfrak{S}_{#1}}}}
\newcommand{\nlp}[2]{\ensuremath{\lVert#2\rVert_{L^{#1}}}}
\newcommand{\nso}[2]{\ensuremath{\lVert#2\rVert_{H^{#1}}}}
\newcommand{\nhi}[1]{\ensuremath{\lvert\lvert#1\rvert\rvert_{E}}}
\newcommand{\nqq}[1]{\ensuremath{\lVert#1\rVert_{\text{Ex}}}}
\newcommand{\nqf}[1]{\ensuremath{\lvert\lvert#1\rvert\rvert_{F}}}
\newcommand{\nb}[1]{\ensuremath{\lVert#1\rVert_{\mathcal{B}}}}
\newcommand{\ncc}[1]{\ensuremath{\lVert#1\rVert_{\mathcal{C}}}}
\newcommand{\nrr}[1]{\ensuremath{\lvert\lvert#1\rvert\rvert_{\mathcal{R}}}}
\newcommand{\nrstar}[1]{\ensuremath{{}^*\lvert\lvert#1\rvert\rvert_{\mathcal{R}}}}
\newcommand{\nlpw}[2]{\ensuremath{\lvert\lvert#2\rvert\rvert_{L^{#1}(|x|^2dx)}}}
\newcommand{\norm}[1]{\ensuremath{\lVert#1\rVert}}

\newcommand{\nqbf}[1]{\ensuremath{\lVert#1\rVert_{\mathbf{Q}}}}
\newcommand{\nqkin}[1]{\ensuremath{\lVert#1\rVert_{\mathrm{Kin}}}}
\newcommand{\nqbfg}[1]{\ensuremath{\lvert\lvert#1\rvert\rvert_{\mathbf{Q}_w}}}
\newcommand{\nqbfu}[1]{\ensuremath{\lVert#1\rVert_{\mathbf{Q}_1}}}
\newcommand{\ncg}[1]{\ensuremath{\lVert#1\rVert_{\mathfrak{C}_w}}}
\newcommand{\ncgp}[1]{\ensuremath{\lVert#1\rVert_{\mathfrak{C}'_w}}}
\newcommand{\nxg}[1]{\ensuremath{\lvert\lvert#1\rvert\rvert_{\mathcal{X}_w}}}

\newcommand{\ov}[1]{\ensuremath{\overline{#1}}}
\newcommand{\un}[1]{\ensuremath{\underline{#1}}}

\newcommand{\lpsi}{\ensuremath{\psi_\lambda}}
\newcommand{\wlpsi}{\ensuremath{\widehat{\psi_\lambda}}}
\newcommand{\upsi}{\ensuremath{\psi_1}}
\newcommand{\wupsi}{\ensuremath{\widehat{\psi_1}}}
\newcommand{\llo}{\ensuremath{\log(\Lambda)}}
\newcommand{\sll}{\ensuremath{\sqrt{\log(\Lambda)}}}
\newcommand{\sal}{\ensuremath{\sqrt{\alpha}}}
\newcommand{\rr}{\ensuremath{\mathfrak{R}}}
\newcommand{\ph}{\ensuremath{\varphi}}
\newcommand{\unp}{\ensuremath{\underline{\psi}}}

\newcommand{\ee}[1]{\ensuremath{E\left(#1\right)}}
\newcommand{\ef}[1]{\ensuremath{E_m\left(#1\right)}}
\newcommand{\wh}[1]{\ensuremath{\widehat{#1}}}
\newcommand{\hlp}{\ensuremath{\widehat{\psi_\lambda}}}
\newcommand{\hlh}{\ensuremath{\widehat{\psi_\lambda}}}
\newcommand{\hlu}{\ensuremath{\widehat{\psi_1}}}
\newcommand{\la}{\ensuremath{\lambda}}
\newcommand{\La}{\ensuremath{\Lambda}}
\newcommand{\W}{\ensuremath{\alpha_r}}
\newcommand{\tr}{\ensuremath{\mathrm{Tr}_{\mathcal{P}^0}}}
\newcommand{\ttr}{\ensuremath{\mathrm{Tr}}}
\newcommand{\pl}{\ensuremath{\frac{p}{\la}}}
\newcommand{\ww}[1]{\ensuremath{\widehat{#1}}}
\newcommand{\wt}[1]{\ensuremath{\widetilde{#1}}}
\newcommand{\kappab}[1]{\ensuremath{\boldsymbol{\kappa}_{#1}}}

\newcommand{\D}{\ensuremath{\mathcal{D}^0}}
\newcommand{\Dbf}{\ensuremath{\mathbf{D}}}
\newcommand{\Dt}{\ensuremath{\widetilde{D}}}
\newcommand{\dd}{\ensuremath{\mathrm{d}}}
\newcommand{\hl}{\ensuremath\mathfrak{H}_\Lambda}
\newcommand{\sign}{\ensuremath{\tfrac{D_0}{|D_0|}}}

\newcommand{\PP}{\ensuremath{\mathcal{P}^0_-}}
\newcommand{\PPP}{\ensuremath{\mathcal{P}^0_+}}
\newcommand{\EE}{\ensuremath{\mathcal{E}}}
\newcommand{\MM}{\ensuremath{\widetilde{M}}}
\newcommand{\eps}{\ensuremath{\varepsilon}}
\newcommand{\oo}[1]{\ensuremath{\omega_#1}}
\newcommand{\om}{\ensuremath{\omega}}
\newcommand{\ed}[1]{\ensuremath{\widetilde{E}\left(#1\right)}}
\newcommand{\eed}[1]{\ensuremath{\widetilde{E}_{#1}}}

\newcommand{\Ebf}[1]{\ensuremath{\overline{E}_{#1}}}
\newcommand{\sbf}[1]{\ensuremath{\mathbf{s}_{#1}}}
\newcommand{\Tbf}{\ensuremath{\mathbf{T}}}
\newcommand{\fla}{\ensuremath{f_{\Lambda}}}
\newcommand{\Fla}{\ensuremath{F_{\Lambda}}}
\newcommand{\Flac}{\ensuremath{\check{F}_{\Lambda}}}
\newcommand{\gla}{\ensuremath{g_{\Lambda}}}
\newcommand{\wit}[1]{\ensuremath{\widetilde{#1}}}
\newcommand{\ala}{\ensuremath{a[\La]}}
\newcommand{\rg}{\ensuremath{R_{\text{g}}}}

\newcommand{\ash}{\ensuremath{\mathrm{arcsinh}}}
\newcommand{\CC}{\ensuremath{\mathbb{C}^4}}
\newcommand{\RR}{\ensuremath{\mathbb{R}^3}}

\newcommand{\weak}{\ensuremath{\rightharpoonup}}

\newcommand{\nqu}[1]{\ensuremath{\lvert\lvert#1\rvert\rvert_{q_1}}}
\newcommand{\nqd}[1]{\ensuremath{\lvert\lvert#1\rvert\rvert_{q_2}}}
\newcommand{\nqt}[1]{\ensuremath{\lvert\lvert#1\rvert\rvert_{q_3}}}

\newcommand{\nqz}[1]{\ensuremath{\lvert\lvert#1\rvert\rvert_{q_0}}}

\newcommand{\rgg}{\ensuremath{\text{R}_g}}

\newcommand{\etabfc}[1]{\ensuremath{\boldsymbol{\eta}_{c\text{R}_g}^{(#1)}}}

\newcommand{\pvacno}{\ensuremath{\boldsymbol{\pi}_{\mathrm{vac}}}}

%%%%%%%%%%%%%%%%%%%%%%%%%%%%%%
%\tableofcontents

\abstract{We study the Bogoliubov-Dirac-Fock model which is a mean-field approximation of QED. It allows to consider relativistic electrons interacting with the Dirac sea. We study the system of two electrons in the vacuum: it has been shown in a previous paper \cite{sok} that an electron alone can bind due to the vacuum polarisation, under some technical assumptions. Here we prove the absence of binding for the system of two electrons:% for sufficently small values of $\alpha,\alpha\llo,\La^{-1}$. 
the response of the vacuum is not sufficient to counterbalance the repulsion of the electrons.}

\section{Introduction and main results}
\begin{center}
 \textsc{The Dirac operator}
\end{center}

The theory of relativistic quantum mechanics is based on the Dirac operator $D_0$, that describes the kinetic energy of a relativistic electron. To simplify formulae, we take relativistic units $\hbar=c=4\pi\eps_0=1$ and set the bare particle mass equal to $1$.

In this case, the Dirac operator is defined by \cite{Th}: $D^0=-i\boldsymbol{\alpha}\cdot \nabla+\beta$ where $\beta,\alpha_j\in\mathcal{M}_4(\mathbf{C})$ are the Dirac matrices:
\begin{subequations}
\begin{equation}
\begin{array}{l  l}
\beta=\begin{pmatrix} \mathrm{Id}_2 & 0 \\ 0 & -\mathrm{Id}_2 \end{pmatrix}, &\ \ \alpha_j= \begin{pmatrix} 0 & \sigma_j \\ \sigma_j & 0 \end{pmatrix},j=1,2,3
\end{array}
\end{equation}

\begin{equation}
\begin{array}{lll}
\sigma_1=\begin{pmatrix} 0 & 1\\ 1 & 0 \end{pmatrix}, & \sigma_2=\begin{pmatrix} 0 & -i \\ i & 0 \end{pmatrix}, & \sigma_3=\begin{pmatrix} 1 & 0 \\ 0 & -1 \end{pmatrix}.
\end{array}
\end{equation}
\end{subequations}
%[
It acts on the Hilbert space $\mathfrak{H}=L^2(\mathbb{R}^3,\mathbb{C}^4)$ with domain $H^1(\mathbb{R}^3,\mathbb{C}^4)$. Its spectrum is not bounded from below: $\sigma(D_0)=(-\infty,-1]\cup [1,+\infty)$, which implies the existence of states with arbitrarily small negative energy. Dirac postulated that all the negative energy states are already occupied by "virtual" electrons forming the so-called Dirac sea: by Pauli principle a real electron can only have positive energy.

According to this interpretation, the vacuum, filled by the Dirac sea, is a polarizable medium that reacts to the presence of an electromagnetic field.%For instance, if one turns on such a potential and make it stronger enough, a "virtual" electron can reach a positive energy state, leaving a hole in the Dirac sea, which is interpreted as an electron-positron pair.  
%]
%$\alpha=e^2/(4\pi)$
\begin{center}
 \textsc{BDF model}
\end{center}
In this paper we study the Bogoliubov-Dirac-Fock (BDF) model which is a no-photon, mean-field approximation of Quantum Electrodynamics (QED) which was introduced by Chaix and Iracane \cite{CI}. It enables us to consider a system of relativistic electrons interacting with the vacuum in the presence of an electrostatic field.
This paper is a continuation of previous works by Hainzl, Gravejat, Lewin, S\'er\'e, Siedentop \cite{HaiSied,ptf,Sc,at,gs} and Sok \cite{sok,sokd}. 

The derivation of the BDF model from QED is explained in \cite{CI} and \cite[Appendix]{ptf}: we refer the reader to these papers for full details.

\medskip
In QED, an electronic system is described by a state in the fermionic Fock space $\mathcal{F}_{el}$ \cite[Chapter 10]{Th} on which (formally) acts the Hamiltonian $\mathbb{H}_{\text{QED}}$ \cite[Appendix]{ptf}. The mean-field approximation consists to restricting the study to Hartree-Fock type states, called BDF states. They are fully characterized by their one-body density matrix (1pdm) which are orthogonal projectors of $\mathfrak{H}$. 

For instance, the projector $P^0_-:=\chi_{(-\infty,0)}(D_0)$ is the 1pdm of the vacuum state $\Omega_0\in\mathcal{F}_{el}$: it must be thought of as the infiniter Slater determinant $f_1\wedge f_2\wedge\cdots$ where $(f_i)_{i\ge 1}$ is an orthonormal basis (BON) of $\text{Ran}(P^0_-)$. A projector $P$ defines a BDF state \emph{iff} $P-P^0_-$ is Hilbert-Schmidt (\emph{i.e.} its integral kernel is square integrable).
 
 We take $P^0_-$ as a reference state and define a renormalized Hamiltonian $:\mathbb{H}_{\text{QED}}:$ by a procedure of normal ordering relative to $P^0_-$ \cite{CI,ptf}. The energy $\psh{\Omega_P}{:\mathbb{H}_{\text{QED}}: \Omega_P}$ of a state $\Omega_P$, turns out to be a function of the reduced density matrix (r1pdm) $Q:=P-P^0_-$. Formally this function is
%The (formal) difference of the energy $\mathcal{E}_{\text{QED}}(P)$ of a state $P$ with that of $P^0_-$ considered as a reference state turns out to be a function of the reduced density matrix (r1pdm) $Q:=P-P^0_-$. Formally this function is
\begin{equation}\label{no_bind_bdf_1}
 \wt{\mathcal{E}}_{\text{BDF}}^\nu(Q):=\ttr\big(D_0 Q\big)-\alpha D(\nu,\rho_Q)+\frac{\alpha}{2}\big(D(\rho_Q,\rho_Q)-\nqq{Q}^2\big),
\end{equation}
where $\alpha>0$ is the fine structure constant, $\nu$ is the external density of charge, $\rho_Q(x):=\ttr_{\CC}\big(Q(x,x)\big)$ is the density of $Q$, with $Q(x,y)$ the integral kernel of $Q$, and:
\begin{equation}
 D(\nu,\nu)=\ncc{\nu}^2:=4\pi\dint_{\RR} \frac{|\wh{\nu}(k)|^2}{|k|^2}dk\text{\ and\ }\nqq{Q}^2:=\underset{\RR\times \RR}{\diint}\frac{|Q(x,y)|^2}{|x-y|}dxdy.
\end{equation}
The hat in $\wh{\nu}$ denotes the Fourier transform and $D(\nu,\nu)<+\infty$ is the Coulomb energy of $\nu$: it coincides with $\iint \frac{\nu(x)^*\nu(y)}{|x-y|}dxdy$ whenever this integral makes sense. We also write
\begin{equation}
\mathcal{C}:=\Big\{\nu\in  \mathcal{S}'(\mathbb{R}^3),\ \wh{\nu}\ \text{measurable\ and\ }\dint\frac{|\wh{\nu}(k)|^2}{|k|^2}dk<+\infty\Big\}.
\end{equation}

In \eqref{no_bind_bdf_1} we recognize the kinetic energy, the interaction energy with $\nu$, the direct term $\tfrac{\alpha}{2}D(\rho_Q,\rho_Q)$ and the exchange term $-\tfrac{\alpha}{2}\nqq{Q}^2$. \emph{A priori} this formula makes sense only when $Q$ and $D_0 Q$ are trace-class and the variational problem is ill-defined.

An ultraviolet cut-off $\La>0$ is necessary. Following \cite{gs}, we replace $D_0$ by

\[
\Dbf:=D_0\big(1-\frac{\Delta}{\La^2}\big)\text{\ with\ domain\ }H^{3/2}(\mathbf{R}^3,\mathbf{C}^4),
\]
and only consider states $Q$ such that $\ttr\big(|\Dbf|\,|Q|^2\big)<+\infty$.

By adapting \eqref{no_bind_bdf_1}, we get a well-defined energy $\mathcal{E}^\nu_{\text{BDF}}$ (defined in the next section).

\begin{remark}
We use the terms Direct space and Fourier space: a function that depends on position variables (such as a wave function $\psi(x)$ or a 1pdm $Q(x,y)$) is in Direct space, while its Fourier transform that depends on momentum variables is in Fourier space (such as $\wh{\psi}(p)$ or $\wh{Q}(p,q)$).
\end{remark}

\begin{remark}\label{no_bind_sharp_smooth}
Other choices of cut-off are possible. This one, the \emph{smooth} cut-off, is convenient for the study of functions in Direct space. In \cite{ptf,Sc,at} Hainzl \emph{et al.} have chosen the \emph{sharp} cut-off, replacing $L^2(\RR,\CC)$ by its subspace $\mathfrak{H}_\La$ made of square-integrable functions whose Fourier transform vanishes outside the ball $B_{\mathbb{R}^3}(0,\La)$.
\end{remark}

%From a mathematical point of view, the BDF energy is a function with three paremeters: the fine structure constant $\alpha$($=e^2/(4\pi\hbar \eps_0)$), the cut-off $\La>0$ and the external density $\nu$ \cite{CI,ptf}.

\begin{remark}
We still have $\chi_{(-\infty,0)}(\Dbf)=P^0_-$. We also write $P^0_+:=\chi_{(-\infty,0)}(D^0)=\text{Id}-P^0_-$ the projector on its positive spectral subspace.
\end{remark}

\begin{notation}
 For an operator $Q$, we define $R_Q$ by its integral kernel:
 \begin{equation}
  R_Q(x,y):=\frac{Q(x,y)}{|x-y|},\ x,y\in\RR\times \RR,\ x\neq y.
 \end{equation}
 
 Moreover for any $\rho\in\mathcal{C}$ we write 
 \begin{equation}
  v_\rho:=\rho*\tfrac{1}{|\cdot|}.
 \end{equation}
\end{notation}

\begin{center}
 \textsc{Existence of minimizers}
\end{center}
For a r1pdm $Q=P-P^0_-$, the charge of the system is given by its so-called $P^0_-$-trace $\ttr_{P^0_-}(Q)$, defined by
\begin{equation}
    \ttr_{P^0_-}(Q):=\ttr\big(P^0_- Q P^0_-\big)+\ttr\big( P^0_+ Q P^0_+\big).
\end{equation}
It coincides with the usual trace for trace-class operators and is well-defined for r1pdm because of their structure. Indeed as a difference of orthogonal projectors $Q$ satisfies:
\begin{equation}
 P^0_+ (P-P^0_-)P^0_+-P^0_- (P-P^0_-)P^0_-=(P-P^0_-)^2.
\end{equation}

A minimizer for $\mathcal{E}_{\text{BDF}}^\nu$ among states with charge $M\in\mathbb{N}$ is interpreted as a ground state of the system with $M$ electrons in the presence of $\nu$. For $q\in\mathbb{R}$, the infimum of the BDF energy on the charge sector $\mathcal{Q}_\La(q):=\{Q:\ \ttr_{P^0_-}(Q)=q\}$ is written $E^\nu(q)$.

A sufficient condition for the existence of a minimizer for $E^\nu(q)$ is the validity of binding inequalities at level $q$ \cite[Theorem 1]{at}. This result is stated for the \emph{sharp} cut-off, however it is possible to adapt its proof to get this Theorem:
\begin{theorem}\label{no_bind_HVZ}
 Let $0\le \alpha<\tfrac{4}{\pi}$, $\La>0$, $\nu\in\mathcal{C}$ and $q\in\mathbb{R}$. Then the following assertions are equivalent:
 \begin{enumerate}
  \item the binding inequalities hold: $\forall\,k\in\mathbb{R}\backslash\{0\},\ E^\nu(q)<E^\nu(q-k)+E^0(k)$,
  \item each minimizing sequence $(Q_n)_{n\ge 1}$ for $E^\nu(q)$ is precompact in $\mathcal{Q}_\La(q)$ and converges, up to a subsequence, to a minimizer for $E^\nu(q)$.
  If $\nu=0$, this result holds up to translation.
 \end{enumerate}
If $q$ is an integer, then we can only consider $k\in\mathbb{Z}\backslash \{0\}$ in the first assertion.
\end{theorem}

Checking binding inequalities is a difficult task. Hainzl \emph{et al.} checked them in some cases with non-vanishing $\nu$ \cite[Theorems 2 and 3]{at}. \cite[Theorem 3]{at} states that for $\nu\in L^1(\mathbb{R}^3,\mathbb{R}_+)\cap \mathcal{C}$, there exists a minimizer for $E^\nu(M)$ provided that $M-1<\int \nu$ under technical assumptions on $\alpha,\La$.

In \cite{sok}, the existence of a ground state for $E^0(1)$ is proved, still under technical assumptions on $\alpha,\La$. It is remarkable that an electron can bind alone without any external potential: this is due to the vacuum polarisation. The electron creates a hole in the Dirac sea that allows it to bind. This effect causes a charge screening: from far away the charge of the electron appears smaller as it is surrounded by the hole. 

Let $Q$ be a minimizer for $E^0(1)$, then its density $\rho_Q$ is integrable \cite{sokd}, and we have the \emph{charge renormalisation formula}:
\begin{equation}\label{no_bind_z_3}
 \dint \rho_Q=1\times Z_3\approx 1\times \frac{1}{1+\frac{2}{3\pi}\alpha\llo}\neq 1.
\end{equation}
Here $Z_3$ is the \emph{renormalisation constant}. This inadequacy is possible because the minimizer is \emph{not} trace-class (hence the mere fact that $\rho_Q$ is integrable is non-trivial). %It was proved in a simplified BDF model in \cite{gs}.

We emphasize that these results were proved with the \emph{sharp} cut-off, but the proofs can be adapted in the present case.

%In \cite{sok_dens}, this result was enhanced 

Our purpose in this paper is to study the variational problem $E^0(2)$, that is two electrons in the vacuum. We recall that an electron does not see its own field, but in the case of two electrons any electron feel the field induced by the other resulting to a repulsive force. If the vacuum polarisation is not strong enough to counterbalance this repulsion, then there is no minimizer for $E^0(2)$. This constitutes our main Theorem.

\begin{theorem}\label{no_bind_main}
There exist $\alpha_0, \La_0,L_0$ such that if $\alpha \le \alpha_0$, $\La\ge \La_0$ and $\alpha\llo\le L_0$, then there is no minimizer for $E^0(2)$.
\end{theorem}
\begin{remark}
 This result is proved in the case of the \emph{smooth} cut-off, and we expect it to be true for the \emph{sharp} one but we were unable to show it.
\end{remark}

We prove it \emph{ad absurdum}. Let us give the main ideas.

Along this paper we suppose that there exists a minimizer $Q$ for $E^0(2)$. Such a minimizer satisfies a self-consistent equation \cite[Proposition 1]{at}, \cite{gs} and can be decomposed as follows:
%We know \cite{at,gs} that such a minimizer satisfies a self-consistent equation. In particular, we can decompose it as follows:
\begin{equation}
 Q=\ket{\psi_1}\bra{\psi_1}+\ket{\psi_2}\bra{\psi_2}+\g,
\end{equation}
where the $\psi_j$'s are eigenvectors of the so-called mean-field operator:
\begin{equation}\label{no_bind_m-f}
\begin{array}{l}
 D_Q:=\Dbf+\alpha\big(v_{\rho_Q}-R_Q \big),
\end{array}
\end{equation}
where for a density $\rho\in\mathcal{C}$ and an operator $Q$, we define
 \begin{equation}
  R_Q(x,y):=\frac{Q(x,y)}{|x-y|},\,x,y\in\RR \text{\ and\ }v_\rho:=\rho*\tfrac{1}{|\cdot|}.
 \end{equation}
 For short we will also write 
 \begin{equation}
  B_Q:=v_{\rho_Q}-R_Q.
 \end{equation}

By studying $E^0(2)\le 2 E^0(1)$, we get \emph{a priori} information on the $\psi_j$'s. In particular we show that the subspace $\text{Span}(\psi_1,\psi_2)$ splits as follows
\[
 \text{Span}(\psi_1,\psi_2)=\mathbb{C}h_1\overset{\perp}{\oplus}\mathbb{C}h_2,\ \nlp{2}{h_j}=1,
\]
where $h_1$ and $h_2$ are essentially two bump functions which are some distance $\rgg$ away from each other. The operator $\g$ is also localised around each $h_j$ such that the energy $\mathcal{E}^0_{\text{BDF}}(Q)$ can be written
\[
 \mathcal{E}^0_{\text{BDF}}(Q)=2 E^0(1)+\theta_{12},
\]
where $\theta_{12}>0$ in our range of parameters $(\alpha,\La)$.

Roughly speaking the BDF energy should be the sum of the BDF energy of these two parts plus the interaction energy. This interaction energy is too big to ensure $E^0(2)$ is attained.

\begin{remark}\label{no_bind_regime}
Throughout this paper, we work in the regime where $\alpha$ and $\La$ satisfy these conditions: $\alpha\le \alpha_0,\alpha\llo:=L\le L_0$ and $\La\ge \La_0>0$ for small constants $\alpha_0,L_0,\La_0^{-1}$. $K$ is some constant independent of those numbers while $K(\la)$ means a constant depending on the quantity $\la$. Symbols $o(\cdot)$,$\mathcal{O}(\cdot)$ and $\Theta(\cdot)$ are to be understood in this regime.
\end{remark}

The paper is organised as follows. In the next section we properly define our model and give \emph{a priori} estimates about $E^0(2)$ and its hypothetical minimizer in Lemma \ref{no_bind_estim_minim}. This Lemma is proved in Section \ref{no_bind_minimini}. 

Then in Section \ref{no_bind_the_PT_fun}, we study the Pekar-Tomasevitch functional to exploit these results (Propositions \ref{no_bind_eloig}, \ref{no_bind_eloig1} and \ref{no_bind_RR}). These Propositions are proved in Appendix \ref{no_bind_ptt}.

Section \ref{no_bind_techn} is devoted to introduce important tools of the proof: the Cauchy expansion (part \ref{no_bind_the_cauchy}) and useful inequalities (part \ref{no_bind_some_ineq}). We recall in part \ref{no_bind_eq_dens} the form of the density of a minimizer.

Section \ref{no_bind_loc_dir} is dedicated to prove Theorem \ref{no_bind_main}. We show how the energy is distributed in Direct space (Proposition \ref{no_bind_place}). This enables us to prove Theorem \ref{no_bind_main} (part \ref{no_bind_proof_main_no}). To this end we first study the localisation of the "real" electrons' wave functions (Lemma \ref{no_bind_decaylemma}, proved in Appendix \ref{no_bind_elds_wf}). We then show how this enables us to get localisation of the energy of a minimizer (Lemma \ref{no_bind_titim}, proved in this Section but using Appendix \ref{no_bind_elds}). For the sake of clarity we explain in Remark \ref{no_bind_imbriquer} how Appendix \ref{no_bind_elds} is used to prove Lemma \ref{no_bind_titim}.

We have postponed the most technical proofs in the appendices. In Appendix \ref{no_bind_appA}, we prove Proposition \ref{no_bind_conti} and Lemma \ref{no_bind_H2}. This last Lemma shows estimates on a minimizer by bootstrap arguments. Maybe the most difficult results lie in Appendices \ref{no_bind_elds_wf} and \ref{no_bind_elds}, dedicated to prove localisation estimates in Direct space.

%as foll

\noindent\textit{Acknowledgment}: The author wishes to thank \'Eric S\'er\'e and Mathieu Lewin for useful discussions and helpful comments. This work was partially supported by the Grant ANR-10-BLAN0101 of the French Ministry of research.

%%%%%%%%%%%%%%%%%%%%%%%%%%%%%%%%%%%%%%%%%%%%
%%%%%%%%%%%%%%%%%%%%%%%%%%%%%%%%%%%%%%%%%%%%

\section{Presentation of the model}\label{no_bind_pres_model}

\begin{remark}[Fourier transform]
In this paper, the Fourier transform is defined on $L^1(\RR)$ by the formula:
\[
\forall\ f\in L^1(\RR),\ \wh{f}(p):=\frac{1}{(2\pi)^{3/2}}\dint_{\mathbf{R}^3}f(x)e^{-ip\cdot x}dx.
\]
\end{remark}

\begin{notation}[Splitting w.r.t. $P^0_{\pm}$]
For an operator $Q$ and $e_1,e_2\in\{+,-\}$ we write $Q^{e_1\,e_2}:= P^0_{e_1} Q P^0_{e_2}$.
\end{notation}

\begin{notation}[Schatten classes]\label{no_bind_schatten}
 We recall that for $1\le p\le \infty$, the set of compact operators whose singular values form a sequence in $\ell^p$ is denoted by $\mathfrak{S}_p(\hl)$ \cite{Sim,Sim}. The case $p=2$ (resp. $p=1$) corresponds to Hilbert-Schmidt operators (resp. trace-class operators).
 
 Those Banach spaces satisfy Hölder-type inequalities \cite{ReedSim}. We also recall the Kato-Seiler-Simon inequalities \cite{Sim}:
 \begin{equation}\label{no_bind_kss}
  \forall\,2\le p\le \infty,\ \forall\,f,g\in\,L^p(\RR),\ \ns{p}{f(x)g(-i\nabla)}\le (2\pi)^{-3/p}\nlp{p}{f}\nlp{p}{g}.
 \end{equation}
Furthermore we write $\mathcal{B}(\hl)$, the set of bounded linear endomorphisms on $\hl$.
\end{notation}

\begin{notation}[On $D_0$ and $\Dbf$]\label{no_bind_sbf}
We write $\mathbf{s}_p$ for $\frac{\wh{D_0}(p)}{\sqrt{1+|p|^2}}$ the action of $\text{sign}(D_0)$ in the Fourier space. The function $\sqrt{1+|p|^2}$ is also written $E(p)$ and $\Ebf{p}:=\sqrt{1+|p|^1}(1+|p|^2/\La^2)$. 

Throughout this paper 
\begin{equation}\label{ea_la}
 \eps[\La]=\eps_\La:=\frac{1}{\llo}\text{\ and\ }\ala:=\frac{1+\eps[\La]}{2}.
\end{equation}
We have
\begin{equation}\label{no_bind_trick_ala}
|D_0|^{1+\eps_\La}\le E(\La)^\eps|\Dbf|\le (1+e)|\Dbf|,\ \La\ge e=\text{exp}(1).
\end{equation}
\end{notation}

\subsection{The BDF energy}\label{no_bind_bdf_energ}
%\begin{center}
% \textsc{BDF energy}
%\end{center}

Let $\nu$ be an external charge density in $\mathcal{C}$ and $\alpha,\La>0$ be given. We want to extend \eqref{no_bind_bdf_1}: the result is the BDF energy \eqref{no_bind_bdf_2} below.

Following \cite{gs} we define the set:
\begin{equation}
\mathcal{Q}_{\text{Kin}}:=\big\{Q\in\mathfrak{S}_2, |\Dbf|^{1/2}Q,Q|\Dbf|^{1/2}\in\mathfrak{S}_2,|\Dbf|^{1/2}Q^{++}|\Dbf|^{1/2},|\Dbf|^{1/2}Q^{--}|\Dbf|^{1/2}\in\mathfrak{S}_1\big\}.
\end{equation}
The kinetic energy functional is defined on $\mathcal{Q}_{\text{Kin}}$ by the following formula
\begin{equation}
\ttr_{P^0_-}(\Dbf Q):=\ttr(|\Dbf|^{1/2}(Q^{++}-Q^{--})|\Dbf|^{1/2}).
\end{equation}
It coincides with $\ttr(\Dbf Q)$ when $\Dbf Q$ is trace-class. We will work in the subset of this space defined by:
\begin{equation}
\mathcal{K}:=\{Q\in \mathcal{Q}_{\text{Kin}},\ -P^0_-\le Q\le P^0_+\}\subset\big\{Q\in \mathcal{Q}_{\text{Kin}},\ Q^*=Q\big\},
\end{equation}
the closed convex hull (under that norm) of the difference of two orthogonal projectors: $P-P^0_-$.

We also define $\mathbf{Q}_1$ the Hilbert space of $Q(x,y)\in L^2(\mathbf{R}^3\times \mathbf{R}^3,\mathbf{C}^4)$ such that
\begin{equation}
\lVert Q\rVert_{\mathbf{Q}_1}^2:= \diint (\Ebf{p}+\Ebf{q})|\wh{Q}(p,q)|^2dpdq<+\infty.
\end{equation}

\noindent The definition of the density $\rho_Q$ must coincide with the usual one when $Q$ is (locally) trace-class and $\rho_Q$ must be of finite Coulomb norm: $\ncc{\rho_Q}<+\infty$. For $Q$ in $\mathfrak{S}_1^{P^0_-}$, $\rho_Q$ is defined by duality:
\begin{equation}
\forall\ V\in\mathcal{C}',\ QV\in\mathfrak{S}_1^{P^0_-}\ \mathrm{and}\ \ttr_{P^0_-}(QV)=\psh{V}{\rho_Q}_{\mathcal{C}'\times \mathcal{C}}.
\end{equation} 
We have the following proposition (proved in Appendix \ref{no_bind_appA}). % or Lemma. (mettre le lemme en appendice)
\begin{proposition}\label{no_bind_conti}
The map $Q\in \mathfrak{S}_1^{P^0_-}\mapsto \rho_Q\in \mathcal{C}$ is continuous and:
\begin{equation}
\begin{array}{rl}
\ncc{\rho_Q}&\apprle \ns{1}{|D_0|^{\ala} Q^{++} |D_0|^{\ala} }+\ns{1}{|D_0|^{\ala} Q^{--} |D_0|^{\ala} }\\
 &\quad\quad\quad\quad+\sqrt{\llo}\ns{2}{|D_0|^{\ala}Q}.
\end{array}
\end{equation}
\end{proposition}

%%%%%%% with norms 

\noindent Thanks to Kato's inequality \eqref{no_bind_kato}, the exchange term is well-defined \cite{stab}
\begin{equation}
\begin{array}{rl}
\dfrac{2}{\pi}\diint \frac{|Q(x,y)|^2}{|x-y|}dxdy&\le \ttr(|D_0|Q^2)=\ttr\{|D_0|^{1/2}Q^2|D_0|^{1/2}\}\\
    \mathrm{and\ for}\ Q\in\mathcal{K}:      &\le \ttr \{  |D_0|^{1/2}(Q^{++}-Q^{--})|D_0|^{1/2}\}\le \ttr_{P^0_-}(\Dbf Q),
\end{array}
\end{equation}
\noindent The BDF energy is defined as follows:
\begin{equation}\label{no_bind_bdf_2}
\mathcal{E}^{\nu}_{\text{BDF}}(Q):=\ttr_{P^0_-}(\Dbf Q)-\alpha D(\nu,\rho_Q)+ \frac{\alpha}{2}\Big(D(\rho_Q,\rho_Q)-\diint\frac{|Q(x,y)|^2}{|x-y|}dxdy\Big),\ Q\in\mathcal{K}.
\end{equation}

Any charge sector $\mathcal{Q}(q):=\{Q\in\mathcal{K},\ \ttr_{P^0_-}(Q)=q\}$ leads to a variational problem
\begin{equation}\label{no_bind_BDFq}
E^{\nu}_{\mathrm{BDF}}(q):=\underset{Q\in \mathcal{Q}(q)}{\inf}\mathcal{E}_{\mathrm{BDF}}(Q).
\end{equation}
By Lieb's variational principle \cite[Proposition 3]{at}, a minimizer $Q$ for $E^\nu(M)$ with $M\in\mathbb{Z}$ is necessarily a difference of two projectors $P-P^0_-$.

\subsection{Form of a minimizer}\label{no_bind_form_of_minim}
%\begin{center}
% \textsc{Form of a minimizer}
%\end{center}
To simplify, from this point we assume that $\nu=0$. For an integer $M\in\mathbb{N}$, let $Q$ be a ground state for $E^0(M)$, then necessarily $Q=\ov{P}-P^0_-$, where $\ov{P}$ is an orthogonal projector. 

The study of the first and second derivative gives more information: we have $\big[D_Q,\ov{P}\big]=0$, and \cite[Proposition 1]{at}
\begin{equation}\label{no_bind_form_no}%[
 \ov{P}=\chi_{(-\infty,\mu]}\big(D_Q\big),\ 0<\mu <1,
\end{equation}%)
where we recall the mean-field operator is defined in \eqref{no_bind_m-f}. We decompose $Q$ with respect to the positive and negative spectrum:
\begin{equation}\label{no_bind_form_no_1}%[
 N:=\chi_{(0,\mu]}(D_Q)\text{\ and\ }\pvacno=\g+P^0_-:=\chi_{(-\infty,0)}(D_Q),
\end{equation}%)
where $\pvacno$ (resp. $n$) is interpreted as the polarized vacuum (resp. as the real electrons). If $\alpha M$ is small enough, then we can show that $\ttr_{P^0_-}(\g)=0$ and thus $N$ has rank $M$ \cite{at,sokd}. We will recall the proof below.

In the present case, a minimizer for $E^0(2)$ can be written as in \eqref{no_bind_form_no}-\eqref{no_bind_form_no_1}. For small enough $\alpha$, we have
\begin{equation}\label{no_bind_form_no_2}
 N=\ket{\psi_1}\bra{\psi_1}+\ket{\psi_2}\bra{\psi_2},\ D_Q\psi_j=\mu_j\psi_j,\ 0<\mu_2\le \mu_1=\mu<1,\ j\in\{1,2\}.
\end{equation}

These equations constitutes the starting point of our proof: they enable us to get estimates on the Sobolev norms of the $\psi_j$'s. More precisely we will prove Lemma \ref{no_bind_estim_minim}.

Before stating it, let us recall the Pekar-Tomasevitch functional:
\[
 \mathcal{E}_{\text{PT}}(\psi):=\nlp{2}{\nabla \psi}^2-\diint \frac{|\psi(x)|^2|\psi(y)|^2}{|x-y|}dxdy,\ \forall\,\psi\in H^1.
\]
It describes the energy of a single electron in its own hole. In the case of $M$ electrons, the energy is \cite{bip}:
\begin{equation}\label{no_bind_pekar_def}
 \forall\,0\le \G\le 1,\ \ttr\,\G=M,\ \mathcal{E}_{\text{PT}}^U(\G):=\ttr\big(-\Delta\big)-\ncc{\rho_\G}^2+U\Big(\ncc{\rho_\G}^2-\nqq{\G}^2 \Big),
\end{equation}
where $U>0$ is some number. By scaling we can assume $U=1$ but $-\ncc{\rho_\G}^2$ has to be replaced by $U^{-1}$: this last number measures the strength of the polarisation.

In this paper, a specific value $U=U_0(\alpha,\La)$ is considered: $U_0^{-1}=1-Z_3(\alpha,\La)$ where $Z_3$ is the renormalisation constant that we have mentionned in the introduction. Its precise expression is given below \eqref{no_bind_def_Z_3}. %partie densité

We write $E_{\text{PT}}^U(M)$ the infimum of the Pekar-Tomasevitch energy on the set $\{0\le \G\le 1,\ \ttr\,\G=M\}$, with $U=U_0$.
\begin{remark}\label{no_bind_no_minim_PT}
We assume that $U_0> 2U_{\text{c}}$, where $U_{\text{c}}$ is the critical value above which, there is no minimizer for $E_{\text{PT}}^U(M)$ for any integer $M\ge 2$. This important result is proved in \cite{bip}.

For unitary wave functions $\phi_1\perp\phi_2$, we also write
\[
 \mathcal{E}_{\text{PT}}^U(\phi_1\wedge \phi_2):=\mathcal{E}^U_{\text{PT}}\Big(\sum_{j=1}^2\ket{\phi_j}\bra{\phi_j}\Big).
\]
\end{remark}

\begin{lemma}\label{no_bind_estim_minim}
In the regime of Remark \ref{no_bind_regime}, let $Q=N+\g$ be a minimizer for $E^0(2)$, decomposed as in \eqref{no_bind_form_no}-\eqref{no_bind_form_no_2}. %For each $j\in\{1,2\}$, we split $\psi_j$ into an upper spinor $\ph_j$ and a lower one $\chi_j$, both in $L^2(\RR,\mathbb{C}^2)$.
 
Let $c$ be $\big\{ \alpha(1-Z_3(\alpha,\La))\big\}^{-1}$ where $Z_3$ is defined in \eqref{no_bind_def_Z_3}. We write $\un{\psi_j}$ the scaling of $\psi_j$ by $c$:
\[
 \un{\psi_j}(x):=c^{3/2}\psi_j(cx),\ x\in\RR,
\]
Then we have the following:
 \begin{equation}
  \left\{
    \begin{array}{rcl}
     E^0_{\text{BDF}}(1)&=&1+\tfrac{1}{2c^2}E_{\text{PT}}(1)+\mathcal{O}(\alpha c^{-2}),\\
     E^0_{\text{BDF}}(2)=\mathcal{E}^0_{\text{BDF}}(Q)&=&2+\tfrac{1}{2c^2}\mathcal{E}_{\text{PT},U_0}(\un{\psi_1}\wedge \un{\psi_2})+\mathcal{O}(\alpha c^{-2}).
    \end{array}
  \right.
 \end{equation}
We split each $\psi_j$ into an upper spinor $\ph_j$ and a lower one $\chi_j$, both in $L^2(\RR,\mathbb{C}^2)$. We write $n_j=|\psi_j|^2$ (resp $\un{n_j}=|\un{\psi_j}|^2$) and $n=n_1+n_2$ (resp $\un{n}=\un{n_1}+\un{n_2}$). Then we have
 \begin{equation}\label{no_bind_mumu}
\mu_j=1+\frac{\nlp{2}{\nabla \un{\varphi_j}}^2}{2c^2}-\frac{1}{c^2}D(\un{n_j},\un{n})+\mathcal{O}(\alpha c^{-2}),
\end{equation}
in particular:
\begin{equation}\label{no_bind_away_mu}
(1-\mu_j)c^2\apprge 1.
\end{equation}
\end{lemma}

Estimate \eqref{no_bind_away_mu} follows from \eqref{no_bind_psipsi}-\eqref{no_bind_psipsi_1}. This quantitative error $\mathcal{O}(\alpha c^{-2})$ gives \emph{a priori} information about the $\un{\psi_j}$'s thanks to \cite{L,bip} (see the next Section).

\begin{notation}\label{no_bind_gen_nota}
 Throughout this paper, we will use the following notations. 
 \begin{equation}
    \begin{array}{rcl| rcl}
     N_j&=&\ket{\psi_j}\bra{\psi_j}& N&=&N_1+N_2,\\
     n_j&=&|\psi_j|^2 & n&=&n_1+n_2,\\
     \g'=Q&=&=\g+N,&\rho'_\g&=&\rho_\g+n.
    \end{array}
 \end{equation}
When we add an underline $\un{N_j}$ etc. we mean the scaled object by $c=(\alpha(1-Z_3))^{-1}$. Writing
\[
 O_c:\phi(x)\in L^2\mapsto c^{3/2} \phi(cx),
\]
we have $\un{\psi_j}=O_c \psi_j$, $\un{N_j}:=O_c N_j O_c^{-1}$, $\un{\g}=O_c \g O_c^{-1}$.
\end{notation}

%%%%%%%%%%%%%%%%%%%%%%%%%%%%%%%%%%%%%%%%%%%%%%%%%%%%%%%%%%%%%%%%%%%%%%%%%%%%%%%%%%%%%%%%%%%%%%%%%%%%%%%%%%%%%%%%%ù

\section{The Pekar-Tomasevitch functional}\label{no_bind_the_PT_fun}

\subsection{Decoupling of almost minimizers for $E_{\text{PT}}^{U_0}(2)$}

Thanks to \cite{L}, one knows that there exists but one minimizer for $E_{\text{PT}}(1)$ up to a phase and to translation in $L^2(\mathbb{R}^3,\mathbb{C})$. This minimizer can be chosen positive radially symmetric and decreasing. It is also smooth and with exponential falloff. As $\int \big|\nabla |\phi|\big|^2\le \int |\nabla \phi|^2$ \cite{LL}, there holds the same in $L^2(\mathbb{R}^3,\mathbb{C}^4)$. The set of minimizers is a manifold $\mathscr{P}\simeq \mathbb{S}^7\times\mathbb{R}^3$ where $\mathbb{S}^7$ is the unit sphere of $\mathbb{C}^4$. There also holds coercivity inequality \cite{lenz}:
\begin{proposition}\label{no_bind_coer}
Let $\phi \in H^1$ with $\nlp{2}{\phi}=1$ and let $\ov{\phi}\in\mathscr{P}$ such that: 

$\nso{1}{\phi-\ov{\phi}}=\underset{f\in \mathscr{P}}{\inf}\nso{1}{\phi-f}$, then there exists 
$\kappa>0$ such that (at least in a neighborhood of $\mathscr{P}$):
\[
\mathcal{E}_{\text{PT}}(\phi)-E_{\text{PT}}(1)\ge \kappa \nso{1}{\phi-\ov{\phi}}^2.
\]
\end{proposition}

\begin{notation}
 We write $\mathscr{P}_0\subset \mathscr{P}$ the submanifold of $\mathscr{P}$ made of minimizers with center $0\in\RR$: it is isomorphic to $\mathbb{S}^7$.
\end{notation}

We are interested in $E_{\text{PT}}^U(2)$, with $U=U_0> 2U_{\text{c}}$, where $U_{\text{c}}$ is the critical value above which there is no mminimizers for $E_{\text{PT}}(2)$ \cite{bip}: in particular $E_{\text{PT}}(2)=2E_{\text{PT}}(1)$ (the proof of \cite{bip} also applies for spinor-valued functions). If we \emph{choose} $U_0> 2\text{U}_c$:
\begin{equation}
\forall \Psi\in L^2_a(\mathbb{R}^3\times\mathbb{R}^3),\,\nlp{2}{\Psi}=1:\,\mathcal{E}_{\text{PT}}(\Psi)-2E_{\text{PT}}(1)\ge \frac{U_0}{2}(D(\rho_\Psi,\rho_\Psi)-\ttr(\g_\Psi R[\g_\psi]))
\end{equation}
where we recall $\rho_\Psi$ is the density of $\Psi$ and $\g_\Psi$ is its one-body density matrix.

There holds Lieb's variational principle: $E_{\text{PT}}^U(2)$ is also the infimum of $\mathcal{E}_{\text{PT}}^U$ over Slater determinant $h_1\wedge h_2$ with $h_j\in H^1$ and $\psh{h_j}{h_k}=\delta_{jk}$. 

Let us consider such a state $\Psi=h_1\wedge h_2$. The plane $\text{Span}(h_1,h_2)$ can be defined with other orthonormal families: $\mathbf{U}(2)$ acts on the set $\text{S}[\Psi]$ of those families:
\begin{equation}\label{no_bind_action}
\Big(\begin{pmatrix} a & c \\ b & d \end{pmatrix},\begin{pmatrix}h_1\\ h_2\end{pmatrix}\Big)\in \mathbf{U}(2)\times \text{S}[\Psi]\mapsto \begin{pmatrix}ah_1+bh_2 \\ ch_1+dh_2\end{pmatrix}\in\text{S}[\Psi],
\end{equation}
The first vector is written $(\mathbf{m}\cdot \mathbf{h})_1$ and the second is written $(\mathbf{m}\cdot \mathbf{h})_2$.

\paragraph{Characteristic length}
For $\Psi=h_1\wedge h_2$ we define the inverse $d_\Psi$ of the \emph{characteristic length} $R_{12}(\Psi)$:
\begin{equation}
d_\Psi:=\underset{\mathbf{m}\in \mathbf{SU}(2)}{\inf}D(|(\mathbf{m}\cdot \mathbf{h})_1|^2,|(\mathbf{m}\cdot \mathbf{h})_2|^2)=R_{12}(\Psi)^{-1}.
\end{equation}
%The characteristic length is written $R_{12}(\Psi)$: $R_{12}(\Psi):=d_{\Psi}^{-1}$.
%\begin{remark}
%In fact we can only consider the $\mathbf{m}$'s in $\mathbf{SU}(2)\simeq \mathbb{S}^3$ (unit sphere of $\mathbb{R}^4$).
%\end{remark}

Let $\phi_0\in\mathscr{P}_0$ be the radially symmetric and positive function (with $\phi_0(x)$ parallel to $\begin{pmatrix} 1 & 0 & 0 & 0\end{pmatrix}^*$ for instance). Let $\phi_{\mathbf{x}_0}=\tau_{\mathbf{x}_0}\phi_0$ be its translation by $\mathbf{x}_0\in\mathbb{R}^3$. We have:
\begin{equation}
\forall x_0,\, |\mathbf{x}_0|\ge 1:\ |\mathbf{x}_0|\times D(|\phi_0|^2,|\phi_{\mathbf{x}_0}|^2)\le \underset{|\mathbf{z}|\ge 1}{\sup}|\mathbf{z}|\sqrt{\diint \frac{|\phi_0(x)|^2|\phi_{\mathbf{z}}(y)|^2}{|x-y|^2}dxdy}:=\text{Y}_0<+\infty.
\end{equation}

\paragraph{Geometric length} For a Slater determinant $\Psi=h_1\wedge h_2$ where $h_1$ and $h_2$ satisfy $D(|h_1|^2,|h_2|^2)=d_\Psi$, we define the \emph{geometric length} $\rgg$ as follows.

 Let $\phi_{(j)}\in\mathscr{P}$ be the closest function of $\mathscr{P}$ to $h_j$ in $H^1$. Each $\phi_{(j)}$ is radial with respect to some vector $z_j\in\mathbb{R}^3$, we set $\rgg(\Psi):=|z_1-z_2|$ (or the smallest of such $|z_1-z_2|$): it should be seen as the \emph{interparticle distance}. 
\begin{remark}
The geometric length $\rgg$ does not appear in the energy and $R_{12}=d_{\Psi}^{-1}$ may be much smaller.
\end{remark}

\begin{proposition}\label{no_bind_eloig}
%Let $\mathcal{E}_{\text{PT}}^U$ be the PT functional with parameter $U>2\text{U}_c$ where $\text{U}_c$ is the critical value for the existence of binding \cite{bip}. 
There exist $a_0>0$ and $\text{b}=\text{b}(a_0)>0$ such that
\begin{equation}\label{no_bind_ptd}
\forall \Psi=h_1\wedge h_2:\ \Delta_2\mathcal{E}=\mathcal{E}_{\text{PT}}^U(\Psi)-2E_{\text{PT}}(1)<a_0\ \Rightarrow\,\frac{\Delta_2 \mathcal{E}}{d_\Psi} \ge \text{b}.
\end{equation}
\end{proposition}

\begin{proposition}\label{no_bind_eloig1}
There exist $a_0'>0$ and $\text{b}'>0$ such that:
\begin{equation}
\forall \Psi=h_1\wedge h_2:\ \Delta_2\mathcal{E}<a_0'\ \Rightarrow\,\diint \frac{|\Psi(x,y)|^2}{|x-y|}dxdy\ge \frac{\text{b}'}{\rgg}.
\end{equation}
More precisely:

For any $0<\la$ let $B_j^{\la}$ be $B(z_j, \la \rgg)$ and $\mathcal{B}^{\la}:=B_1^{\la}\times B_2^{\la}\cup B_2^{\la}\cup B_1^{\la}$. Then there exist $a_\la>0,\text{k}_\la>0$ such that
\begin{equation}\label{no_bind_eloig12}
\forall \Psi=h_1\wedge h_2:\ \Delta_2\mathcal{E}<a_\la\ \Rightarrow\,\underset{(x,y)\in \mathcal{B}^{\la}}{\diint} \frac{|\Psi(x,y)|^2}{|x-y|}dxdy\ge \frac{\text{k}_\la}{\rgg}
\end{equation}
\end{proposition}

\begin{remark}
It is not possible to replace $\rgg^{-1}$ by $d_{\Psi}$.
\end{remark}

To prove Proposition \ref{no_bind_eloig1}, we need to compare $R_{12}(\Psi)$ and $\rgg$.

\subsection{On the relation between $R_{12}(\Psi)$ and $\rgg$}

Let us consider an almost minimizer for $E_{\text{PT}}^U(2)$:
\begin{equation}\label{no_bind_as_above}
\Psi=h_1\wedge h_2,\ \mathcal{E}_{\text{PT}}^U(2)-E_{\text{PT}}^U(2)\apprle a_0\ll 1, U\text{\ big\ enough}.
\end{equation}
We suppose that $D(|h_1|^2,|h_2|^2)=d_{\Psi}$ and write $\phi_j$ the closest function to $h_j$ in $\mathscr{P}$. We write $\delta_j=h_j-\phi_j$. By Propositions \ref{no_bind_coer} and \ref{no_bind_eloig} we have:
\[
\begin{array}{rl c rl}
d_{\Psi}=\frac{1}{R_{12}}&\apprle \eps_0&\mathrm{and}& \nso{1}{\delta_1}^2+\nso{1}{\delta_2}^2\apprle a_0.
\end{array}
\]
We will here compare $R_{12}$ and $\rgg$ (defined as $|z_1-z_2|$ where $z_j$ is the center of $\phi_j$).

As $\phi_j(\cdot-z_j)$ is radial and smooth then:
\begin{equation}
0<\underset{x\in\mathbb{R}^3}{\inf}\frac{(|\phi_j|^2*\tfrac{1}{|\cdot|})(x)}{\big( (|\phi_j|^2*\tfrac{1}{|\cdot|^2})(x)\big)^{1/2}}\le \underset{x\in\mathbb{R}^3}{\sup}\frac{(|\phi_j|^2*\tfrac{1}{|\cdot|})(x)}{\big( (|\phi_j|^2*\tfrac{1}{|\cdot|^2})(x)\big)^{1/2}}<+\infty.
\end{equation}
By Newton's Theorem \cite{LL}, writing $|\phi_0|^2=|\phi_j(\cdot-z_j)|^2$ we have:
\begin{equation}
\forall\,x\in\mathbb{R}^3,\ (|\phi_0|^2*\tfrac{1}{|\cdot|})(x)=\frac{1}{|x|}\dint_{|y|\le |x|}|\phi_0(y)|dy+\dint_{|y|\ge |x|}\frac{|\phi_0(y)|^2}{|y|}dy\le \frac{1}{|x|}.
\end{equation}
As a consequence, for sufficiently small $a_0$:
\begin{equation}
\begin{array}{rl  rl}
|D(\mathfrak{Re}(\delta_1^*\phi_1),|\delta_2|^2)|&\apprle \nlp{2}{\delta_1}D(|\phi_1|^2,|\delta_2|^2),& |D(\mathfrak{Re}(\delta_1^*\phi_1),|\phi_2|^2)|&\apprle \frac{\nlp{2}{\delta_1}}{\rgg},
\end{array}
\end{equation}
where we used Cauchy-Schwarz inequality: 

\[
 \int_x |\delta_1(x)^*\phi_1(x)|\tfrac{dx}{|x-y|}\le \nlp{2}{\delta_1}\{ \int_x |\phi_1(x)|^2\tfrac{dx}{|x-y|^2}\}^{1/2}.
\]
Thus there holds the following.

\begin{proposition}\label{no_bind_RR}
Let $\Psi$ be as in \eqref{no_bind_as_above}. We write $\lVert \delta\rVert=\sum_j \lVert \delta_j\rVert$: there exists $\kappa>0$ such that for sufficiently small $a_0>0$:
\begin{equation}
\begin{array}{l}
d_\Psi\ge (1-\kappa \sqrt{a_0})\big(D(|\phi_1|^2,|\phi_2|^2)+D(|\delta_1|^2,|\phi_2|^2)+D(|\phi_1|^2,|\delta_2|^2)\big)+D(|\delta_1|^2,|\delta_2|^2),\\
\iint \frac{|h_1(x)|^2|h_2(y)|^2}{|x-y|^2}dxdy\apprle \dfrac{1}{\rgg^2}+\dfrac{\nlp{2}{\delta}\nso{1}{\delta}}{\rgg}+\nlp{2}{\delta}^2\nso{1}{\delta},
\end{array}
\end{equation}
\end{proposition}
 %The second point is proven with the same methods.
\begin{remark}\label{no_bind_remRR}
In particular $R_{12}=\mathcal{O}(\rgg)$. Moreover for sufficiently small $a_0$, we have
\[
\Delta_1 \mathcal{E}:=\sum_j\big(\mathcal{E}_{\text{PT}}(h_j)-E_{\text{PT}}(1)\big)=\Theta(\nso{1}{\delta}^2).
\] 

With the help of Proposition \ref{no_bind_eloig}, we get the following estimates:
\begin{equation}
\diint \frac{|h_1(x)|^2|h_2(y)|^2}{|x-y|^2}dxdy\apprle a_0^3.
\end{equation}
\end{remark}

\subsection{On the decomposition of \underline{$\psi_1$} $\wedge$  \underline{$\psi_2$}}
In our problem, we consider a couple $(a_0,\text{b})$ described in Lemma \ref{no_bind_eloig}, and we \emph{choose} $(\alpha,\La)$ such that $U_0\ge (2+1)U_{\text{c}}$. 

We consider $\un{\Psi}=\un{\psi_1}\wedge \un{\psi_2}$ of Lemma \ref{no_bind_estim_minim}. We have: $\mathcal{E}_{\text{PT}}^U(\un{\psi_1}\wedge \un{\psi_2})\apprle \alpha$ and $d_{\Psi}\apprle \alpha$. 

This result and the estimate of Remark \ref{no_bind_remRR} lead to the following Lemma.

\begin{lemma}\label{no_bind_lem_on_decomp}
For $(k,k')=(1,2)$ or $(2,1)$ and $\psi_k(x)=c^{-3/2}\un{\psi_k}(x/c)$, we have
\[
\nlp{2}{|\psi_{k'}|^2*\tfrac{1}{|\cdot|}\times \psi_k-(\psi_{k'}^*\psi_{k})*\tfrac{1}{|\cdot|}\times \psi_{k'}}^2\apprle \frac{1}{c^2}\diint \frac{|h_1(x)|^2|h_2(y)|^2}{|x-y|^2}dxdy\apprle \frac{\alpha^3}{c^2}.
\]
\end{lemma}
\begin{dem}
Indeed the quantity in the l.h.s. of \eqref{no_bind_lem_on_decomp} corresponds to the squared $L^2$-norm of $(\rho_\Psi*\tfrac{1}{|\cdot|} \psi_k-R[\g_\Psi]\psi_k)$ where $\Psi:=\psi_1\wedge \psi_2$. Then we decompose $\psi_k$ with respect to an orthonormal family $(h_1,h_2)$ with $h_1\wedge h_2=\Psi$ and $D(|h_1|^2,|h_2|^2)=d_{\Psi}$.
\end{dem}

We recall that $\psi_1$ and $\psi_2$ are eigenvectors of the mean-field operator with eigenvalues $\mu_1$ and $\mu_2$. In the case $\mu_1\neq \mu_2$ we cannot choose $\un{\psi_1}=h_1$ and $\un{\psi_2}=h_2$. 

From the estimation of the $\mu_j$'s \eqref{no_bind_mumu} we may ask whether the quantity%lplplp
\begin{equation}\label{no_bind_psipsi}
F_\mathcal{E}(\un{\psi_k}):=\mathcal{E}_{\text{PT}}(\un{\psi_k})-D(|\un{\psi_k}|^2,|\un{\psi_{k'}}|^2)
\end{equation}
is negative and away from $0$ or not. As $h_k=\phi_k+\delta_k$ with $\phi_k\in\mathscr{P}$ and $\nso{1}{\delta_k}=\mathcal{O}(\sqrt{\Delta_2\mathcal{E}})$ a simple computation shows that:
\begin{equation}\label{no_bind_psipsi_1}
\forall (a,b)\in \mathbb{C}^2\cap\mathbb{S}^3:\ F_\mathcal{E}(ah_1+bh_2)=\frac{3}{2}E_{\text{PT}}(1)+\mathcal{O}((\Delta_2\mathcal{E})^{1/4}).
\end{equation}

%Starting from $\begin{pmatrix} h_1\\ h_2\end{pmatrix}$ we apply elements of $\mathbf{U}(2)$. 

%%%%%%%%%%%%%%%%%%%%%%%%%%%%%%%%%%%%%%%%%%%%%%%%%%%%%%%%%%%%%%%%%%%%%%%%%%%%%%%%%%%%%%%%%%%%%%%%%%%%%%%%%%%%%%%%%%%%%%%%%%%%%%%%%%%%%%%%%%%%%%%%%%%%%%%%%%%%%%%%%%%%%%ù

\section{Technical tools}\label{no_bind_techn}

%In the next section we recall an important technical ingredient: the Cauchy expansion.
%beberk

%\begin{remark}
%There lies a contradiction for there always hold large binding inequalities as shown in \cite{at,these}:
%\[
%\forall q\in\mathbb{R},\forall k\in\mathbb{R}:\ E_{\text{BDF}}(q)\le E_{\text{BDF}}(q-k)+E_{\text{BDF}}(k).
%\]
%\end{remark}

\subsection{The Cauchy expansion}\label{no_bind_the_cauchy}
%\begin{center}
% \textsc{The Cauchy expansion}
%\end{center}
In this part we use the functions $\sbf{\cdot}$, $E(\cdot)$ and $\Ebf{\cdot}$ and numbers $\eps_\la,\ala$ defined in Notation \ref{no_bind_sbf}. We recall Ineq. \eqref{no_bind_trick_ala}. The results stated here follow from \cite{sok,sokd}.

Let $\wt{\g}$ be the operator defined by:
\[
 \wt{\g}=\chi_{(-\infty,0)}(\Dbf +\alpha (v_{\wt{\rho}}-R_{\wt{Q}}))-P^0_-,\ (\wt{Q},\wt{\rho})\in \mathbf{Q}_1\times \mathcal{C}.
\]
For instance we can take $\g$ of \eqref{no_bind_form_no_1}. Provided that $\nqkin{\wt{Q}},\ncc{\wt{\rho}}$ are small enough, by Lemma \ref{no_bind_BB} we have%minimizer for $E^0(M)$
\[
 |D+\alpha (v_{\wt{\rho}}-R_{\wt{Q}})|\ge |\Dbf|\big(1-\alpha(\ncc{\rho_Q}+\nqq{Q})\big)=|\Dbf|(1+o(1)).
\]
As a result we can expand $\wt{g}$ in power of $\alpha$, this is the Cauchy expansion \cite{ptf}: 
\begin{equation}\label{no_bind_cauchy_no}
 \left\{
  \begin{array}{rcl}
   \wt{\g}&=&\ssum_{j=1}^{+\infty} \alpha^j Q_j\big[\wt{Q},\wt{\rho} \big],\\
   Q_j\big[\wt{Q},\wt{\rho} \big]&:=&-\frac{1}{2\pi}\dint_{-\infty}^{\infty}\frac{d\om}{\Dbf+i\om}\Big(\big(R_{\wt{Q}}-v[\wt{\rho}]\big)\frac{1}{\Dbf+i\om}\Big)^j.
  \end{array}
 \right.
\end{equation}
We can further expand each $Q_j$ into $\sum_{j=0}^j Q_{k,j-k}\big[\wt{Q},\rho_{\wt{Q}}\big]$ where each $Q_{k,j-k}$ is polynomial in $R_{\wt{Q}}$ (resp. $v[\rho_{\wt{Q}}]$) of degree $k$ (resp. $j-k$). 

The respective densities of $Q_{k,j-k}$ and $Q_j$ are written $\rho_{k,j-k}$ and $\rho_j$.

\paragraph{Convergence of the series \eqref{no_bind_cauchy_no}} In \cite{ptf,gs}, Hainzl \emph{et al.} proved that this series is well-defined and in \cite{sok,sokd} the functions $(Q_{k,j-k},\rho_{k,j-k})[\cdot,\cdot]$ are studied in several norms.

It is possible to adapt the proofs to show that these functions are multilinear continuous in $\mathbf{Q}_1\times \mathcal{C}$ or more generally in the banach spaces $\mathcal{X}_w=\mathbf{Q}_w\times \mathfrak{C}_w$, defined by the following norms:
\begin{equation}
 \nqbfg{Q}^2:=\diint (\Ebf{p}+\Ebf{q})w(p-q)|\wh{Q}(p,q)|^2dpdq\text{\ and\ }\ncg{\rho}^2:=\diint \frac{w(k)}{|k|^2}|\wh{\rho}(k)|^2dk,
\end{equation}%(
where $\sqrt{w}:\RR\to [1,+\infty)$ is a weight function satisfying some sub-additive assumptions. 

Furthermore the growth of the norms $\lVert(Q_{k,j-k},\rho_{k,j-k}) \rVert_{\mathcal{B}(\mathcal{X}_w)}$ is also polynomial: it follows that there exists some radius $A(\alpha,\La,w)$ such that
\[%]
 (\wt{Q},\wt{\rho})\in B_{\mathcal{X}_w}(0,A)\mapsto \Big( \wt{\g}:=\sum_{j=1}^{+\infty}\alpha^j Q_j\big[\wt{Q},\wt{\rho}\big],\rho_{\wt{\g}}\Big)\in B_{\mathcal{X}_g}(0,A),
\]
is well-defined and contractant.

The main ingredients of the proof are the following inequalities:
\begin{equation}\label{no_bind_main_ingred}
 \begin{array}{rcl|rcl}
  \lVert  P^0_{\pm} v_{\wt{\rho}} P^0_{\mp}\frac{1}{|D_0|^{\ala}}\rVert_{\mathfrak{S}_2}&\apprle& \sqrt{\llo}\ncc{\wt{\rho}} & \lVert R_{\wt{Q}} \frac{1}{|\nabla|^{1/2}}\rVert_{\mathfrak{S}_2}&\apprle & \nqq{\wt{Q}},\\
  \lVert v_{\wt{\rho}}\frac{1}{|D_0|^{\ala}}\rVert_{\mathfrak{S}_6}&\apprle& \ncc{\wt{\rho}} & \lVert v_{\wt{\rho}}\tfrac{1}{|\nabla|^{1/2}}\rVert_{\mathcal{B}}&\apprle& \ncc{\wt{\rho}}
 \end{array}
\end{equation}
In the l.h.s. the first estimate follows from a simple computation in Fourier space \cite{ptf,sok}, and the second one is an application of the KSS inequality \eqref{no_bind_kss}.

In the r.h.s. the first is proved below (Lemma \ref{no_bind_BB}) and the last follows from an homogeneous Sobolev inequality \eqref{no_bind_sob}. We will say no more about these results and refer the reader to the cited articles and to \cite{these}.

\subsection{On the minimizers: equation and density}\label{no_bind_eq_dens}

The results of this part are proved in \cite{sokd}.

Let $Q=\g+N$ be a minimizer for $E^0(M)$ with $M\in\{1,2\}$. It satisfies Eq. \eqref{no_bind_form_no}-\eqref{no_bind_form_no_1} and $\text{rank}\,N=M$ for $\alpha$ sufficiently small. We recall:
\begin{equation}\label{no_bind_eq_gamma}
 \g=\chi_{(-\infty,0)}(D_Q)-P^0_-.
\end{equation}
In \cite{ptf,sok,sokd}, a fixed-point scheme is used to see $\g$ as a fixed point of some function $F^{(1)}$ (with parameter $N$). This scheme enables us to get estimates on $\g$ and $N$. By the Cauchy expansion, Eq. \eqref{no_bind_eq_gamma} is rewritten as follows:
\[
 \big(\text{Id}-\alpha Q_{1,0}[\cdot]\big)\big[\g'\big]=N+\alpha Q_{0,1}\big[\rho_\g'\big]+\ssum_{j=2}^{+\infty}\alpha^jQ_j\big[\g',\rho'_\g\big].
\]
In \cite{sokd}, it is proved that the linear operator $\big(\text{Id}-\alpha Q_{1,0}[\cdot]\big)$ is a continuous endomorphism for $\mathbf{Q}_g$ and $\mathfrak{S}_p$ ($1\le p\le 2$) provided that $\alpha\llo\le L_0$ is small enough. 

Its inverse $\Tbf$ is written and it has a uniform bound for all those Banach spaces.

This gives
\begin{equation}
 \g=\alpha \Tbf[Q_{1,0}(N)]+\alpha \Tbf[Q_{0,1}(\rho'_\g)]+\ssum_{j=2}^{+\infty}\alpha^j\Tbf\big[Q_j[\g',\rho'_\g]\big]. 
\end{equation}

In \cite{sokd}, the density $\alpha\rho\big[Q_{0,1}(\rho'_\g)\big]$ is computed and we have:
\[
 \alpha\rho\big[Q_{0,1}(\rho'_\g)\big]=-\check{f}_\La*\rho'_\g,
\]
where $\check{f}_\La\in L^1$ with norm $\nlp{1}{\check{f}_\la}\apprle L$. 
\begin{remark}
For the \emph{smooth} cut-off, the same proof applies for $|\cdot|^\ell \check{f}_\La$. For any fixed integer $\ell$, there exists $K(\ell)>0$ such that, if $ \alpha\le K(\ell)$ then
\begin{equation}\label{no_bind_est_fla_no}
\begin{array}{rcl}
  \nlp{1}{\,|\cdot|^\ell \check{f}_\La}&\le& \Big\{\dint |x|^{2(1+\ell)}(1+|x|^2)|\check{f}_\La(x)|^2dx\dint\frac{dx}{|x|^2(1+|x|^2)} \Big\}^{1/2},\\
                                        &\apprle& \alpha.
\end{array}
\end{equation}
The same results hold for
\begin{equation}\label{no_bind_est_fla_no_1}
 \check{F}_\La:=\mathscr{F}^{-1}\Big(\frac{\fla}{1+\fla}\Big)=\ssum_{j=1}^{+\infty}(-1)^{j+1}\check{f}_\La^{*j}
\end{equation}
provided that $\alpha\le K'(\ell)$ with a smaller bound $K'(\ell)\le K(\ell)$.
\end{remark}

We write $\tau_j[\cdot]:=\rho\big[\Tbf Q_j[\cdot]\big]$ and $\tau_{k,j-k}[\cdot]:=\rho\big[\Tbf Q_{k,j-k}[\cdot]\big]$. There holds:
\begin{equation}\label{no_bind_form_dens}
\begin{array}{rcl}
 \rho_\g&=&-\check{F}_\La*n+(\delta_0-\check{F}_\La)*\big(\alpha \tau_{1,0}[N]+\ssum_{j=2}^{+\infty}\alpha^j \tau_j[\g',\rho'_\g]\big),\\
        &=&-\check{F}_\La*n+(\delta_0-\check{F}_\La)*\big(\alpha \tau_{1,0}[N]+\alpha^2\wt{\tau}_2[\g',\rho'_\g]\big).
\end{array}
\end{equation}
We have $\rho_\g\in L^1$ with $\int \rho_\g=-\Fla(0)\times M$. The renormalisation constant $Z_3$ is
\begin{equation}\label{no_bind_def_Z_3}
 Z_3:=1-\Fla(0)=\frac{1}{1+\fla(0)}\approx \frac{1}{1+\frac{2}{3\pi}\alpha\llo}\text{\ and\ }U_0:=\frac{1}{\Fla(0)}.
\end{equation}

We also recall \cite{sokd}
\begin{equation}\label{no_bind_Fla2}
\forall\,k,k'\in B_{\mathbb{R}^3}(0,2):\ |\Fla(k)-\Fla(k')|\apprle \alpha |k-k'|
\end{equation}
we will use below with $k'=0$.

%{
\subsection{Some inequalities}\label{no_bind_some_ineq}
\noindent -- Let us recall some Sobolev inequalities in $\mathbb{R}^3$:
\begin{equation}\label{no_bind_sob}
\begin{array}{l  l  l}
\nlp{6}{f}\apprle \nlp{2}{\nabla f},&
\nlp{4}{f}\apprle \nlp{2}{|\nabla|^{3/4}f},&
\nlp{3}{f}\apprle \nlp{2}{|\nabla|^{1/2}f}
\end{array}
\end{equation}

The last one gives $\nb{v_{\wt{\rho}} \tfrac{1}{|\nabla|^{1/2}}}\apprle \ncc{\wt{\rho}}$ for $\wt{\rho}\in\mathcal{C}$.

\noindent -- We also recall Kato's inequality and Hardy's inequality:
\begin{equation}\label{no_bind_kato}
\left\{
	\begin{array}{rcl}
		\dint_{\RR}\frac{|\phi(x)|^2}{|x|}dx&\le& \frac{\pi}{2}\psh{|\nabla|\phi}{\phi},\\
		\dint_{\RR}\frac{|\phi(x)|^2}{|x|^2}dx&\le&4\psh{(-\Delta)\phi}{\phi}.
	\end{array}
\right.
\end{equation}

\noindent -- The following Lemma gives estimates about the operator $R_Q$.
\begin{lemma}\label{no_bind_BB}
Let $Q(x,y)$ be an operator of finite exchange term and $\rho$ of finite Coulomb energy, then:
\[
\left\{
\begin{array}{rl}
\ns{2}{\tfrac{1}{|\nabla|^{1/2}}R_Q}&=\ttr(R_Q^*\tfrac{1}{|\nabla|} R_Q)\le \big( \dint \frac{dy}{|y|^2|y-\mathbf{e}|^2}\big)^2\ttr(Q^*R_Q),\\
\diint \frac{|Q(x,y)|^2}{|x-y|}dxdy&=\ttr(Q^*R_Q)\le \frac{\pi}{2(2\pi)^3}\diint |u||\wh{Q}(u+k/2,u-k/2)|^2dudk,\\
\nb{v_\rho \tfrac{1}{|\nabla|^{1/2}}}&\apprle \ncc{\rho}.
\end{array}
\right.
\]
\end{lemma}
In particular $\nlp{2}{(v_\rho-R_Q)f}\apprle (\ncc{\rho}+\nqq{Q})\nlp{2}{|\nabla|^{1/2} f}$.
\begin{dem}
The proof for $\ns{2}{\tfrac{1}{|\nabla|^{1/2}}R_Q}$ is just an application of the Cauchy-Schwarz inequality once we remark that $|\nabla|^{-1}$ is the convolution by $\text{Const}/|\cdot|^2$ \cite{LL}. For the last inequality we write $s=\tfrac{x+y}{2}$ and $t=x-y$ and $A(s,t):=Q(s+t/2,s-t/2)$ \emph{a.e.} By Kato's inequality:
\[
\begin{array}{rl}
\diint \frac{|Q(x,y)|^2}{|x-y|}dxdy&=\diint \frac{|A(s,t)|^2}{|t|}dsdt\\
         &\le \dfrac{\pi}{2}\dint ds \psh{|\nabla| A(s,\cdot)}{A(s,\cdot)}\\
      &\le \dfrac{\pi}{2}\diint |u| |\wh{Q}(u+k/2,u-k/2)|dudk.
\end{array}
\]
Those inequalities are true at least for $Q(x,y)$ in the Schwartz class $\mathcal{S}(\mathbb{R}^3\times \mathbb{R}^3)$, we conclude by density.
\end{dem}
%\end{enumerate}
%\begin{remark}\label{no_bind_important}
%Thanks to the estimate of $\tfrac{1}{|\nabla|^{1/2}}R_Q$, we can change estimates in the fixed point method of \cite{ptf,sokd} in $\mathcal{X}=\mathbf{Q}_1\times\mathcal{C}$, the natural Banach space of the problem. As a consequence, we get estimate of $\lvert\lvert F^{(1)}(Q,\rho) \rvert \rvert_{\mathcal{X}}$ in terms of $\ncc{\rho}$ and $\nqq{Q}$.
%\end{remark}

\noindent -- To end this part we give estimates about $\Dbf$.

%\[
%\sbf{p}:=\wh{\mathrm{sign}(D_0)}(p)=\wh{\tfrac{D_0}{|D_0|}}(p)=\frac{\beta+\boldsymbol{\alpha}\cdot p}{E(p)}.
%\]
We have
\[
\mathrm{Id}-\sbf{p}\sbf{q}=\sbf{p}(\sbf{p}-\sbf{q})=(\sbf{p}-\sbf{q})\sbf{q}
\]
and
\begin{equation}\label{no_bind_trick1}
|\mathrm{Id}-\sbf{p}\sbf{q}|\le |\sbf{p}-\sbf{q}|=\Big|\frac{\wh{D_0}(p)}{E(p)}-\frac{\wh{D_0}(p)}{E(q)}+\frac{\wh{D_0}(p)-\wh{D_0}(q)}{E(q)}\Big|\le \frac{2|p-q|}{\max(E(p),E(q))}.
\end{equation}

\begin{notation}
The symbol $\mathbf{e}$ will always stand for any unitary vector in $\mathbb{R}^3$.
\end{notation}

\begin{remark}
There holds (\emph{cf} \cite{LL} for the expression of $(a^2-\Delta)^{-1}$):
\[
\begin{array}{rl}
\dfrac{1}{|D_0|}(x-y)&=\dfrac{2}{\pi}\dint_0^{+\infty} \frac{d\omega}{|D_0|^2+\omega^2}(x-y)\\
                                 &=\sqrt{\dfrac{2}{\pi}}\dint_0^{+\infty}\frac{e^{-E_{\omega}|x-y|}}{|x-y|}d\omega\\
                                 &=\mathrm{Cnst}\frac{\mathrm{K}_1(|x-y|)}{|x-y|}
\end{array}
\]
where $\mathrm{K}_1$ is the modified Bessel function \cite{watson}.
\end{remark}

%%%%%%%%%%%%%%%%%%%%%%%%%%%%%%%%%%%%%%%%%%%%%%%%%%%%%%%%%%%%%%%%%%%%%%%%%%%%%%%%%%%%%%%%%%%%%%%%%%%%%%%%%%%%%%%%%%%%%%%%%%%%%%%%%%%%%%%%%%%%%%%%%%%%%%%%ù
%\section{Idea of the proof}

%%%%%%%%%%%%%%%%%%%%%%%%%%%%%%%%%%%%%%%%%%%%%%%%%%%%%%%%%%%%%%%%%%%%%%%%%%%%%%%%%%%%%%%%%%%%%%%%%%%%%%%%%%%%%%%%%%%%%%%%%%%%%%%%%%%%%%%%%%%%%%%%%%%%%%%%ù

%\section{$E^0(1)$ and $E^0(2)$}

\section{Proof of Proposition \ref{no_bind_estim_minim}}\label{no_bind_minimini}

%\subsection{$E^0(2)$}
%\begin{center}	
%	\textsc{\emph{A priori} estimates on a minimizer for $E^0(2)$}
%\end{center}
\subsection{\emph{A priori} estimates on a minimizer for $E^0(2)$}
This part is devoted to prove \eqref{no_bind_c-2}.

Let us say $\g'=\g+N$ is a minimizer for $E^0(2)$ written as in \eqref{no_bind_form_no}-\eqref{no_bind_form_no_1}. 

First we prove \eqref{no_bind_form_no_2}. There holds \emph{a priori} estimates \cite{sokd}: 
\[
\frac{1}{2}\ttr\big(\frac{-\Delta(1-\tfrac{\Delta}{\La^2})}{|D_0|}(\g')^2\big)+\frac{\alpha}{2}\ncc{\rho'_\g}^2\le \mathcal{E}(\g')-2+\frac{\alpha}{2}\ttr(\g'R[\g'])\le \frac{\alpha\pi}{4}\ttr(|\nabla|(\g')^2)
\] 
where we have used $|\Dbf|-1\ge \frac{1}{2}\frac{-\Delta\big(1-\tfrac{\Delta}{\La^2}\big)}{|D_0|}.$ It follows that:
\[
\ttr\big(\frac{-\Delta(1-\tfrac{\Delta}{\La^2})}{|D_0|}(\g')^2\big)+\alpha \ncc{\rho'_\g}^2\le K\alpha.
\]
As in \cite{sokd}, we can apply a fixed point scheme on $(\g,\rho_\g)$ with the help of the self-consistent equation (in $\mathbf{Q}_1\times\mathcal{C}$ for instance). This gives:
\[
\nqbfu{\g}\apprle \sqrt{L\alpha}\ncc{\rho'_\g}+\alpha \ns{2}{|\nabla|^{1/2} \g'}\text{\ and\ }\ncc{\rho_\g}\apprle L\ncc{\rho'_\g}+\sqrt{L\alpha}\ns{2}{|\nabla|^{1/2}\g'}.
\]
Hence $|\ttr_0(\g)|\le \ns{2}{\g}<1$ and $\ttr_0(\g)=0$ as shown in \cite{ptf}. This proves $\ttr(N)=\ttr_0(N)=\ttr_0(\g')-\ttr_0(\g)=2$.

Let $(\psi_i)_{1\le i\le 2}$ be a basis of orthonormal eigenvectors of $\chi_{0,\mu}(D_{\g'})$ with eigenvalues $0<\mu_1\le \mu_2<1$. We write $N_j:=\ket{\psi_j}\bra{\psi_j}$ and $|n_j:=\psi_j|^2$. From the equation satisfied by $\psi_j$
\begin{equation}\label{no_bind_equaj}
(\Dbf+\alpha (v[\rho_\g+n]-R[\g+N]))\psi_j=\mu_j\psi_j
\end{equation}
 we get the following.

\begin{lemma}\label{no_bind_resdelta} Let $\g'$ and $(\psi_j)_j$ be as above in the regime of Remark \ref{no_bind_regime}. Then there holds:
\[\left\{
\begin{array}{l}
\frac{1}{(2\pi)^3}\dint |p|^2\big(\La^{-2}(2+\tfrac{|p|^2}{\La^2})+(1+\tfrac{|p|^2}{\La^2})\big)|\wh{\psi}_j(p)|^2dp\le\nlp{2}{\Dbf \psi_j}^2-1\text{\ and\ }\\
\nlp{2}{\Dbf \psi_j}^2-1\le \alpha \ncc{\rho_\g}\ncc{n_j}+\alpha\ns{2}{\g}\ns{2}{R[N_j]}+\big(\alpha\nb{B_{\g'}\tfrac{1}{|\nabla|^{1/2}}}\nlp{2}{|\nabla|^{1/2}\psi_j}\big)^2.
\end{array}\right.
\]
As a consequence we also have:
\begin{equation}\label{no_bind_c-2}
\ttr(-\Delta(1-\tfrac{\Delta}{\La^2}+\tfrac{\Delta^2}{\La^4})N)\apprle c^{-2}.
\end{equation}
\end{lemma}
It suffices to use the inequalities in the r.h.s. of \eqref{no_bind_main_ingred} in Eq. \eqref{no_bind_equaj}.

\begin{remark}
Compared to the case of $E^0(1)$ there is an additional term $(v_n-R_N)\psi_j$ that has been neglected in $-2\alpha \mathfrak{Re}\psh{B_N\psi_j}{\psi_j}$: this term is non-positive. %Indeed let us write $v_{jk}=(\psi_j^*\psi_k)*\tfrac{1}{|\cdot|}$ and $v_j:=v_{jj}$: 
%\[
%\psh{(v_n-R_N)\psi_1}{\psi_1}=\psh{v_{2}\psi_1-v_{21}\psi_2}{\psi_1}=D(n_1,n_2)-D(\psi_2^*\psi_1,\psi_2^*\psi_1)\ge 0.
%\]
\end{remark}
\begin{notation}
From now on, we write $v_{jk}=(\psi_j^*\psi_k)*\tfrac{1}{|\cdot|}$ and $v_j:=v_{jj}$ and define $a_{jk}:=\nlp{2}{v_k\psi_j-v_{kj}\psi_k}$.
\end{notation}

%With the estimates of Section \ref{no_bind_minimini} and the method of \cite{sok,sokd} we get:
%\begin{lemma}\label{no_bind_resE(2)}
%\begin{equation}
%\mathcal{E}_{\text{BDF}}(\g+N)=2+\frac{1}{2c^2}\mathcal{E}_{\text{PT}}(\un{\psi_1}\wedge \un{\psi_2})+\mathcal{O}(\alpha c^{-2})
%\end{equation}
%where $\mathcal{E}_{\text{PT}}$ is the Pekar-Tomasevitch energy \cite{bip}:
%\begin{equation}
%\left\{
%\begin{array}{l}
%\mathcal{E}_{\text{PT}}(\Psi)=\ttr(-\Delta \g_\Psi)+\frac{1}{\Fla(0)}(D(\rho_\Psi,\rho_\Psi)-\ttr(\g_\Psi R[\g_\Psi]))-D(\rho_\Psi,\rho_\Psi)\\
%\text{\ with\ }\g_\Psi\text{\ one-body\ density\ matrix\ and\ }\rho_\Psi\text{\ density\ of\ }\Psi,
%\end{array}\right.
%\end{equation}
%and $\un{\psi_j}(x):=c^{3/2}\psi_j(cx)$.
%\end{lemma}

%Proposition \ref{no_bind_eloig} (Section \ref{no_bind_ptt}) enables us to say that 

%\[
%M^2(\un{\psi_1}\wedge\un{\psi_2})=\iint |\un{\psi_1}\wedge \un{\psi_2}(x,y)|^2\tfrac{dxdy}{|x-y|}\apprle c^{-1}.
%\]
%Moreover there exists a decomposition:
%\[
%\un{\psi_1}\wedge\un{\psi_2}=h_1\wedge h_2
%\]
%such that $D(|h_1|^2,|h_2|^2)=\mathcal{O}\big(M^2(\un{\psi_1}\wedge\un{\psi_2})\big)$. Each $h_j$ is close to $\mathscr{P}$ and we even get some decay estimates (Propositions \ref{no_bind_eloig} and \ref{no_bind_RR}). To some extent Lemma \ref{no_bind_h2l2} enables us to say $\un{h_j}$ is close to $\un{\psi_j}$.

%\section{Estimates on the minimizers}

\subsection{Proof of Lemma \ref{no_bind_estim_minim}: estimate of $E^0(1)$}

%\begin{center}
%\textsc{Estimate of $E^0(1)$}
%\end{center}
We compute the energy of a particular test function $Q'_0=Q_0+N_0$, defined as follows \cite{sok}. First, we take $\phi_{\text{CP}}=\phi_1$ a minimizer for $E_{\text{PT}}(1)$ in $L^2(\mathbb{R}^3,\mathbb{C})$ (\emph{e.g.} real-valued and positive centered in $0$, \textit{cf} \cite{L}). Then let $\psi_{1}$ be:
\begin{equation}
\psi_1:={}^{t}\begin{pmatrix}\phi_1 & 0 & 0 & 0 \end{pmatrix}\in L^2(\mathbb{R}^3,\mathbb{C}^4).
\end{equation}
Then, we define $\psi_{\tfrac{1}{c}}:=c^{-3/2}\psi_1(c^{-1}(\cdot))$ where $c^{-1}:=\alpha^2 \Fla(0)$ and
\[
\begin{array}{rl | rl}
\ov{N}_0&:=\ket{\psi_{\tfrac{1}{c}}}\bra{\psi_{\tfrac{1}{c}}}, & Q_0+P^0_-=\Pi_0&:=\chi_{-\infty,0}\big\{\Dbf+\alpha\big((\rho_{Q_0}+\ov{n}_0)*\tfrac{1}{|\cdot|}- (R_{Q_0}+R_{\ov{N}_0})\big) \big\}, \\
\ov{n}_0&:=|\psi_{\tfrac{1}{c}}|^2,& \psi_0 &:= \frac{1}{\sqrt{1-\nlp{2}{\Pi_0 \psi_{\tfrac{1}{c}}}^2}}(\psi_{\tfrac{1}{c}}-\Pi_0 \psi_{\tfrac{1}{c}}).\\
\end{array}
\]
We have used the fixed point scheme of Section \ref{no_bind_the_cauchy} to define $Q_0$. We also write
\[
\begin{array}{rl | rl}
N_0&:=\ket{\psi_0}\bra{\psi_0},& Q'_0&:=Q_0+N_0,\\
B_0&:=(\rho_{Q_0}+\ov{n}_0)*\tfrac{1}{|\cdot|}-\alpha (R_{Q_0}+R_{\ov{N}_0}), & \Dbf_{Q_0}&:=\Dbf+\alpha B_0.
\end{array}
\]
The test function $Q'_0$ is the difference between the orthogonal projections $\Pi_0+N_0$ and $P^0_-$. Following the same method as in \cite{sok}, the following estimates hold.%It can be proved (\textit{cf} \cite{sok}):
\begin{equation}\label{no_bind_testestim}
\begin{array}{rl | cl}
\norm{Q_0}_{\mathbf{Q}_{w_2}}&\apprle \alpha & \norm{\ov{n}_0}_{\mathfrak{C}_{w_2}}&\apprle c^{-1/2}\\
\norm{Q_0}_{\mathbf{Q}_{w_1}}&\apprle c^{-1} & \norm{\rho_{Q_0}}_{\mathfrak{C}_{w_2}}&\apprle L c^{-1/2}\\
 \ns{2}{Q_0}&\apprle \alpha c^{-1/2}& \ns{2}{R_{\ov{N}_0}}&\apprle c^{-1}
\end{array}
\end{equation}
where $w_1(p-q)=E(p-q)$ and $w_2(p-q)=E(p-q)^2$.
%\begin{remark}\label{no_bind_remdiff}
%We recall \emph{an electron does not see its own field}:we have 

%\noindent $(n*\tfrac{1}{|\cdot|}-R_N)\psi=0$. As $\Dbf_{\g'}\psi=\Dbf_{\g}\psi=\mu\psi$ these equalities can be rewritten as:
%\[
%\Dbf_{\g'}\psi=|\Dbf_{\g'}|\psi=|\Dbf_{\g}\psi|=\Dbf_{\g}\psi=\mu\psi.
%\]
%\end{remark}

As shown previously in \cite{sok,sokd} there holds
\begin{equation}\label{no_bind_estim_q0}
\begin{array}{l}
\mathcal{E}_{\text{BDF}}(Q'_0)=\psh{\Dbf \psi_0}{\psi_0}-\frac{\alpha}{2}\ttr_0(B[Q_0]Q_0)-\frac{1}{2}(\ttr(|\Dbf+\alpha B_0|Q_0^2)-\ttr(|\Dbf| Q_0^2))\\
\ \ \ \ \ \ \ \ \ \ \ \ \ \ \ \ \ \ +\frac{\alpha}{2}\big(D(\rho[Q_0]+n_0,\rho[Q_0]+n_0)-\ttr(Q'_0R[Q'_0])\big)
\end{array}
\end{equation}

\paragraph{Estimate of the density $\rho_{Q_0}$} By Section \ref{no_bind_eq_dens}, we write

\begin{align}
\rho_{Q_0'}&=(\delta_0-\check{F}_\La)*(\ov{n}_0+\mathfrak{t}[\ov{N}_0]+\alpha^2\wit{\tau}_2),\\
                    &=(\delta_0-\check{F}_\La)*\ov{n}_0+\tau_{rem}.
\end{align}We have
\[
\nlp{\infty}{(\delta_0-\check{F}_\La)*\ov{n}_0*\tfrac{1}{|\cdot|}}\le \frac{\pi}{2}(1+\nlp{1}{\check{\Fla}})\psh{|\nabla| \psi_{\tfrac{1}{c}}}{\psi_{\tfrac{1}{c}}}\apprle \nlp{2}{\nabla \psi_{\tfrac{1}{c}}}=\mathcal{O}(c^{-1}).
\]
We use Ineq. \eqref{no_bind_main_ingred} to estimate the norm $\ncc{\tau_{rem}}$ of the remainder $\tau_{rem}$.

\paragraph{The traces in \eqref{no_bind_estim_q0}} By Lemma \ref{no_bind_BB}, we can estimate $\big| \Dbf+\alpha B_0\big|-\big| \Dbf\big|$ and get the following \cite{sok}.

\begin{lemma}\label{no_bind_lemtr}
There holds:
\begin{equation}
\begin{array}{rl}
|\delta \ttr|&:=\Big|\ttr\big\{|\Dbf+\alpha B[Q_0']|\g^2_0-|\Dbf|Q_0^2\big\} \Big|\\
                &\apprle \{\nb{Q_0}^2+\alpha (\nqkin{Q_0}+\ncc{\tau_{rem}})\}\nqkin{Q_0}^2+\alpha\{\ncc{\tau_{rem}}+\nlp{2}{\nabla \psi_{\tfrac{1}{c}}}\}\ns{2}{Q_0}^2\\
                &\apprle \alpha c^{-3}+\alpha c^{-1}\times \alpha^2 c^{-1}\apprle \alpha c^{-3}.
 \end{array}
\end{equation}
\end{lemma}

\paragraph{$\psh{\Dbf \psi_0}{\psi_0}$ in \eqref{no_bind_estim_q0}} There holds $(1-\Pi_0)\psi_{\tfrac{1}{c}}=-Q_0\psi_{\tfrac{1}{c}}+P^0_+\psi_{\tfrac{1}{c}}$. Then
\[
\begin{array}{rl}
\psh{\Dbf \psi_{\tfrac{1}{c}}}{\psi_{\tfrac{1}{c}}}&=\psh{\Dbf Q_0 \psi_{\tfrac{1}{c}}}{Q_0 \psi_{\tfrac{1}{c}}}-2\mathfrak{Re}\psh{P^0_+Q_0\psi_{\tfrac{1}{c}}}{P^0_+\psi_{\tfrac{1}{c}}}+\psh{|\Dbf|P^0_+\psi_{\tfrac{1}{c}}}{\psi_{\tfrac{1}{c}}}\\
\psh{|\Dbf|P^0_+\psi_{\tfrac{1}{c}}}{\psi_{\tfrac{1}{c}}}&=1+\frac{1}{2}\nlp{2}{\nabla \psi_{\tfrac{1}{c}}}^2+\mathcal{O}(c^{-4}).
\end{array}
\]
Then thanks to Lemma \ref{no_bind_BB}: $\nlp{2}{|\Dbf|^{1/2} Q_0 \psi_{\tfrac{1}{c}}}\le \nb{|\Dbf|^{1/2} Q_0 \tfrac{1}{|\nabla|^{1/2}}}\nlp{2}{|\nabla|^{1/2}\psi_{\tfrac{1}{c}}}$ and
\[
\nlp{2}{|\Dbf|^{1/2} Q_0 \psi_{\tfrac{1}{c}}}\apprle \alpha c^{-1}.
\]
As $Q_0=\alpha Q_{1}[Q_0',\rho_{Q_0}']+\alpha^2\wit{Q}_2[Q_0',\rho_{Q_0}']$ and that $Q_1=Q_1^{+-}+Q_1^{-+}$:
\[
P^0_+Q_0\psi_{\tfrac{1}{c}}=\alpha Q_1^{+-}P^0_-\psi_{\tfrac{1}{c}}+\alpha^2P^0+ \wit{Q}_2 \psi_{\tfrac{1}{c}}.
\]
Therefore:
\[
\begin{array}{rl}
\alpha^2\psh{|\Dbf| \wit{Q}_2\psi_{\tfrac{1}{c}}}{P^0+\psi_{\tfrac{1}{c}}}&\le \alpha^2\nlp{2}{|\nabla|^{1/2}\psi_{\tfrac{1}{c}}}^2\nb{\tfrac{|\Dbf|}{|\nabla|^{1/2}}\wit{Q}_2\tfrac{1}{|\nabla|^{1/2}}}\\
        &\apprle \alpha^2c^{-1}\times c^{-1}=\mathcal{O}(\alpha^2c^{-2})\\
\alpha \psh{|\Dbf| Q_1^{+-} P^0_-\psi_{\tfrac{1}{c}}}{P^0_+\psi_{\tfrac{1}{c}}}& \le \alpha \nb{|\Dbf|^{1/2} Q_1^{+-}\tfrac{1}{|\nabla|^{1/2}}} \nlp{2}{|\nabla|^{1/2}P^0_-\psi_{\tfrac{1}{c}}} \nlp{2}{|\Dbf|^{1/2}\psi_{\tfrac{1}{c}}}\\
 &\apprle \alpha c^{-1/2}\times c^{-3/2}=\mathcal{O}(\alpha c^{-2}).
\end{array}
\]
Hence:
\begin{equation}
\psh{\Dbf (1-\Pi_0)\psi_{\tfrac{1}{c}}}{(1-\Pi_0)\psi_{\tfrac{1}{c}}}/(1-\nlp{2}{\Pi_0 \psi_{\tfrac{1}{c}}}^2)=1+\frac{1}{2}\nlp{2}{\nabla \psi_{\tfrac{1}{c}}}^2+\mathcal{O}(\alpha c^{-2}).
\end{equation}

\paragraph{The potential energy in \eqref{no_bind_estim_q0}} By the same methods we prove:
\begin{equation}
\begin{array}{l}
\frac{\alpha}{2}\big(2D(\rho[Q_0],n_0)-D(\rho[Q_0],\ov{n}_0)-\mathfrak{Re}(2\ttr(Q_0 R[N_0])-\ttr(Q_0 R[\ov{N}_0]))\big)\\
\ \ \ =-\frac{\alpha}{2}D(\check{\Fla}*\ov{n}_0,\ov{n}_0)+\mathcal{O}(\alpha^2c^{-3/2}).
\end{array}
\end{equation}

For instance by Cauchy-Schwarz inequality followed by Hardy inequality:
\[
\big|D\big(\rho[Q_0],(P^0_+\psi_{\tfrac{1}{c}})^*(Q_0\psi_{\tfrac{1}{c}})\big)\big|\le \ncc{\rho[Q_0]}\times 4^{1/4}\nlp{2}{\nabla \psi_{\tfrac{1}{c}}}^{1/2}\nlp{2}{Q_0 \psi}=\mathcal{O}(c^{-3}).
\]
By Ineq. \eqref{no_bind_Fla2}, there holds:
\[
-\frac{\alpha}{2}D(\check{\Fla}*\ov{n}_0,\ov{n}_0)=-\frac{1}{2c}D(\ov{n}_0,\ov{n}_0)+\mathcal{O}(\alpha^2 c^{-2}+c^{-1}\nlp{2}{\ov{n}_0}^2)=\mathcal{O}(\alpha^2 c^{-2});
\]
indeed: $\nlp{2}{\ov{n}_0}=\nlp{4}{\psi_{\tfrac{1}{c}}}^2\apprle \nlp{2}{|\nabla|^{3/4}\psi_{\tfrac{1}{c}}}^2$. As a consequence:

\begin{equation}
E^0_{\text{BDF}}(1)\le \mathcal{E}_{\text{BDF}}(Q_0+N_0)=1+\frac{\mathcal{E}_{\text{PT}}(\phi_1)}{2c^2}+\mathcal{O}(\alpha c^{-2}).
\end{equation}
We have proved the inequality the $\le$ part. For the $\ge$ part, it suffices to take a \emph{real} minimizer and with the same estimates as above and \cite{sok} we prove similar estimates.

That there exists a minimizer for $E^0(1)$ follows from Theorem \ref{no_bind_HVZ}, using the same method as in \cite{sok}. We have proved $E^0(1)<1$, then by Lieb's variational principle we get that for any $0<q<1$, $E^0(q)>qE^0(1)$, hence the binding inequalities holds for $0<q<1$. For $q\in [0,1]^c$, binding inequalities hold for sufficiently small $\alpha$. We refer to \cite{sok} for more details.

Similar estimates apply for $E^0(2)$, in particular we have $E^0(2)\le 2E^0(1)\le 2+\tfrac{\mathcal{E}_{\text{PT}}(\phi_1)}{2c^2}+\mathcal{O}(\alpha c^{-2})$.%, we have trivially $E^0(2)\le 2E^0(1)$. %Furthermore, if $\g'$ is a minimizer for $E^0(2)$, then it also satisfy 

\subsection{Study of a minimizer $\g'$ for $E^0(2)$}
%\begin{center}
%\textsc{Study of a minimizer for $E^0(2)$}
%\end{center}

%\subsection{Proofs of Lemmas \ref{no_bind_resdelta} and \ref{no_bind_estim_minim}}
\paragraph{Bootstrap argument} We write $x^2:=\ttr(-\Delta(1-\tfrac{\Delta}{\La^2}+\tfrac{\Delta^2}{\La^4})N)$. By Lemma \ref{no_bind_resdelta}, we have $x^2\apprle c^{-2}$. This fact enables us to use the method of \cite{lim,sok}.

%by summing inequalities of Lemma \ref{no_bind_resdelta}:
%\[
%\begin{array}{rl}
%x^2&\le \alpha\sqrt{\pi}\ncc{\rho_\g}x^{1/2}+\alpha\sqrt{2}\big(2\ns{2}{\g}+\alpha\nb{B[\g']\tfrac{1}{|\nabla|^{1/2}}}\big)x
%\end{array}
%\]
%One already knows $x,\ns{2}{\g},\ncc{\rho}\le k_0$ then it is clear that 

%\[x^2=\mathcal{O}(c^{-1}),\  \ncc{\rho_\g}=\mathcal{O}(Lc^{-1/4})\text{\ and\  }\nqbfu{\g}=\mathcal{O}(c^{-3/4}).\]

%As $\rho_\g=-\check{\Fla}*n+(\delta_0-\check{\Fla})*(\mathfrak{t}_N+\alpha^2\wit{\tau}_2)$ and $\g=\alpha \Tbf[Q_{0,1}(\rho_\g')]+(\Tbf-1)[N]+\alpha^2\Tbf[\wit{Q}_2]$:
%\[
%x^2\le k_1 c^{-1}x+k_{3/2}\alpha^2x^{3/2}+k_2\alpha^2x^{2}\text{\ and\ so\ }x^2\apprle c^{-2}.
%\]

%The fact that $x^{2}=\mathcal{O}(c^{-2})$ enables to follow the methods of \cite{lim,sokd}: 

We scale $\psi_j$ by $c$: $\un{\psi_j}(x)=c^{3/2}\psi_j(cx)$ and scale $\g$ accordingly: $\un{\g}(x,y)=c^3\g(cx,cy)$. Then writing $\mathcal{L}_{A}:=(1-\Delta/A^2)$, the wave function $\un{\psi_j}$ satisfies:
\begin{equation}\label{no_bind_equn}
(c^2\beta-ic\boldsymbol{\alpha}\cdot \nabla)\un{\psi_j}+\alpha c\mathcal{L}_{c\La}^{-1}(v[\rho[\un{\g}]+\un{n}]-R[\un{g}+\un{N}])\un{\psi_j}=c^2\mu_j\mathcal{L}_{c\La}^{-1}\un{\psi_j}.
\end{equation}
Splitting $\un{\psi_j}$ between upper spinor $\un{\varphi_j}$ and lower spinor $\un{\chi_j}$ both in $L^2(\RR,\mathbb{C}^2)$, this gives:
\[
\nlp{2}{\un{\chi_1}}+\nlp{2}{\un{\chi_2}}\apprle c^{-1}.
\]

Going back to $\psi_j$ one gets $\psh{\Dbf \psi_j}{\psi_j}=1+\mathcal{O}(c^{-2})$ and it shows that for $j=1,2$: $0< (1-\mu_j)c^2\le K$ thanks to the equation \eqref{no_bind_form_no_2}. As 
\begin{equation}
0\le c^2(1-\mathcal{L}_{c\La}^{-1})=\frac{-c^2\Delta}{c^2\La^2-\Delta}\le \frac{-\Delta}{\La^2},\ \text{then}
\end{equation}
\[
\begin{array}{rl}
c^2(\mu_j \mathcal{L}_{c\La}-1)\un{\varphi_j}&=c^2(\mu_j-1)\un{\varphi_j}+\frac{c^2\Delta}{c^2\La^2-\Delta}\un{\varphi_j}\\
                 &=c^2(\mu_j-1)\un{\varphi_j}+\mathcal{O}_{L^2}\big(\frac{c}{\La}\big)
\end{array}
\]
thanks to Lemma \ref{no_bind_estim_minim} ($\mathcal{O}_{L^2}$ means in $L^2$-norm). We can get another estimate: in the spirit of \cite{sok,sokd} we can use bootstrap argument with the norms
\[
\nqbfg{Q}^2=\diint E(p-q)^{2k}(E(p+q))|\wh{Q}(p,q)|^2dpdq\text{\ and\ }\ncg{\rho}^2=\dint\frac{E(k)^2|\wh{\rho}(k)|^2}{|k|^2}dk,
\]
to get the following statement: 

\begin{lemma}\label{no_bind_H2}
For any fixed $k\in\mathbb{N}^*$, there exists $\alpha_{(k)}>0$ such that for $\alpha\le \alpha_{(k)}$, $\psi_j$ with $j=0,1,2$ is in $H^{k/2}$ with norms $\mathcal{O}(1)$ and
\[
\nqbfg{\g_0},\nqbfg{\g},\ncg{\rho[\g]},\ncg{\rho[\g_0]}\apprle 1.
\]
It is supposed $\alpha\llo\le L_0$. There also holds:
\[
\nlp{2}{\Delta \psi}\apprle \min(c^{-1}(c^{-1}+\La^{-1}),c^{-3/2}),\ 
\nlp{2}{\un{\chi}}\apprle c^{-1}\mathrm{\ and\ } \nlp{2}{\nabla\un{\chi}}\apprle c^{-1},%\nlp{2}{\Delta\psi_1}^2=\alpha c^{-1}a_{12}+c^{-3},
\]

\end{lemma}
%\begin{remark}
%This is not sufficient, we would like to have $\nlp{2}{\Delta \psi_j}\apprle c^{-2}$ which is the case if $c\apprle \La$ that is if $\frac{1}{\sqrt{\La\llo}}\le \alpha \le \frac{L_0}{\llo}.$
%\begin{equation}\label{no_bind_newregime}
%\frac{1}{\sqrt{\La\llo}}\le \alpha \le \frac{L_0}{\llo}.
%\end{equation}
%\end{remark}
The estimation of $\mathcal{E}_{\text{BDF}}(\g')$ is proven with the help of the estimate $\nlp{2}{\Delta\psi}\apprle c^{-3/2}$ as shown in the (technical) proof of Lemma \ref{no_bind_H2} in Appendix \ref{no_bind_ph2}.

\begin{remark}\label{no_bind_hypestim}
By Estimate \eqref{no_bind_c-2} we can prove that $n,\g,\rho_\g$ have estimates of the same kind of those stated in \eqref{no_bind_testestim} \cite{sok,sokd}: we have
\begin{equation}
\ncc{n}\apprle c^{-1/2},\ \ncc{\rho_\g}\apprle Lc^{-1/2},\ \ns{2}{R_{N_j}} \apprle c^{-1},\ns{2}{|\Dbf|^{1/2} \g}\apprle c^{-1},\ns{2}{\g}\apprle \alpha c^{-1/2}.
\end{equation}
There also holds $\nlp{2}{n_j}\apprle c^{-3/2}$. 

By Lemma \ref{no_bind_H2}, we get:
\[
\nlp{2}{\rho_\g}\apprle Lc^{-3/2}.
\]

Following \cite{sokd} we can prove $\rho_\g\in L^1$ and $\nlp{1}{\rho_\g}\apprle L$.
\end{remark}

\paragraph{Estimate on $c^2(1-\mu_j)$} Using estimates on $\nabla \ph_j$ and $\nabla \chi_j$ (Lemma \ref{no_bind_H2}) together with Ineq. \eqref{no_bind_Fla2}, we get the following estimate from \eqref{no_bind_form_no_2}:
\begin{equation}\label{no_bind_mumu_proof}
\mu_j=1+\frac{\nlp{2}{\nabla \un{\varphi_j}}^2}{2c^2}-\frac{1}{c^2}D(\un{n_j},\un{n})+\mathcal{O}(\alpha c^{-2}).
\end{equation}

With \eqref{no_bind_psipsi}-\eqref{no_bind_psipsi_1}, we get:
\begin{equation}
(1-\mu_j)c^2\le -\tfrac{3}{2}E_{\text{PT}}(1)+\mathcal{O}(\alpha^{1/4})\apprge 1.
\end{equation}

%%%%%%%%%%%%%%%%%%%%%%%%%%%%%%%%%%%%%%%%%%%%%%%%%%%%%%%%%%%%%%%%%%%%%%%%%%%%%%%%%
%%%%%%%%%%%%%%%%%%%%%%%%%%%%%%%%%%%%%%%%%%%%%%%%%%%%%%%%%%%%%%%%%%%%%%%%%%%%%%%%%

\section{Localisation of minimizers in Direct space}\label{no_bind_loc_dir}

\subsection{Decay estimates on the \underline{$\psi_j$}'s}\label{no_bind_decc}

It is known $\un{\psi_1}\wedge\un{\psi_2}$ can be split into two almost minimizers of Choquard-Pekar energy $h_1$ and $h_2$: $h_1\wedge h_2=\un{\psi_1}\wedge\un{\psi_2}$. For $j\in\{1,2\}$, we write $\phi_j\in\mathscr{P}$ the closest Pekar minimizer to $h_j$ and its center is written $z_j$. We write
\begin{equation} \rgg:=|z_1-z_2|.
\end{equation}
By Section \ref{no_bind_the_PT_fun}, we have:
\begin{equation}\label{no_bind_Mwed2}
 M^2(\un{\psi_1}\wedge\un{\psi_2}):=\diint \frac{|\un{\psi_1}\wedge \un{\psi_2}(x,y)|^2}{|x-y|}dxdy    \apprge\dfrac{1}{\rgg}.
\end{equation}

Our aim is to show decay estimates far away from $z_1$ and $z_2$. Up to translations, we assume the mean $z_m=\tfrac{z_1+z_2}{2}$ is $0$. 

\paragraph{Localisation functions} Let $\xi_1\ge 0$ be some \emph{radial} Schwartz function in $\mathcal{S}(\mathbb{R}^3)$ satisfying
\[
|x|\le 1\,\Rightarrow\ \xi_1(x)=1\text{\ and\ }|x|\ge 2\,\Rightarrow\ \xi_1(x)=0.
\] 
We define $\xi_A(x):=\xi_1(\tfrac{x}{A})$ for any $A>0$ and $\theta_A:=\sqrt{1-\xi_A^2}$. For any $x\in\mathbb{R}^3$ we write
\begin{equation}
d(x):=\min \{ |x-z_1|,|x-z_2|\}.
\end{equation}
Let $H$ be the plane $ \{x:\ |x-z_1|=|x-z_2| \}$ ; the function $d(\cdot)$ is differentiable in $\mathbb{R}^3\backslash \big(\{z_1,z_2 \}\cup H\big)$. For any $A\gg \rgg$ and $0<\la<2$ we define
\begin{equation}
\boldsymbol{\eta}^{\la}_{ \rgg}(x):=\big(1-\xi_{\la \rgg}^2(x-z_1)-\xi_{\la \rgg}^2(x-z_2)\big)^{1/2}.
\end{equation}
We define $\la_0>0$, defined by the formula

\begin{equation}\label{no_bind_def_la_0}
\la_0 \rgg=\tfrac{C_0}{L}\text{\ where\ }C_0(L,\rgg)>1\text{\ is\ chosen\ large.}
\end{equation}
The function $\boldsymbol{\eta}^{\la}_{ \rgg}$ can be seen as the dilation of $\boldsymbol{\eta}^{\la}_{1}:=\sqrt{1-\xi_{\la}^2(\cdot -e_1)-\xi_{\la}^2(\cdot-e_2)}$ by $\rgg$ where $e_j:=\frac{z_j-z_m}{\rgg}$.

At last we define:
\begin{equation}
\etabfc{\la}(x):=\sqrt{1-\xi_{c\la\rgg}^2(x-cz_1)-\xi_{c\la\rgg}^2(x-cz_2)},
\end{equation}
we use it in Section \ref{no_bind_estmoche}.

\begin{lemma}\label{no_bind_decaylemma}
\noindent\textbullet\ For each $\la_0\le \la<2^{-1}$, there exists $K_\la$ such that:
\begin{equation}\label{no_bind_unun}
\begin{array}{rl}
\forall\,A>0,\ \dint d(x)^2\xi_A^2(x)(\boldsymbol{\eta}_{\rgg}^{\la}(x))^2\Big(\big||D_0|^{1/2}\un{\psi_1}(x)\big|^2+\big||D_0|^{1/2}\un{\psi_2}(x)\big|^2\Big)dx&\le  K_\la
\end{array}
\end{equation}
Moreover we can choose $(K_\la)_\la$ to be nonincreasing and $K_{\la_0}$ is (uniformly) bounded in the regime $\alpha,L,\La^{-1}$ small.

\noindent\textbullet\ For any $2\la_0\le \la <2^{-1}$ the same holds for $d^{(2)}_{A,\la}:=d(x)^2\xi_A\etabfc{\la}$:
\begin{equation}\label{no_bind_deuxdeux}
\dint d(x)^4\xi_A^2(x)(\etabfc{\la})^2(x) \Big(\big||D_0|^{1/2}\un{\psi_1}(x)\big|^2+\big||D_0|^{1/2}\un{\psi_2}(x)\big|^2\Big)dx\le  K_\la',
\end{equation}
where $K'_\la>K_\la$ depends on $\la,K_\la$, $\xi_1$.

\noindent\textbullet\ We can replace $|D_0|^{1/2}\un{\psi_j}$ by $\un{\psi_j}$ above.
\end{lemma}

\begin{remark}
This is a weak estimate due to the presence of $\un{v}_{k}\un{\psi_j}-\un{v}_{kj}\un{\psi_k}$. %In the case of $E^0(1)$ we can be more precise \cite{these}.
\end{remark}
This proposition is proved in Appendix \ref{no_bind_decaydecay}.

\subsection{Localisation operators}

We want to prove that minimizers are localised in space around the centers $z_1,z_2$ of the electrons. To this end we use localisation operators of \cite{at,bindpol} with respect to the functions $\xi_{c\la \rgg}$ and $\etabfc{\la}$ introduced in the previous Section (\ref{no_bind_decc}). 

By Lemma \ref{no_bind_decaylemma} we know that the wave functions $\un{\psi_1}$ and $\un{\psi_2}$ are localized near $z_1$ and $z_2$. By scaling, it follows that $\psi_1$ and $\psi_2$ are localized near $cz_1$ and $cz_2$. We consider:%This suggests the following localisation:
\[
\begin{array}{lcl}
\xi_1^{(\la)}(x):=\xi_{c\la\rgg}(x-cz_1)&\mathrm{\ and\ }&\xi_2^{(\la)}(x):=\xi_{c\la\rgg}(x-cz_2),\\%\ \etabfc{\la}(x)=(1-(\xi_1^{(\la)})^2(x)-(\xi_2^{(\la)})^2(x))^{\tfrac{1}{2}},\\
X_1^{(\la)}:=(\xi_1^{(\la)})^{++}+(\xi_1^{(\la)})^{--}&\mathrm{\ and\ }&X_2^{(\la)}:=(\xi_2^{(\la)})^{++}+(\xi_2^{(\la)})^{--},%\ H^{\la}:=(\etabfc{\la})^{++}+(\etabfc{\la})^{--},
\end{array}
\]
and localise $\g'$:
\[
\xi_1^{(\la)}\cdot[\g']:= X_1^{(\la)}(\g')X_1^{(\la)},\ \xi_2^{(\la)}\cdot[\g']=X_2^{(\la)}(\g')X_2^{(\la)}.
\]
%We recall $\etabfc{\la}(x)=(1-(\xi_1^{(\la)})^2(x)-(\xi_2^{(\la)})^2(x))^{\tfrac{1}{2}}$.

We define the set
\begin{equation}
B_{\la}:=\big\{B(cz_1, c\la\rgg)\times B(cz_2, c\la\rgg)\big\}\cup\big\{B(cz_2, c\la\rgg)\times B(cz_2, c\la\rgg)\big\}\subset\mathbb{R}^3\times\mathbb{R}^3.
\end{equation}
Our aim in this section is to prove:
\begin{proposition}\label{no_bind_place}
If $\g'$ is a minimizer of $E^0(2)$ in the regime $\alpha,L,\La^{-1}$ small then:
\begin{equation}
\mathcal{E}_{\text{BDF}}^0(\g')=\mathcal{E}_{\text{BDF}}^0(\xi_1^{3^{-1}}\cdot[\g'])+\mathcal{E}_{\text{BDF}}^0(\xi_2^{3^{-1}}\cdot[\g'])+\frac{\alpha}{2}\underset{(x,y)\in B_{3^{-1}}}{\diint}\dfrac{|\psi_1\wedge \psi_2(x,y)|^2}{|x-y|}dxdy+\mathcal{O}\Big(\frac{1}{c^2\rgg}\Big).
\end{equation}
Moreover:
\begin{equation}\label{no_bind_ttr0place}
\left\{
\begin{array}{l}
\ttr_0(\xi_j^{(\tfrac{1}{3})}\cdot[\g'])=1+\eps_j,\ \eps_j=o(1),j=1,2,\\
\ttr_0(\xi_1^{(\tfrac{1}{3})}\cdot[\g'])+\ttr_0(\xi_2^{(\tfrac{1}{3})}\cdot[\g'])=2+\mathcal{O}\Big(\dfrac{1}{c^2\rgg}\Big).
\end{array}
\right.
\end{equation}
\end{proposition}

%Let us show how this enables us to prove Theorem \ref{no_bind_main}.

Assuming this Proposition -- proved in Subsection \eqref{no_bind_prop6} -- we can prove Theorem \ref{no_bind_main}.

\subsection{Proof of Theorem \ref{no_bind_main}}\label{no_bind_proof_main_no}
By Proposition \ref{no_bind_eloig1}, for sufficiently small $\alpha,L$, there holds:
\[
\frac{\alpha}{2}\underset{(x,y)\in B_{3^{-1}}}{\diint}\dfrac{|\psi_1\wedge \psi_2(x,y)|^2}{|x-y|}dxdy\ge \frac{L^{-1}}{c^2K_{\text{g}}\rgg},
\]
for some constant $K_{\text{g}}>1$ independent of $\alpha,\La$ in the regime of Remark \ref{no_bind_regime}. This gives:
\[
\mathcal{E}_{\text{BDF}}(\g')\ge E_{\text{BDF}}^0(1+\eps_1)+E_{\text{BDF}}^0(1+\eps_2)+\frac{L^{-1}}{K_{\text{g}}c^2\rgg}+\mathcal{O}\Big(\dfrac{1}{c^2\rgg}\Big).
\]
We know that the function $E_{\text{BDF}}^0(\cdot):\mathbb{R}\mapsto \mathbb{R}$ is uniformly Lipschitz with constants $1$ and this function is concave on each interval $[M,M+1]$ where $M\in\mathbf{Z}$ \cite[Corollary 3 \emph{mutatis mutandis}]{at}. Furthermore we may assume $\eps_1=-\eps_2>0$ up to an error $\mathcal{O}\Big(\dfrac{1}{c^2\rgg}\Big)$. The case $\eps_1,\eps_2<0$ is easily excluded by concavity of $E^0_{\text{BDF}}$ in $[0,1]$ because $E_{\text{BDF}}^{0}(0)= 0$ and $2E_{\text{BDF}}^{0}(1)\ge E_{\text{BDF}}^{0}(2)$. Then:
\[
\begin{array}{l}
E_{\text{BDF}}^0(1+\eps_1)+E_{\text{BDF}}^0(1-\eps_1)\ge \eps_1 E_{\text{BDF}}^{0}(2)+(1-\eps_1)E_{\text{BDF}}(1)+ (1-\eps_1)E_{\text{BDF}}(1)\\
 \ \ \ \ \ \ \ \ \ \ \ \ \ \ \ \ \ \ \ge \eps_1 E_{\text{BDF}}^{0}(2)+(1-\eps_1)(2 E_{\text{BDF}}^{0}(1))\ge (1-\eps_1+\eps_1)E_{\text{BDF}}^{0}(2)=E_{\text{BDF}}^{0}(2).
\end{array}
\]
Thus taking $\Fla(0)=\Theta(\alpha\llo)$ sufficiently small, the quantity $L^{-1}$ is big enough to compensate the error term $\mathcal{O}\Big(\dfrac{1}{c^2\rgg}\Big)$. We get the desired contradiction:
\[
E_{\text{BDF}}^{0}(2)=\mathcal{E}_{\text{BDF}}(\g')\ge E_{\text{BDF}}^{0}(2)+\dfrac{1}{c^2\rgg K_{\text{g}}'}> E_{\text{BDF}}^{0}(2).
\]
%llllllllllll

\subsection{Localisation of the energy of the vacuum $\g$}
%\subsection{Estimation of $\ncc{\etabfc{\la} \rho_\g},\ns{2}{\etabfc{\la} |\Dbf|^{1/2}\g}$}

\begin{lemma}\label{no_bind_titim}
For $\la_0\le \la<2^{-1}$ big enough (\emph{e.g.} $\la=\tfrac{1}{12},\tfrac{1}{6},\tfrac{1}{3}$) there holds:
\begin{equation}\label{no_bind_correct_behav}
\ncc{\etabfc{\la} \rho_\g}\apprle \frac{L}{\sqrt{c\la\rgg}}\mathrm{\ and\ } \ns{2}{\etabfc{\la} |\Dbf|^{1/2} \g},\ns{2}{\etabfc{\la} |D_0|^{1/2} \g}\apprle \dfrac{1}{c\sqrt{\la\rgg}}.
\end{equation}
\end{lemma}

This part comes after lots of technicalities: we put together results of Lemma \ref{no_bind_decaylemma}, Propositions \ref{no_bind_estdens}, \ref{no_bind_ns2kin}, \ref{no_bind_estint}, Remark \ref{no_bind_remestdens} and the known estimates of Remark \ref{no_bind_hypestim}. We refer the reader to Remark \ref{no_bind_imbriquer} for explanation.

Here we assume that $L$ is small enough in such a way that 

\noindent $\la_0 \rgg=\mathcal{O}(L^{-1})$ is big enough. Lemma \ref{no_bind_titim} gives that for all $\la_0\le \la<2^{-1}$:
\begin{equation}\label{no_bind_aa}
\ncc{\etabfc{\la} \rho_\g}\le \frac{\epsilon_1}{\sqrt{c\la\rgg}}+\epsilon_2\ncc{\etabfc{\la/2} \rho_\g},\ \epsilon_1,\epsilon_2=\mathcal{O}(L).
\end{equation}
We recall that $\la_0\rgg:=\tfrac{C_0}{L}$ with $C_0(L,\rgg)>1$ to be chosen. Up to taking a bigger $C_0$: $C_0\le \wit{C}_0<6C_0$ we assume $\la_0=2^{-J_0},J_0\in\mathbb{N}$. Taking $\ell_0:=c\tfrac{\wit{C}_0}{3L}$ as unity of length, we define the sequences $(u_m), (v_m), (w_m)$ by the formulae:
\begin{equation}
\left\{
\begin{array}{l}
u_0=v_0=w_0=\ncc{\etabfc{\la_0} \rho_\g},\\
u_m:=\ncc{\etabfc{2^m\la_0} \rho_\g},\ v_m=2^{m/2}u_m,\\
w_{m+1}:=\epsilon_1\sqrt{\dfrac{2}{\ell_0}} +\epsilon_2\sqrt{2}w_{m}
\end{array}
\right.
\end{equation}
It is clear from \eqref{no_bind_aa} that $v_{m+1}\le \epsilon_1\sqrt{\dfrac{2}{\ell_0}} +\epsilon_2\sqrt{2}v_{m}$. Thus we have:
\[
\forall m\in\mathbb{N}^*:\ v_m\le w_m=w_{\infty}+(2^{1/2}\epsilon_2)^{m}(w_0-w_{\infty})
\]
where $w_{\infty}=\epsilon_1(2/\ell_0)^{1/2}(1-\epsilon_2\sqrt{2})^{-1/2}$ is well defined provided $\epsilon_2<2^{-1/2}$. In particular:

\[
\forall m\in\mathbb{N}^*:\ u_m\le \dfrac{\epsilon_1\sqrt{2}}{\sqrt{2^{m} \ell_0}}+\dfrac{(\sqrt{2} \epsilon_2)^{m}}{\sqrt{2^{m}}}\Big\{\ncc{\etabfc{\la_0} \rho_\g}-\dfrac{\epsilon_1\sqrt{2}}{\sqrt{\ell_0}(1-\epsilon_2\sqrt{2})}\Big\}
\]
It remains to evaluate at $m=J_0$: this gives $\ncc{\etabfc{3^{-1}} \rho_\g}$. Similarly the case $m=J_0-1$ corresponds to $6^{-1}$ etc. By Hardy-Littlewood-Sobolev inequality \cite[Theorem 4.3]{LL}:
\[
\ncc{\etabfc{\la_0} \rho_\g}\le \ncc{|\rho_\g|}\apprle \nlp{6/5}{\rho_\g}\apprle \nlp{1}{\rho_\g}^{\tfrac{2}{3}}\nlp{2}{\rho_\g}^{\tfrac{1}{3}}\apprle Lc^{-1/2}.
\]
%This gives the right estimate in the case $(\sqrt{2} \epsilon_2)^{m}\sqrt{\ell_0}$ is an $\mathcal{O}(1)$. In fact thanks to Section \ref{no_bind_ptt} it is known $\rgg L\apprge \alpha^{-1}$ which solves the problem: $(k_1L)^{-\log(\alpha)+\log(k_2)}\times L^{-1/2}\apprle 1.$

For $\ns{2}{\etabfc{\la} |\Dbf|^{1/2} \g}$ it suffices to use this result, Proposition \ref{no_bind_ns2kin} with Lemma \ref{no_bind_decaylemma}.

%We know that $\ttr_{P^0_-}(\g')=2$. We easily see then that
%\begin{align*}
%0&=\ssum_{\eps\in\{\pm\}}\suum_{\zeta\in\{\etabfc{\tfrac{1}{3}},\xi_j^{\tfrac{1}{3}}\}}\ttr\Big(\zeta (\g')^{\eps,\eps}\zeta\Big),\\
 % &=\ssum_{\eps\in\{\pm\}}\suum_{\zeta\in\{\etabfc{\tfrac{1}{3}},\xi_j^{\tfrac{1}{3}}\}}\ttr\Big( \zeta^{\eps,\eps} (\g)^{\eps,\eps} \zeta^{\eps,\eps}\Big)+\mathcal{O}((\la c R_g)^{-1}\ns{2}{\g}^2)\\
 % &\ \ \ \ +\ssum_{\eps\in\{\pm\}}\suum_{\zeta\in\{\xi_j^{\tfrac{1}{3}}\}}\ttr\Big( \zeta^{\eps,\eps} N^{\eps,\eps} \zeta^{\eps,\eps}\Big)+\mathcal{O}()
%\end{align*}

\begin{remark}\label{no_bind_imbriquer}
The following holds.
\begin{enumerate}
 \item Lemma \ref{no_bind_decaylemma} states that each $\psi_j$ is localized around its center $cz_j$,
 \item we give in Remark \ref{no_bind_hypestim} estimates on the norms of $\g$,$N,\rho_\g$ and $n$. In particular the densities have the "correct behaviour" in $L^1$, $L^2$ and Coulomb norms.
   We call these estimates: "non-localized estimates".
\end{enumerate}

The other cited results are used of as follows. We remark that $\etabfc{\la}=\etabfc{\la}\etabfc{\tfrac{\la}{2}}$. 

Proposition \ref{no_bind_ns2kin} gives an estimate of $\ns{2}{\etabfc{\la}|\Dbf|^{1/2}\g}$ and $\ns{2}{\etabfc{\la}|D_0|^{\wt{a}}\g}$ 

\noindent (where $\wt{a}\in\{2^{-1},\ala\}$) in terms of
\[
\nlp{2}{\etabfc{\la}v[\rho'_\g]},\nqq{\etabfc{\la}\g},\ns{2}{\etabfc{\la}R_N}\mathrm{\ and\ }\nlp{6}{\etabfc{\la}v[\rho'_\g]},
\]
and in terms of the non-localized estimates (with the "correct behaviour" with respect to $c\la \rgg$, that is as in \eqref{no_bind_correct_behav}). Below, we shorten: non. loc. est. w. the c. b.

Proposition \ref{no_bind_estint} gives an estimate of $\nlp{2}{\etabfc{\la}\nabla v[\rho_\g]}$ in terms of
\[
\ns{2}{\etabfc{\la}|D_0|^{1/2}\g}\mathrm{\ and\ }\ncc{\etabfc{\la}\rho_\g}=\ncc{\rho[\etabfc{\la}\g\etabfc{\tfrac{\la}{2}}]},
\]
and in terms of the non. loc. est. w. the c. b. 

Furthermore, it gives an estimate of $\nlp{6}{\etabfc{\la}v_{\rho_\g}}$ in terms of $\nlp{2}{\etabfc{\la}\nabla v_{\rho_\g}}$ and of the non. loc. est. w. the c. b. The term $\nqq{\etabfc{\la}\g}$ is controlled by $\ns{2}{\etabfc{\la} |D_0|^{1/2}\g}$ and by the non. loc. est. w. the c. b.

Thanks to Lemma \ref{no_bind_decaylemma}, the term $\nqq{\etabfc{\la}R_N}$ (resp. $\ncc{\etabfc{\la}n}$) is proved to be of order $(c^2\la \rgg)^{-1}$ (resp. $(c \la \rgg)^{-1/2}$).

Finally Proposition \ref{no_bind_estdens} together with Remark \ref{no_bind_remestdens} gives an estimate of
$\ncc{\rho[\etabfc{\la}\g\etabfc{\tfrac{\la}{2}}]}$
in terms of
$
\ns{2}{\etabfc{\tfrac{\la}{2}} P^0_{\pm}\g},\ \ns{2}{\etabfc{\la}P^0_{\pm}\g},
$
and in terms of the non. loc. est. w. the c. b. The presence of $P^0_{\pm}$ is harmless as we can check from the proofs.
\end{remark}

\subsection{Proof of Proposition \ref{no_bind_place}}\label{no_bind_prop6}

We consider each term of the BDF energy and write $1=(\etabfc{\tfrac{1}{3}})^2+(\xi_1^{(\tfrac{1}{3})})^2+(\xi_2^{(\tfrac{1}{3})})^2.$

We use once again Lemma \ref{no_bind_decaylemma}, Proposition \ref{no_bind_titim} and Remark \ref{no_bind_hypestim}. We treat one after the other the case of $N$ and $\g$. We write 
\[(\xi^{(\la)})^2:=(\xi_1^{(\la)})^2+(\xi_2^{(\la)})^2.\] The function $\zeta$ refers to $\xi^{(\la)}$ or $\etabfc{\la}$.
%\subsubsection{Error term due to $\etabfc{\la}$}
\subsubsection{Kinetic energy}
\paragraph{Kinetic energy for $\g$}: 
\[
\begin{array}{rcl}
\ttr\big( (\etabfc{\tfrac{1}{3}})|\Dbf|^{1/2} \g^2 |\Dbf|^{1/2} \big)&\le& \ns{2}{(\etabfc{\tfrac{1}{3}}) |\Dbf|^{1/2} \g}^2\apprle \frac{1}{c^2 \rgg}\\
\ttr\big( \zeta^{\pm\mp} |\Dbf|^{1/2} \g^2 |\Dbf|^{1/2} \big)&\le& \nb{\zeta^{\pm\mp}}\ns{2}{|\Dbf|^{1/2}\g}^2\apprle \frac{1}{c^3 \la\rgg}
\end{array}
\]

\paragraph{Kinetic energy for $N$}:
We recall the following equalities: $\Dbf \psi_j=\mu_j-\alpha B\psi_j$ and

\noindent $(v_n-R_N)\psi_1=v_{2}\psi_1-v_{21}\psi_2=\mathcal{O}_{L^2}(\alpha^{3/2}c^{-1})$ . Thus, we have:
\[
\begin{array}{rl}
\psh{\etabfc{\tfrac{1}{3}} \Dbf \psi_j}{\etabfc{\tfrac{1}{3}}\psi_j}&=\psh{\etabfc{\tfrac{1}{3}}(\mu_j-\alpha B)\psi_j}{\etabfc{\tfrac{1}{3}}\psi_j}\\
|\psh{\etabfc{\tfrac{1}{3}} \Dbf \psi_j}{\etabfc{\tfrac{1}{3}}\psi_j}|&\le (1+\alpha \nlp{\infty}{v[\rho'_\g]})\nlp{2}{\etabfc{\tfrac{1}{3}}\psi_j}^2+\alpha \nlp{2}{\etabfc{\tfrac{1}{3}}\psi_j}(\nlp{2}{(\etabfc{\tfrac{1}{3}}) R_\g \psi_j}\\
&\ \ +\alpha \nlp{\infty}{v_{kj}} \nlp{2}{\etabfc{\tfrac{1}{3}} \psi_k})\apprle \frac{1}{\rgg^4}+\frac{\alpha \nqq{\g}}{c\rgg^3 }=o(c^{-2} \rgg^{-1}).
  \end{array}
\]
%For $\xi^{(\tfrac{1}{3})}$ 
We write : 
\[
(\xi^{(\tfrac{1}{3})})^2=(P^0_++P^0_-)(\xi^{(\tfrac{1}{3})})(P^0_++P^0_-)(\xi^{(\tfrac{1}{3})})(P^0_++P^0_-),
\]
we have to show that $\psh{\xi^{\eps_1 \eps_2} \xi^{\eps_2 \eps_3} \Dbf \psi_j}{\psi_j}$ is $\mathcal{O}(c^{-2}\rgg^{-1})$ whenever $\eps_1\neq \eps_2$ or $\eps_2\neq \eps_3$.

We recall that $\nlp{2}{P^0_-\psi_j}$ and $\alpha \nlp{2}{B\psi_j}$ are $\mathcal{O}(c^{-1})$.

The operator $(\xi^{(\tfrac{1}{3})})^{+-}(\xi^{(\tfrac{1}{3})})^{-+}$ is $\mathcal{O}(c^{-2} \rgg^{-2})$ in $\nb{\cdot}$-norm. Except for the corresponding term, we have $\eps_1=-$ or $\eps_3=-$, leading to an upper bound:
\[
\mathcal{O}\big( \nb{(\xi^{(\tfrac{1}{3})})^{+-}}(\nlp{2}{P^0_-\psi_j}+\alpha \nlp{2}{B\psi_j})\big)=\mathcal{O}\Big(\dfrac{1}{c^2 \rgg} \Big).
\]

Similar estimates lead to \eqref{no_bind_ttr0place}. The estimates $\eps_1,\eps_2=o(1)$ follow from the fact that $n=|\un{\psi_1}|^2+|\un{\psi_2}|^2=|h_1|^2+|h_2|^2$, where the $h_j$'s satisfy $h_1\wedge h_2=\un{\psi_1}\wedge \un{\psi_2}=\un{\Psi}$ and
\[
D\big(|h_1|^2,|h_2|^2\big)=d_{\un{\Psi}}.
\]
In fact, this $o(1)$ is an $\mathcal{O}(\alpha +e^{-K\rgg})$.

\subsubsection{Direct term}
\paragraph{On the outside: $\etabfc{\la}$}.\,
By Lemma \ref{no_bind_decaylemma} and Kato's inequality (Appendix \ref{no_bind_appA}): 
\[
\ncc{(\etabfc{\la})^2 n}\apprle \frac{1}{c^{1/2} \la^2\rgg^2}.
\]

\paragraph{On the inside: $\xi^{(\tfrac{1}{3})}$}. We remark the following:
\begin{equation}\label{no_bind_tritri1_3}
\begin{array}{rl}
(\xi^{(\tfrac{1}{3})})^2&=(\xi^{(\tfrac{1}{3})})^2\big((\etabfc{\tfrac{1}{12}})^2+(\xi^{(\tfrac{1}{12})})^2\big)=(\etabfc{\tfrac{1}{12}})^2-(\etabfc{\tfrac{1}{12}})^2(\etabfc{\tfrac{1}{3}})^2+(\xi^{(\tfrac{1}{12})})^2(\xi^{(\tfrac{1}{3})})^2\\
 &=(\etabfc{\tfrac{1}{12}})^2-(\etabfc{\tfrac{1}{3}})^2+(\xi^{(\tfrac{1}{12})})^2.
 \end{array}
\end{equation}
Thus:
\[
\begin{array}{rl}
\Big|D\big((\xi^{(\tfrac{1}{3})})^2\rho_\g,(\etabfc{\tfrac{1}{3}})^2\rho_\g'  \big)\Big|&\le \ncc{(\etabfc{\tfrac{1}{3}})^2\rho_\g}(\ncc{(\etabfc{\tfrac{1}{3}})^2\rho_\g}+\ncc{(\etabfc{\tfrac{1}{12}})^2\rho_\g'})\\
   &+|D((\xi^{(\tfrac{1}{12})})^2\rho_\g,(\etabfc{\tfrac{1}{3}})^2\rho_\g')|\apprle \dfrac{\nlp{1}{\rho_\g}\nlp{1}{\rho'_\g}}{c\rgg}+o\big( \frac{L}{c\rgg}\big).
\end{array}
\]
We treat $D\big((\xi_1^{(\tfrac{1}{3})})^2\rho_\g, (\xi_1^{(\tfrac{1}{3})})^2\rho_\g' \big)$ in a similar way: it is $\mathcal{O}\big( \frac{L}{c\rgg}\big)$.
We have proved so far:
\[
\begin{array}{rl}
D(\rho'_\g,\rho'_\g)&=D((\xi_1^{(\tfrac{1}{3})})^2 \rho_\g',(\xi_1^{(\tfrac{1}{3})})^2 \rho_\g')+D((\xi_2^{(\tfrac{1}{3})})^2 \rho_\g',(\xi_2^{(\tfrac{1}{3})})^2 \rho_\g')\\
&+2D((\xi_1^{(\tfrac{1}{3})})^2 n, (\xi_2^{(\tfrac{1}{3})})^2n)+\mathcal{O}\big( \dfrac{L}{c\rgg}\big).
\end{array}
\]
In appendix \ref{no_bind_elds} we prove the following Lemma.
\begin{lemma}\label{no_bind_ultimchiant}
For $j=1,2$, we have:
\[
D\big((\xi_1^{(\tfrac{1}{3})})^2 \rho_\g',(\xi_1^{(\tfrac{1}{3})})^2\rho_\g'\big)=D\Big( \rho\Big[ \xi_j^{(\tfrac{1}{3})}\cdot\big[\g'\big]\Big],\rho\Big[ \xi_j^{(\tfrac{1}{3})}\cdot[\g']\Big]\Big)+\mathcal{O}\big( \dfrac{L}{c\rgg}\big).
\]
\end{lemma}

\subsubsection{Exchange term}
By Lemma \ref{no_bind_decaylemma} and Kato's inequality \eqref{no_bind_kato}: 
\[
\ttr\Big((\etabfc{\la})^2 NR_N\Big)\apprle \sum_j \nlp{2}{\etabfc{\la}\psi_j}^2\ttr\big( |\nabla| N\big)\apprle \dfrac{1}{c(\la\rgg)^2}=o\big(\frac{\alpha}{\la^2 cR_g}\big).
\]
With the same trick used before, we have:
\[
\diint \dfrac{|\g'(x,y)|^2}{|x-y|}dxdy=\diint ((\etabfc{\tfrac{1}{3}}(x))^2+(\xi^{\tfrac{1}{3}}(x))^2)\dfrac{|\g'(x,y)|^2}{|x-y|}((\etabfc{\tfrac{1}{3}}(y))^2+(\xi^{\tfrac{1}{3}}(y))^2)dxdy.
\]
We use Kato's inequality as usual to get:
\begin{align*}
\nqq{\etabfc{\la} \g'}\apprle \ns{2}{|D_0|^{1/2}\etabfc{\la} \g'}&\le \nb{[|D_0|^{1/2},\etabfc{\la}]\tfrac{1}{|D_0|^{1/2}}}\ns{2}{|D_0|^{1/2} \g'}+\ns{2}{\etabfc{\la}|D_0|^{1/2}\g'},\\
       &\apprle \dfrac{1}{c\sqrt{\la\rgg}}.
\end{align*}
Using trick \eqref{no_bind_tritri1_3}, we get%and
%\[
%|\g'(x,y)|^2\le 2|\g(x,y)|^2+2|N(x,y)|^2,
%\]
%we get
\[
\begin{array}{rl}
\diint \dfrac{|\g'(x,y)|^2}{|x-y|}dxdy&=\nqq{\xi_1^{\tfrac{1}{3}} \g'}^2+\nqq{\xi_2^{\tfrac{1}{3}} \g'}^2+2\diint (\xi_1^{\tfrac{1}{3}}(x))^2\dfrac{|\g'(x,y)|^2}{|x-y|}(\xi_2^{\tfrac{1}{3}}(y))^2dxdy\\
&+\mathcal{O}\Big(\dfrac{\ns{2}{\g}^2}{c\rgg}+\ttr\Big((\etabfc{\tfrac{1}{3}})^2 NR_N\Big)+\nqq{\etabfc{\tfrac{1}{12}} \g'}^2 \Big).
\end{array}
\]

Now let us show that for $j=1,2$:
\begin{equation}
\nqq{\xi_j^{\tfrac{1}{3}} \g'}^2=\nqq{\xi_j^{\tfrac{1}{3}}\cdot[\g']}^2+\mathcal{O}\big( \frac{1}{(c\rgg)^2}\big).
\end{equation}
It suffices to use Kato's inequality and Eq. \eqref{no_bind_comm}, we have:
\[
\begin{array}{rl}
\ns{2}{|D_0|^{1/2} \xi^{+-} Q}&\le \dfrac{1}{2\pi}\dint_{-\infty}^{+\infty}d\omega \Big\lvert\Big\lvert\dfrac{|D_0|^{1/2}}{D_0+i\omega} \boldsymbol{\alpha}\cdot \nabla \xi \dfrac{1}{D_0+i\omega} Q\Big\rvert\Big\rvert_{\mathfrak{S}_2}\\
     &\apprle \nlp{\infty}{\nabla (\xi_{c\la\rgg})}\ns{2}{Q}\dint_{-\infty}^{+\infty}\dfrac{d\omega}{E(\omega)^{3/2}}\apprle \dfrac{\ns{2}{Q}}{c\la\rgg}.
\end{array}
\]

%%%%%%%%%%%%%%%%%%%%%%%%%%%%%%%%%%%%%%%%%%%%%%%%%%%%%%%%%%%%%%%%%%%%%%%%%%%%%%%%%%%%

\begin{appendix}

\section{Estimates}\label{no_bind_appA}

%\subsection{Some inequalities}

%\textbullet\ We recall 
%\begin{equation}
%
%\end{equation}

%\begin{enumerate}

%\begin{notation}
%Let us define for any $a>0$:
%\[
%\phi_a(x-y):=|D_0|^{-a}(x-y),
%\]
%in particular for $0<a < 1$ there holds
%\[
%\phi_a(x-y)=\frac{1}{\Gamma(a)}\dint_{s=0}^{+\infty}\frac{ds}{s^{1-a}}e^{-s\sqrt{1-\Delta}}(x-y)
%\]
%which is \textit{well defined and positive} whenever $x-y\neq 0$. (\textit{cf} \cite{LL} for an expression of $e^{-s\sqrt{1-\Delta}(x-y)}$ with the modified Bessel function $\mathrm{K}_2$.)
%\end{notation}

\subsection{$[V,P^0_-]$ and proof of Proposition \ref{no_bind_conti}}
For any smooth complex valued function $V$, there holds \cite{gs}:
\begin{equation}\label{no_bind_comm}
[V,P^0_-]=-\frac{i}{2\pi}\dint_{-\infty}^{+\infty}\frac{1}{D_0+i\eta}\boldsymbol{\alpha}\cdot \nabla V\frac{d\eta}{D_0+i\eta}.
\end{equation}
Thanks to the KSS inequality as shown in \cite{locdef}, provided smoothness of $V$ ($\nabla V \in L^p$) then this operator is $\mathfrak{S}_p(L^2(\mathfrak{R}^3,\mathfrak{C}^4))$ for $p> 3$.

The integral kernel of its Fourier transform \cite{ptf} is:
\begin{equation}\label{no_bind_impft}
\mathscr{F}\big([V,P^0_-];p,q \big)=\frac{i}{2(2\pi)^{3/2}}\frac{1}{E(p)+E(q)}(\alpha_j \wh{\partial_j V}(p-q)-\sbf{p}\alpha_j \wh{\partial_j V}(p-q)\sbf{q}).
\end{equation}

We prove Proposition \ref{no_bind_conti} by duality, following \cite{gs}. Let $V$ be in $\mathcal{S}(\RR)$, $Q\in\mathfrak{S}_1^{P^0_-}$ (we recall that $2\ala=1+\tfrac{1}{\llo}$), then
\[
\begin{array}{rl}
\ttr_0(Q V)&=\ttr(P^0_+ Q (P^0_++P^0_-) V P^0_+)+\ttr(P^0_- Q (P^0_++P^0_-) V P^0_-).
\end{array}
\]
The operator $Q^{+-}|D_0|^{\ala} \tfrac{1}{|D_0|^{\ala}}[P^0_-, V]$ is in $ \mathfrak{S}_1$: indeed thanks to \eqref{no_bind_impft} we have
\[
\diint \frac{|\wh{V}(p-q)|^2|p-q|^2dpdq}{E(p)^{1+\tfrac{1}{\llo}}(E(p)+E(q))}\apprle \llo\nlp{2}{\nabla V}^2
\]
showing $\ns{2}{\tfrac{1}{|D_0|^{\ala}}[P^0_-, V]}\apprle \sqrt{\llo}\nlp{2}{\nabla V}$. This also treats the case 

\noindent $Q^{-+}V^{+-}\in\mathfrak{S}_1.$ Then we have $Q^{++} V^{++}=Q^{++} |D_0|^{\ala} \tfrac{1}{|D_0|^{\ala}} V^{++}\in \mathfrak{S}_1$. 

Indeed $|D_0|^{\ala}Q^{++} |D_0|^{\ala}\in \mathfrak{S}_1$ and $\tfrac{1}{|D_0|^{\ala}} V^{++}\in\mathfrak{S}_6$ with norm $\mathcal{O}((\llo)^{1/6}\nlp{2}{\nabla V})$. 

Then $\tfrac{1}{|D_0|^{\ala}} V^{++}\tfrac{1}{|D_0|^{\ala}}\in\mathfrak{S}_6$ with norm $\mathcal{O}(\nlp{2}{\nabla V})$. So:
\[
\begin{array}{rl}
\ttr(Q^{++} V^{++})&=\ttr\big(\dfrac{|D_0|^{\ala}}{|D_0|^{\ala}}Q^{++}\dfrac{|D_0|^{\ala}}{|D_0|^{\ala}} V^{++}\big)\\
                                &=\ttr\big( \{|D_0|^{\ala} Q^{++} |D_0|^{\ala}\} \{\frac{1}{|D_0|^{\ala}} \} V^{++}\frac{1}{|D_0|^{\ala}}\big)\\
                                &=\mathcal{O}\big(\ns{1}{|D_0|^{\ala} Q^{++} |D_0|^{\ala}} \nlp{2}{\nabla V} \big).
\end{array}
\]
The same holds for $Q^{--} V^{--}$. This ends the proof.

\begin{remark}
In Appendix \ref{no_bind_elds} we do analogous estimates but with an additional localisation operator.
\end{remark}

We adapt \cite[Lemma 5]{locdef}:
\begin{lemma}\label{no_bind_commd}%[
Let $p$ be in $(3,+\infty]$ and $V$ a smooth function with $\nabla V\in L^p$. Then for any $0<a<1$:
\begin{equation}
[|D_0|^a, V]\tfrac{1}{|D_0|^a}\in\mathfrak{S}_p.
\end{equation}
\end{lemma}%)
To prove it we use \cite[p. 87]{stabilitymatter}
\begin{equation}\label{no_bind_sauve}
\forall\,x>0,\,0<a<1:\ x^a=\frac{\sin(a\pi)}{\pi}\dint_0^{+\infty}\frac{ds}{s^{1-a}}\frac{x}{x+s}.
\end{equation}

\subsection{Proof of Lemma \ref{no_bind_H2}}\label{no_bind_ph2}
\begin{dem}
Let us explain the bootstrap argument.

\noindent -- We show that $\ttr((-\Delta)^{a+1}N)\apprle 1$. As a consequence:
\[
\begin{array}{rl}
\nlp{2}{|\nabla|^a n_j}&\le \sum_{\ell=0}^aK(\ell,a)\nlp{4}{|\nabla|^\ell\mathscr{F}^{-1}(|\wh{\psi}_j|)}\nlp{4}{|\nabla|^{a-\ell}\mathscr{F}^{-1}(|\wh{\psi}_j)|}\\
                                       &\apprle \sum_{\ell=0}^aK(\ell,a)\nlp{2}{|\nabla|^{\ell+3/4}\mathscr{F}^{-1}(|\wh{\psi}_j|)}\nlp{2}{|\nabla|^{a-\ell+3/4}\mathscr{F}^{-1}(|\wh{\psi}_j|)}\\
                                       &\apprle K(a).
\end{array}
\]

\noindent -- As shown in \cite{sokd}, $(\g',\rho'_\g)$ is the fixed point of some function $F^{(1)}$ in a ball of $\wit{\mathcal{X}}_a$:
\[
\wit{\mathcal{X}}_a=\{(Q,\rho)\in \mathfrak{S}_2\times \mathcal{S}': \diint E(p-q)^{2a}E(p+q)|\wh{Q}(p,q)|^2<+\infty\text{\ and\ } \dint\frac{E(k)^a}{|k|^2}|\wh{\rho}(k)|^2<+\infty\}.
\]
\noindent -- We multiply by $|D_0|^{(a+3)/2}$ the equation $D_0\psi_j=\mathcal{L}_\La^{-1}(\mu_j\psi_j-\alpha B_{\g'}\psi_j)$ and we show that $\ttr((-\Delta)^{a+2}N)\apprle 1$. We have to deal with $[|D_0|^{(a+3)/2},v]\psi_j$ and $[|D_0|^{(a+3)/2},R]\psi_j$: it suffices to compute in Fourier space and to use Taylor's formula on the function $E(\cdot)^{(a+3)/2}$.

\paragraph{Proof of the estimates} Here as $\ttr(-\Delta N)\apprle 1$, the fixed point method can be applied on $\wit{\mathcal{X}}_{a=1}$. Indeed $\nlp{2}{n}\apprle \nlp{2}{|\nabla|^{3/2}\sqrt{n}}\apprle 1$. We get that 
\[
\diint |p-q|E(p+q)|\wh{\g}(p,q)|^2dpdq\apprle 1.
\]

Let us show the assumption on the $H^2$-norm of $\psi_j$. 

\noindent There holds $f(-i\nabla)\Dbf \psi_j=f(-i\nabla)(\mu_j-\alpha B[\g])\psi_j$ for any $f\ge 0$. Taking the $L^2$-norm we have to deal with $[f(-i\nabla),R_{\g'}]$ and $[f(-i\nabla),v[\rho(\g')]]$. For $f(-i\nabla)=|\nabla|^{1/2}$ there holds
\[
\begin{array}{rl}
\nlp{2}{[|\nabla|^{1/2},v_\rho]\psi}^2&\apprle \diint\frac{|\wh{\rho}(p-q)|^2}{|p-q|^2}\frac{dpdq}{|q|^2E(q)^2}\dint dq E(q)^2|q||\wh{\psi}(q)|^2\\
\nlp{2}{[|\nabla|^{1/2},R_Q]\psi}^2&\apprle \diint |p-q| |\wh{Q}(p,q)|^2dpdq\,\nlp{2}{|\nabla|^{1/2}\psi}^2\\
|\nabla|^{1/2}D_0\psi&=\mu\frac{|\nabla|^{1/2}}{\mathcal{L}_\La}\psi-\alpha \frac{|\nabla|^{1/2}}{\mathcal{L}_\La}B\psi=\mathcal{O}_{L^2}(1)\text{\ a\ priori}\\
|\nabla|^{1/2}B\psi&=[|\nabla|^{1/2},B]\psi+B\tfrac{1}{|\nabla|^{1/2}}|\nabla| \psi\text{\ and:}
\end{array}
\]
\[
\psh{|\nabla|(1-\Delta)\psi_1}{\psi_1}-\psh{|\nabla|\psi_1}{\psi_1}\apprle \alpha c^{-1}\nlp{2}{v_2 \psi_1-v_{21}\psi_2}+c^{-3}+\alpha^2c^{-2}=\mathcal{O}(c^{-3}+\alpha c^{-1}a_{12}).
\]

We get $\ttr(|D_0|^3N)\apprle 1$ and by the fixed-point Theorem: 

\[
\nqbf{\g}^2=\diint E(p-q)^2E(p+q)|\wh{\g}(p,q)|^2dpdq\apprle 1.
\] 
\begin{notation}
The star in $\nqbf{\cdot}^*$ means that we replace $E(p-q)^2E(p+q)$ by $|p-q|^2|p+q|$.
\end{notation}
Using the methods of \cite{ptf,sok} we have:%that this squared norm is $\mathcal{O}(\alpha^2)$ and that:
\[\left\{
\begin{array}{rl}
\nqbf{\g}^*&\apprle c^{-1/2}\nlp{2}{\rho'_\g}+\alpha (\nqbf{\g'}^*)+\alpha(\nlp{2}{\rho'_\g}+\nqbf{\g'}^*)\ssum_{k=1}^{+\infty}\sqrt{k}(\alpha K(\ncc{\rho'_\g}+\nqbfu{\g'}))^k,\\
\ns{2}{[\nabla,\g]}&\apprle \alpha (\nlp{2}{\rho'_\g}+\nqbf{N}^*)+\alpha(\nlp{2}{\rho'_\g}+\nqbf{\g'}^*)\ssum_{k=1}^{+\infty}\sqrt{k}(\alpha K(\ncc{\rho'_\g}+\nqbfu{\g'}))^k,\\
\nlp{2}{\rho_\g}&\apprle L\nlp{2}{n}+c^{-1/2}\nqbf{\g'}^*+\alpha(\nlp{2}{\rho'_\g}+\nqbf{\g'}^*)\ssum_{k=1}^{+\infty}\sqrt{k}(\alpha K(\ncc{\rho'_\g}+\nqbfu{\g'}))^k.
\end{array}\right.
\]
Therefore 
\[\nqbf{\g'}^*=\mathcal{O}(c^{-2}),\,\ns{2}{[\nabla,\g]}=\mathcal{O}(\alpha c^{-3/2})\text{\ and\ }\nlp{2}{\rho_\g}=\mathcal{O}(Lc^{-3/2}+c^{-2}+c^{-1}(\sqrt{\alpha a_{12}})).\]

For $f(-i\nabla)=\partial_k$ with $k=1,2,3$ we have:
\[
\begin{array}{rl}
\partial_kR_Q \psi &=[\partial_k,R[Q]]\psi+R_Q\partial_k \psi\text{\ and\ }\partial_k v \psi=(\partial_k v)\psi+v(\partial_k\psi)\\
\nlp{2}{[\partial_k,R_Q]\psi}&=\nlp{2}{R([\partial_k,Q])\psi}\le \ns{2}{[\partial_k,Q]}\nlp{2}{\nabla \psi}\text{\ and\ }\nlp{2}{R_Q\partial_k \psi}\le \ns{2}{Q}\nlp{2}{\Delta\psi}\\
\nlp{2}{v_\rho(\partial_k\psi)}&\le \nlp{6}{v_\rho}\nlp{3}{\partial_k\psi}\apprle \ncc{\rho}\nlp{2}{|\nabla|^{3/2}\psi}\le \ncc{\rho}\sqrt{\nlp{2}{\nabla\psi}\nlp{2}{\Delta\psi}}\\
\nlp{2}{(\partial_k v_\rho)\psi}^2&\apprle \diint\frac{|\wh{\rho}(k)|^2}{|k|^2}\frac{dkdq}{|q|^2(1+|q|^2)}[\nlp{2}{\nabla\psi}^2+\nlp{2}{\Delta\psi}^2] 
\end{array}
\]

\[
\begin{array}{rl}
\ssum_{k=1}^3(\nlp{2}{\partial_k \Dbf\psi}^2)-\nlp{2}{\nabla\psi}^2&\le (\mu^2-1)\nlp{2}{\nabla\psi}^2+6\alpha\mu\nlp{2}{\nabla\psi}\nlp{2}{B[\g']\psi}+\alpha^2\nlp{2}{\nabla B[\g']\psi}^2\\
\ttr(\Delta^2(1-\tfrac{\Delta}{\La^2}+\tfrac{\Delta^2}{\La^4})N)&\apprle \alpha a_{12}c^{-1}+c^{-3}.
\end{array}
\]
This gives $\nlp{2}{\Delta\psi_j}^2\apprle \alpha c^{-2}$ and in particular:
\[
\nlp{2}{c^2(1-\mathcal{L}_{c\La}^{-1})\un{\psi_j}}=\mathcal{O}\big(\frac{\sqrt{\alpha} c}{\La^2}\big).
\]
As a consequence we have:
\begin{equation}\label{no_bind_chichi}
\nlp{2}{\nabla \un{\chi_j}}=\nlp{2}{i\boldsymbol{\sigma}\cdot\nabla \un{\chi_j}}=\mathcal{O}(c^{-1}).
\end{equation}
\end{dem}

Thanks to those estimates, we get:
\begin{equation}
\mathcal{E}_{\text{BDF}}(\g+N)=2+\frac{\mathcal{E}_{\text{PT}}(\un{\psi_1}\wedge \un{\psi_2})}{2c^2}+\mathcal{O}(\alpha^2 c^{-3/2}+c^{-3}).
\end{equation}

We recall that $1-\mathcal{L}_\La^{-1}=\frac{-\Delta}{\La^2-\Delta}.$
%\begin{remark}
%We recall:
%\begin{equation}
%1-\mathcal{L}_\La^{-1}=\frac{-\Delta}{\La^2-\Delta}.
%\end{equation}
%\end{remark}
\medskip

Thanks to Section \ref{no_bind_ptt} there holds
\[
D(\un{n}_1,\un{n}_2)-D(\un{\psi_1}^*\un{\psi_2},\un{\psi_1}^*\un{\psi_2})\apprle c^{-1}\text{\ and\ }a_{12}\apprle \alpha^{3/2}c^{-1}.
\]
From this point we get better estimate on $\nlp{2}{\Delta \psi}^2\apprle c^{-3}$ but this is still unsatisfactory. Let us be more precise about $\mu=\psh{(\Dbf+\alpha B) \psi}{\psi}$ and $\chi$:
\[
\begin{array}{rl}
(1+\mu_1)\chi_1&= -i\sigma\cdot \nabla \phi_1-\frac{\mu\Delta}{\La^2-\Delta}\chi_1+\frac{\alpha}{\mathcal{L}_\La}(v_{\rho_\g}\chi_1+(v_{2}\chi_1-v_{21}\chi_2)-(R_\g\psi_1)_{\downarrow})\\
       &=\frac{1}{1+\mu}(-i\sigma\cdot \nabla \phi_1+X_1^{(r)})=\dfrac{-i\sigma\cdot \nabla}{2} \phi_1+\mathcal{O}_{L^2}(c^{-2}/\La+c^{-2})\\
\psh{\Dbf \psi}{\psi}&=\psh{D_0\psi}{\psi}-\psh{\tfrac{\Delta}{\La^2}\beta \psi}{\psi}+\psh{\tfrac{\Delta}{\La^2}-i\boldsymbol{\alpha}\cdot \nabla \psi}{\psi}\\
           &=1-2\nlp{2}{\chi}^2+2\mathfrak{Re}\psh{-i\sigma\cdot\nabla \varphi}{\chi}+\mathcal{O}\big(\frac{\nlp{2}{\nabla\psi}^2}{\La^2}+\nlp{2}{\Delta \varphi}\frac{\nlp{2}{\nabla \chi}}{\La^2}\big)\\
           &=1+\frac{2}{1+\mu}\big(1-\frac{1}{1+\mu}\big)\nlp{2}{\nabla\varphi}^2+\mathfrak{Re}\frac{2}{1+\mu}\big(1-\frac{2}{1+\mu}\big)\mathfrak{Re}\psh{-i\sigma\cdot\nabla \varphi}{X^{(r)}}+\mathcal{O}\big(\frac{1+\nlp{2}{\Delta \varphi}}{c^2\La^2}\big)\\
           &=1+\frac{1}{2}\nlp{2}{\nabla\varphi}^2+\mathcal{O}(c^{-4}+c^{-2}\La^{-2}(1+\nlp{2}{\Delta \varphi})).
\end{array}
\]
Then:
\[
\begin{array}{rl}
\nlp{2}{\mathcal{L}_\La^{-1}\psi}^2&=1+\mathcal{O}(c^{-2}\La^{-2}+\nlp{2}{\Delta\psi}^2/\La^4)\\
\nlp{2}{\nabla \mathcal{L}_\La^{-1}\psi}^2&=\nlp{2}{\nabla\psi}^2+\mathcal{O}(\nlp{2}{\Delta\psi}/(c\La^2)+\nlp{2}{\Delta\psi}^2/\La^2)\\
-2\alpha \mu\mathfrak{Re}\psh{\frac{1-\Delta}{\mathcal{L}_{\La}}B\psi}{\psi}&=-2\alpha\mu\psh{B\psi}{\psi}+\mathcal{O}(\alpha\nlp{2}{B\psi}\nlp{2}{\Delta \psi}/\La^2)\\
\nlp{2}{-i\boldsymbol{\alpha}\nabla B\psi}&\apprle \nlp{2}{[\nabla, B]\psi}+\nlp{2}{B\nabla \psi}=\mathcal{O}(c^{-3/2}+\nlp{2}{\Delta \psi}^{1/2}c^{-1}+\nlp{2}{\Delta \psi}c^{-1/2}).
\end{array}
\]
and thus:
\[
\begin{array}{rl}
\psh{(1-\Delta)\psi}{(1-\Delta)\psi}&=\mu^2\psh{\frac{1-\Delta}{\mathcal{L}_{\La}^2}\psi}{\psi}-2\alpha \mu\mathfrak{Re}\psh{\frac{1-\Delta}{\mathcal{L}_{\La}}B\psi}{\psi}+\nlp{2}{\frac{D_0}{\mathcal{L}_{\La}}B\psi}^2\\
 &=1+2(\mu-1-\alpha\psh{B\psi}{\psi})+\nlp{2}{\nabla\psi}^2\\
  &\ \ \ +\mathcal{O}(c^{-2}(c^{-2}+\La^{-2})+\frac{\nlp{2}{\Delta\psi}}{c^{2}\La^{2}}+\nlp{2}{\Delta\psi}^2(\La^{-2}+\alpha^2c^{-1})).
\end{array}
\]
From \eqref{no_bind_equaj} and the expression of $D_0\psi_j$, we have $\nlp{2}{\nabla \psi_j}^2=-2\alpha \mathfrak{Re}\psh{B\psi_j}{\psi_j}$. We conclude $\nlp{2}{\Delta \psi}^2\apprle c^{-2}(c^{-2}+\La^{-2})$ and
\[
\nlp{2}{\Delta \psi}^2\apprle \min\big(c^{-3},c^{-2}(c^{-2}+\La^{-2}\big).
\]

%%%%%%%%%%%%%%%%%%%%%%%%%%%%%%%%%%%%
%%%%%%%%%%%%%%%%%%%%%%%%%%%%%%%%%%%%

\section{Proofs of  Section \ref{no_bind_the_PT_fun}}\label{no_bind_ptt}
%\begin{remark}
%In this part $o,\mathcal{O}$ mean $\underset{j\to+\infty}{o},\underset{j\to+\infty}{\mathcal{O}}$.
%\end{remark}

\subsection{Proof of Proposition \ref{no_bind_eloig}}
\emph{Reductio ad absurdum}. 

We assume this is false and take a non-increasing sequence $(a_j)_{j\ge 0}$ tending to $0$ such that there exists $\Psi_j$ that does not satisfy \eqref{no_bind_ptd} with $\text{b}=a_j$: $\Delta_2\mathcal{E}<a_j$ and $\frac{\Delta_2 \mathcal{E}}{d_\Psi}<a_j$. In particular $(\Psi_j)_j$ is a minimizing sequence for $E_{\text{PT}}(2)$. By geometrical methods \cite{geom} we see that $\Psi_j$ can be decomposed in two pieces of mass one, each piece tending to a minimizer for $E_{\mathrm{PT}}(1)$. Indeed it is clear that $(\ttr(-\Delta \g_{\Psi_j}))_j$ is bounded and that there is no vanishing for $(\rho_{\Psi_j})_{j\ge 0}$. If we follow a bubble \cite{PC} of $\rho_{\Psi_j}$ (one of the biggest) let us show its mass is $1$ at the limit.

\noindent By scaling, for any $0<\la<1$ we have $\ E_{\text{PT}}(\la)\ge \la^3 E_{\mathrm{PT}}(1)$, where $E_{\text{PT}}(\la)$ is defined as the infimum of $\mathcal{E}_{\text{PT}}$ over non-negative one-body density matrix whose trace is $\la$. 

Up to following a bubble and extracting a subsequence there holds with $\Psi_j=h_{1,j}\wedge h_{2,j}:$
\[
\ket{h_{1,j}\wedge h_{2,j}}\bra{h_{1,j}\wedge h_{2,j}}\rightharpoonup _gG_{00}\oplus G_{11}\oplus G_{22},\ \sum_{j=0}^2\ttr(G_{jj})=1\text{\ and\ }\ttr(G_{00})<1.
\]
We recall that each $G_{jj}$ is a density matrix in $(L^2)^{\wedge (j)}$. Following \cite[part 5]{geom}: 

\noindent $G_{jj}=\ttr(G_{jj})\wit{G}_{jj}$
\[
\begin{array}{rl}
\underset{j\to+\infty}{\liminf}\mathcal{E}_{\text{PT}}^U(\Psi_j)&=E_{\text{PT}}^U(2)\ge \ssum_{j=0}^2(\mathcal{E}_{\text{PT}}^U(G_{jj})+\ttr(G_{jj})E_{\text{PT}}^U(2-j))\\
     &\ge \ssum_{j=0}^2\ttr(G_{jj})(\mathcal{E}_{\text{PT}}^U(\wit{G}_{jj})+E_{\text{PT}}^U(2-j))\ge E_{\text{PT}}^U(2).
 \end{array}
\]
As not all particles are lost (we follow a bubble) either $G_{11}\neq 0$ or $G_{22}\neq 0$. In the case $G_{2,2}\neq 0$, \cite{bip} enables us to say $\mathcal{E}_{\text{PT}}^U(\wit{G}_{22})>E_{\text{PT}}(2)$. So $G_{22}=0$ and $G_{11}\neq 0$. Thanks to \cite{L} and Lieb's variational principle (we may assume $G_{11}=\ttr(G_{11})\ket{\phi}\bra{\phi}$) there holds
\[
\mathcal{E}_{\text{PT}}(G_{11})\ge (\ttr(G_{11}))^3E_{\mathrm{PT}}(1),
\]
then necessarily $\ttr(G_{11})=1$.

As a consequence there is exactly two bubbles in $(\rho_{\Psi_j})_j$, there exist a decomposition $\Psi_j=\un{h_{1,j}}\wedge\un{h_{2,j}}$ and a sequence $(z_{1,j};z_{2,j})_j$ of $(\mathbb{R}^3)^2$ such that (up to extraction)
\begin{enumerate}
\item $\psh{\un{h_{k,j}}}{\un{h_{\ell,j}}}=\delta_{k\ell}$ and $|z_{1,j}-z_{2,j}|\underset{j\to\infty}{\to}+\infty$,
\item $\un{h_{k,j}}(\cdot -z_{k,j})\overset{H^1}{\underset{j\to\infty}{\to}}\phi_{j,\infty}$ where $\phi_{j,\infty}\in\mathscr{P}$ is radial.
\end{enumerate}

Then it suffices to compute: $\mathcal{E}_{\text{PT}}^U(\Psi_j)$ with this decomposition:
\[
\begin{array}{rl}
\mathcal{E}_{\text{PT}}^U(\Psi_j)&=\mathcal{E}_{\text{PT}}^U(\un{h_{1,j}})+\mathcal{E}_{\text{PT}}^U(\un{h_{2,j}})-D(|\un{h_{1,j}}|^2,|\un{h_{2,j}}|^2)+\frac{U}{2}\diint|\un{h_{1,j}}\wedge \un{h_{2,j}}(x,y)|^2\frac{dxdy}{|x-y|}\\
 &=\mathcal{E}_1+\mathcal{E}_2+\frac{U}{2}W_{12}-\un{D_{12}}\ge \frac{U}{4}W_{12}+2E_{\mathrm{PT}}(1).
\end{array}
\]
The last equality holds because we have $U> 2\text{U}_c$. Let us write 
\[
\Delta_1 \mathcal{E}:=\mathcal{E}_{\text{PT}}(\un{h_{1,j}})+\mathcal{E}_{\text{PT}}(\un{h_{2,j}})-2E_{\mathrm{PT}}(1).
\]
Then:
\[
-a_j<\Delta_1 \mathcal{E}-\un{D_{12}}<a_j\text{\ and\ }\Delta_1 \mathcal{E}\ge \kappa \ssum_{k=1}^2\nso{1}{\un{h_{k,j}}-\phi_{k,j}}^2
\]
where $\phi_{k,j}\in\mathscr{P}$ is the closest function to $\un{h_{k,j}}$ in $H^1$ (Proposition \ref{no_bind_coer}).
We may assume that $\un{D_{12}}=d_{\Psi_j}$ because minimizing this quantity corresponds to minimizing $\Delta_1 \mathcal{E}$. In particular:
\[
|\Delta_1 \mathcal{E}-\un{D_{12}}|<a_j=\underset{j\to+\infty}{o}(\un{D_{12}})\ \Rightarrow\ \Delta_1 \mathcal{E}\underset{j\to+\infty}{\sim}\un{D_{12}}\gg a_j.
\]
Indeed, let us say that $\un{D_{12}}>d_{\Psi_j}$, then $(\un{f_{k,j}}(\cdot-z_k))_j$ still converges to $\phi_{j,\infty}$, in particular $(\Delta_1\mathcal{E})_j$ converges to $0$. But if $(\un{f'_{1,j}},\un{f'_{2,j}})_j$ is a decomposition with $\un{D_{12}'}=d_{\Psi_j}$, then $\Delta_1'\mathcal{E}\le \Delta_1\mathcal{E}$ and 
\[
\text{dist}(\un{f'_{k,j}},\mathscr{P})\underset{j\to+\infty}{\to} 0.
\]

From now we will drop the subscript $j$ for convenience and suppose $\un{D_{12}}=d_{\Psi_j}$.
\begin{notation}
We introduce $\un{h_k}=(\un{h_k}-\phi_{k})+\phi_{k}=\delta_k-\phi_k$ in $|\un{h_k}|^2$ and in $\un{h_1}^*\un{h_2}$. We use the convention 
\[
\nlp{2}{\delta}:=\nlp{2}{\delta_1}+\nlp{2}{\delta_2},\ \nso{1}{\delta}:=\nso{1}{\delta_1}+\nso{1}{\delta_2}.
\]
\end{notation}

We recall that an element of $\mathscr{P}$ has an exponential falloff with respect to its center. For some constant $\ov{\eps}>0$, there holds:
\[
\begin{array}{rl}
|\un{h_k}|^2&=|\delta_k|^2+|\phi_k|^2+2\mathfrak{Re}(\delta_k^*\phi_k)\\
\un{h_1}^*\un{h_2}^*&=\delta_1^*\delta_2+\phi_1^*\phi_2+\delta_1^*\phi_2+\phi_1^*\delta_2\\
\ncc{\un{h_1}^*\un{h_2}}^2&=\ncc{\delta_1^*\phi_2}^2+\ncc{\phi_1^*\delta_2}^2+\mathcal{O}\big((\nlp{2}{\delta_1}\nlp{2}{\delta_2})(\rgg^{-1}+\nlp{2}{\delta}(1+\nlp{2}{\nabla\delta})+e^{-\ov{\eps} \rgg})\big)\\%+2\mathfrak{Re}D(\delta_1^*\delta_2, \delta_2^*\phi_1+\phi_1^*\delta_2)\\
\un{D_{12}}&=D(|\phi_1|^2,|\phi_2|^2)+D(|\delta_1|^2,|\phi_2|^2)+D(|\phi_1|^2,|\delta_2|^2)\\
 &\ \ \ +\mathcal{O}\big( \frac{\nlp{2}{\delta}}{\rgg}+\nlp{2}{\delta_1}\nlp{2}{\delta_2}(\nlp{2}{\delta}(1+\nlp{2}{\nabla\delta})+e^{-\ov{\eps} \rgg})\big)
\end{array}
\]
\[
\begin{array}{ll}
\text{\ Thus:}& a_j U^{-1}\apprge \un{D_{12}}-\ncc{\un{h_1}^*\un{h_2}}^2\apprge \dfrac{1}{\rgg}+\underset{j\to+\infty}{\mathcal{O}}(\nlp{2}{\delta}^3)
\end{array}
\]
\[
\begin{array}{ll}
\text{and}&\frac{1}{\rgg}=\underset{j\to+\infty}{\mathcal{O}}(a_j U^{-1}+\nlp{2}{\delta}^3).
\end{array}
\]
As $j\to+\infty$, thanks to the coercivity inequality \eqref{no_bind_coer} there holds 
\[
\un{D_{12}}\sim \Delta_1 \mathcal{E}=\Theta(\nso{1}{\delta_1}^2+\nso{1}{\delta_2}^2)\text{\ and\ }\frac{1}{\rgg}=\underset{j\to+\infty}{o}(\un{D_{12}}).
\]
Studying more precisely $M^2(\un{h_1}\wedge \un{h_2}):=\iint |\un{h_1}\wedge \un{h_2}(x,y)|^2\tfrac{dxdy}{|x-y|}$:
\begin{equation}\label{no_bind_imp}
\begin{array}{rl}
M^2(\un{h_1}\wedge \un{h_2})&=M^2(\delta_1\wedge \phi_2)+M^2(\phi_1\wedge \delta_2)+\underset{j\to+\infty}{\mathcal{O}}(\rgg^{-1}+\nlp{2}{\delta}^3)=\underset{j\to+\infty}{o}(\un{D_{12}})\\
\un{D_{12}}&=D(|\delta_1|^2,|\phi_2|^2)+D(|\phi_1|^2,|\delta_2|^2)+\underset{j\to+\infty}{o}(\un{D_{12}})\apprge \nso{1}{\delta_1}^2+\nso{1}{\delta_2}^2.
\end{array}
\end{equation}
We can easily exclude the case $\delta_1,\delta_2=0$ for then it is clear $M^2(\phi_1\wedge\phi_2)\apprge D(|\phi_1|^2,|\phi_2|^2)$ thanks to $\psh{\phi_1}{\phi_2}=0$. Say then that $\nso{1}{\delta_1}\ge \nso{1}{\delta_2}$: $\delta_1\neq 0$. The case $\delta_2=0$ and $\delta_1\neq 0$ is an easy adaptation of what follows, we treat it later. As there holds
\[
|\phi_2|^2*\tfrac{1}{|\cdot|}(x)\le \tfrac{1}{|x-z_2|}
\]
where $z_2$ is the center of $\phi_2$, Estimate \eqref{no_bind_imp} is true only if there lies a mass of $\delta_1$ near $z_2$: the quantity $\ncc{\delta_1^*\phi_2}^2$ must compensate $D(|\delta_1|^2,|\phi_2|^2)$. Eventually the same phenomena occurs for $\delta_2$ around $z_1$ the center of $\phi_1$. Up to extraction:
\[
\frac{\delta_k(\cdot-z_k)}{\nso{1}{\delta_k}}\rightharpoonup_{H^1}\ell_k,
\]
and $(\ell_1,\ell_2)\neq (0,0)$. Indeed up to contraction there is convergence in $L^2_{\text{loc}}$ and if $\ell_k=0$ then for all $r>0$ and $(i_1,i_2)\in\big\{(1,2),(2,1)\big\}$
\[
\underset{j\to+\infty}{\limsup} \dint \frac{|\delta_{i_1}(x)|^2}{\nso{1}{\delta_{i_1}}^2}|\phi_{i_2}|^2*\frac{1}{|\cdot|}(x)dx\le \frac{1}{r}+\underset{j\to+\infty}{\limsup} \dint_{|x-z_{i_2}|\le r}\frac{|\delta_{i_1}(x)|^2}{\nso{1}{\delta_{i_1}}^2}|\phi_2|^2*\frac{1}{|\cdot|}(x)dx=\frac{1}{r},
\]
 this would contradict \eqref{no_bind_imp}. Then as we have:
\[
\underset{j\to+\infty}{\lim}M^2\big(\frac{\delta_1}{\nso{1}{\delta_1}}\wedge \phi_2\big)=\underset{j\to+\infty}{\lim}\frac{1}{\un{D_{12}}}M^2\big(\delta_1\wedge \phi_2\big)=0,
\]
then necessarily $\ell_1=\eps_1 \phi_{2,\infty}$ with $|\eps_1|\le 1$. Furthermore, either $\nso{1}{\delta_2}=\underset{j\to+\infty}{o}(\nso{1}{\delta_1})$ or $\nso{1}{\delta_2}=\underset{j\to+\infty}{\Theta}(\nso{1}{\delta_1})$.

\medskip

\noindent -- In the first case then $\nso{1}{\delta_2}^2=\underset{j\to+\infty}{o}(\un{D_{12}})$ and $\ell_1\neq 0$. We get a contradiction by computing:
\[
\begin{array}{rl}
0=\dint \un{h_1}^*\un{h_2}&=\dint \phi_1^*\phi_2+\dint \delta_1^*\phi_2+\dint \phi_1^*\delta_2+\dint \delta_1^*\delta_2\\
   &=\underset{j\to+\infty}{\mathcal{O}}(e^{-\ov{\eps}\rgg})+\dint \delta_1^*\phi_2+\underset{j\to+\infty}{\mathcal{O}}(\nlp{2}{\delta_2}(1+\nlp{2}{\delta_1}))\\
   &=\dint \delta_1^*\phi_2+\underset{j\to+\infty}{o}(\nso{1}{\delta_1}).
\end{array}
\]
\noindent -- In the second case we also get $\underset{j\to+\infty}{\lim}\nso{1}{\delta_2}^{-2}M^2(\delta_2\wedge \phi_1)$ and $\ell_2=\eps_2 \phi_{1,\infty}$, $|\eps_2|\le 1$. Writing for $k\neq k':\,\un{h_k}=\phi_k+\eps_k\nso{1}{\delta_k} \phi_{k'}+h_{k}^{(r)}$, up to extraction the following holds:
\[
\begin{array}{rl}
0=\dint \un{h_1}^*\un{h_2}&=\underset{j\to+\infty}{\mathcal{O}}(e^{-\ov{\eps}\rgg})+\eps_1^*\nso{1}{\delta_1}+\eps_2\nso{1}{\delta_2}+\dint (h_1^{(r)})^*\un{h_2}+\dint\un{h_1}^*h_2^{(r)}\\
\dint (h_1^{(r)})^*\un{h_2}&=\dint (h_1^{(r)})^*\phi_2+\dint (h_1^{(r)})^*(\eps_1\nso{1}{\delta_1}\phi_1)+\dint (h_1^{(r)})^*h_2^{(r)}\\
     &=\underset{j\to+\infty}{o}(\nso{1}{\delta_1})+\underset{j\to+\infty}{\mathcal{O}}(\nso{1}{\delta_1}^2)+\underset{j\to+\infty}{\mathcal{O}}(\nso{1}{\delta_1}\nso{1}{\delta_2}).
\end{array}
\]
The $\underset{j\to+\infty}{o}(\nso{1}{\delta_1})$ comes from the $L^2_{\text{loc}}$-convergence to $0$ of $\tfrac{h_1^{(r)}(\cdot-z_2)}{\nso{1}{\delta_1}}$ and the uniform shape of the $\phi_2(\cdot-z_2)$'s. In particular: 
\[
\eps_1^*\nso{1}{\delta_2}=-\eps_2\nso{1}{\delta_1}+\underset{j\to+\infty}{o}(\nso{1}{\delta}).
\]
Writing $\eps_1\nso{1}{\delta_1}=a$ and $\eps_2\nso{2}{\delta_2}=b=-a^*+(\delta a):$
\[
\left\{
\begin{array}{rl | rl}
\un{h_1}&=\phi_1+a\phi_2+h_1^{(r)}& h_1^{(r)}&=\delta_1-a\phi_2\\
\un{h_2}&=\phi_2-a^*\phi_1+(\delta a)\phi_1+h_2^{(r)}&  h_2^{(r)}&=\delta_2-b\phi_2.
\end{array}\right.
\]
We apply $\begin{pmatrix} \sqrt{1-|a|^2} & a^*\\ -a & \sqrt{1-|a|^2}\end{pmatrix}$ with $\sqrt{1-|a|^2}=:s$
\[
\begin{pmatrix}\un{g_1} \\ \un{g_2}\end{pmatrix}=\begin{pmatrix} \phi_1(s+|a|^2-a(\delta a))+\phi_2(a(s-1))+sh_1^{(r)}-ah_2^{(r)}\\ \phi_2(s+|a|^2)+\phi_1(a^*(1-s)+(\delta a)s)+sh_2^{(r)}+a^*h_1^{(r)} \end{pmatrix},
\]
replacing $s=1-\tfrac{|a|^2}{2}+\underset{j\to+\infty}{\mathcal{O}}(|a|^4)$ and neglecting the term $\mathcal{O}_{H^1}(|a|^3)$:
\[
\begin{pmatrix}\un{g_1} \\ \un{g_2}\end{pmatrix}=\begin{pmatrix} \phi_1(1+\tfrac{|a|^2}{2}-a(\delta a))+h_1^{(r)}-ah_2^{(r)}+\mathcal{O}_{H^1}(|a|^3)\\ (1+\tfrac{|a|^2}{2})\phi_2+\phi_1((\delta a)(1-\tfrac{|a|^2}{2}))+h_2^{(r)}+a^*h_1^{(r)} +\mathcal{O}_{H^1}(|a|^3)\end{pmatrix}.
\]
By $L^2_{\text{loc}}$-convergence, it is clear that $D(|\phi_k|^2, |h_{k'}^{(r)}|^2)=\underset{j\to+\infty}{o}(\nso{1}{\delta_{k'}}^2)$ for $(k,k')$ equal to $(1,2)$ or $(2,1)$. Using $\delta a=\underset{j\to+\infty}{o}(\nso{1}{\delta})$, at last we have:
\[
D(|\un{g_1}|^2,|\un{g}_2|^2)\apprle D(|\phi_1|^2,|\phi_2|^2)+\underset{j\to+\infty}{o}(\nso{1}{\delta}^2)=\underset{j\to+\infty}{o}(\nso{1}{\delta}^2)=\underset{j\to+\infty}{o}(\un{D_{12}}=d_{\Psi}),
\]
which gives the desired contradiction.

\noindent-- Let us treat at last the case $\delta_1\neq 0$ and $\delta_2=0$. Then as before:
\[
\begin{array}{rl}
D(|\un{h_1}|^2,|\phi_2|^2)&=D(|\delta_1|^2,|\phi_2|^2)+\mathcal{O}\big(\frac{1+\nlp{2}{\delta_1}}{\rgg}\big)=D(|\delta_1|^2,|\phi_2|^2)+\underset{j\to+\infty}{o}(\un{D_{12}}).
\end{array}
\]
Then necessarily there lies some mass of $\delta_1$ near $z_2$ and: 
\[
\frac{\delta_1(\cdot -z_2)}{\nso{1}{\delta_1}}\rightharpoonup_{H^1}\ell_1\neq 0.
\]
As before necessarily: $\ell_1=\eps_1\phi_{2,\infty}$ with $0<|\eps_1|\le 1$. But this contradicts:
\[
0=\dint \un{h_1}^*\phi_2=\dint \delta_1^*\phi_2+\dint \phi_1^*\phi_2=\dint \delta_1^*\phi_2+\underset{j\to+\infty}{\mathcal{O}}(e^{-\ov{\eps}\rgg}).
\]

%\subsection{On the decomposition $(\un{\psi_1},\un{\psi_2})$ and $\mu_1,\mu_2$}\label{no_bind_psipsi}
%In our problem we have a natural decomposition of $N$ into $\psi_1$ and $\psi_2$, eigenvectors of the mean-field operator with eigenvalues $\mu_1$ and $\mu_2$. In the case $\mu_1\neq \mu_2$ we cannot choose $\un{\psi_1}=h_1$ and $\un{\psi_2}=h_2$. In fact we have \eqref{no_bind_mumu}: we may ask whether the quantity%lplplp
%\[
%F_\mathcal{E}(\un{\psi_k}):=\mathcal{E}_{\text{PT}}(\un{\psi_k})-D(|\un{\psi_k}|^2,|\un{\psi_{k'}}|)
%\]
%is negative and away from $0$ or not. Starting from $\begin{pmatrix} h_1\\ h_2\end{pmatrix}$ we apply elements of $\mathbf{U}(2)$. As $h_k=\ph_k+\delta_k$ with $\ph_k\in\mathscr{P}$ and $\nso{1}{\delta_k}=\underset{j\to+\infty}{\mathcal{O}}(\sqrt{\Delta_2\mathcal{E}})$ a simple computation shows that:
%\[
%\forall h'\in \mathbf{S}_{L^2}\cap \text{Span}(h_1,h_2):\ F_\mathcal{E}(h')=\frac{3}{2}E_{\mathrm{PT}}(1)+\underset{j\to+\infty}{\mathcal{O}}((\Delta_2\mathcal{E})^{1/4}).
%\]

\subsection{Proof of Proposition \ref{no_bind_eloig1}}

The proof is similar to that of Proposition \ref{no_bind_eloig}: by contradiction we assume the existence of $(a_j)_j$ decreasing to $0$ together with $(\Psi_j=h_1\wedge h_2)$ with $\mathcal{E}_{\text{PT}}^U(\Psi_j)<a_j$ and $M^2(\Psi_j)<a_j R_{\text{g};j}$. We re-use the same notations of the previous Subsection.
\medskip

Thanks to Proposition \ref{no_bind_RR} we know that $d_{\Psi_j}$ is bounded from below by
\[
(1-\kappa \sqrt{a'_0})\big\{D(|\phi_1|^2,|\phi_2|^2)+D(|\delta_1|^2,|\phi_2|^2)+D(|\phi_1|^2,|\delta_2|^2)\big\}+D(|\delta_1|^2,|\delta_2|^2)
\]

As $(h_{k;j}(\cdot-z_{k;j}))_j$ tends to $\phi_{k,\infty}\in\mathscr{P}$ in $H^1$ for $k=1,2$, then for any $A>0$:
\[
\underset{j\to+\infty}{\lim} \dint_{B(z_{k,j}, A)}|h_{k,j}(x)|^2dx= \dint_{B(z_{k,j}, A)}|\phi_{k,\infty}(x)|^2dx.
\]
For any $2^{-1/2}<\la<1$ let $A_\la>0$ be the number such that the last integral with $A=A_\la$ is equal to $\la$. We have:
\[
\begin{array}{l}
\underset{|x-y|<\rgg+2A_\la}{\diint}\dfrac{|h_1\wedge h_2(x,y)|^2}{|x-y|}dxdy\ge \dfrac{2}{\rgg+2A_\la} \underset{|x-y|<\rgg+2A_\la}{\diint} |h_1(x)|^2|h_2(y)|^2dxdy\\ 
\quad\quad\quad\quad\quad\quad\quad\quad\quad\quad\quad\quad-\dfrac{2}{\rgg+2A_\la}\dint dx h_1^*h_2(x)\underset{y\in B(x,\rgg+2A_\la)}{\dint}h_2^*h_1(y)dy\\
\underset{j\to+\infty}{\liminf}\underset{|x-y|<\rgg+2A_\la}{\diint}\dfrac{|h_1\wedge h_2(x,y)|^2}{|x-y|}dxdy\ge \frac{2}{\rgg+2A_\la} (\la^2 -2^{-1}).
\end{array}
\]
We used the following trick: if $\dint h_1^*h_2=0$ where $\nlp{2}{h_k}=1$, then for any Borelian set $B$:
\[
\Big |\dint_B h_1^*h_2 \Big|\le \frac{1}{2}.
\]

The more precise result has the same proof: in the limit there holds similar inequality: for sufficiently small $a>0$, $\la \rgg>A_\eps$ where
\[
\dint_{|x|\le A_\eps} |\phi(x)|^2dx=\eps,\ \eps>2^{-1/2},\ \phi\in\mathscr{P}_0.
\]
We conclude with the same argument.

\section{Localisation in Direct space: the \underline{$\psi_j$}'s}\label{no_bind_elds_wf}

\subsection{Proof of Lemma \ref{no_bind_decaylemma}}\label{no_bind_decaydecay}
\begin{notation}
For convenience here we write $V\cdot \un{\ph_k}:=v'_{\un{\g}}\ph_k-R_{\un{N}}\un{\ph_k}$ (and a similar expression for $\un{\chi_k}$). The function $r_k:=R_{\un{\g}}\un{\psi_k}$ is split into its upper part $r_{k,\uparrow}:=(R_{\un{\g}}\un{\psi_k})_{\uparrow}$ and its lower part $r_{k,\downarrow}$ both in $L^2(\RR,\mathbb{C}^2)$.

Moreover we write:
\[
P_k(-\Delta):=c^2(1-\mu_k^2\mathcal{L}_{c\La}^{-2})-\Delta\text{\ and\ }y_c:=\mathcal{L}_{c\La}^{-1}=\frac{c^2\La^2}{c^2\La^2-\Delta}.
\]

\noindent The operator $P_k(-\Delta)$ can be rewritten as follows: with $a_k:=c^2(1-\mu_k)$ and $b:=c\La$ then
\begin{equation}
\begin{array}{l}
c^2(1-\mu_ky_c^2)-\Delta=a_k(1+\mu_k)-\Delta \Big[1+\frac{\mu_kc^2-a_k}{c^2\La^2}\frac{b^2}{b^2-\Delta}+\frac{\mu_k^2}{\La^2}\Big( \frac{b^2}{b^2-\Delta^2}\Big)^2 \Big]\\
 \ \ \ \  \ \ \ \ =(a_k(1+\mu_k)-\Delta)\Big\{1+\big(1-\frac{a_k(1+\mu_k)}{a_k(1+\mu_k)-\Delta}\big)\Big[\frac{\mu_kc^2-a_k}{c^2\La^2}\frac{b^2}{b^2-\Delta}+\frac{\mu_k^2}{\La^2}\Big( \frac{b^2}{b^2-\Delta^2}\Big)^2\Big]\Big\}
\end{array}
\end{equation}
%It can be shown \cite{these} $\tfrac{1}{P_k(-\Delta)}$ is a perturbation of $\{a_k(1+\mu_k)-\Delta\}^{-1}$ (for $\La\gg 1$) and that this is indeed a convolution operator with an exponentially decaying function (singular in $0$).
\end{notation}

\paragraph{Proof}

We remark that $\un{n}(x)=|h_1(x)|^2+|h_2(x)|^2=|\un{\psi_1}(x)|^2+|\un{\psi_2}(y)|^2$.

Thanks to \eqref{no_bind_psipsi}-\eqref{no_bind_psipsi_1}, there holds:
\begin{equation}
(\Dbf+\alpha B) \psi_k=(1+\tfrac{3E_{\text{PT}(1)}}{2c^2}+\mathcal{O}(\alpha^{1/4}c^{-2}))\psi_k.
\end{equation}
Up to applying some $\mathbf{m}\in\mathbf{SU}(2)$ to $\begin{pmatrix}\psi_1\\ \psi_2 \end{pmatrix}$, we consider $\un{\psi_k}=h_k$ with the following:
\[
\begin{array}{rl}
(c^2\beta-ic\boldsymbol{\alpha}\cdot \nabla h_k)+\alpha cy_c (V\cdot h_k -R_{\un{\g}}h_k)=(c^2-\tfrac{3 E_{\mathrm{PT}}(1)}{2})y_c h_k+\mathcal{O}(\alpha^{1/4}y_c h)
\end{array}
\]
We write $a=-\tfrac{3 E_{\text{PT}}(1)}{2}$ and the additional term $\mathcal{O}(\alpha^{1/4}y_c h)=\delta_k h$.
\medskip

\noindent -- We now rewrite \eqref{no_bind_equn} once again: by substitution, we get:
\begin{equation}\label{no_bind_eqdecay}
\left\{
\begin{array}{rl}
\un{\ph_k}&=\alpha cy_c\dfrac{1+\mu_ky_c}{P_k(-\Delta)}(V\cdot \un{\ph_k}-r_{k,\uparrow})+\dfrac{\alpha y_c}{P_k(-\Delta)}i\sigma\cdot \nabla\big[ V\cdot \un{\chi_k}-r_{k,\downarrow}\big]\\
\un{\chi_k}&=\alpha \dfrac{y_c}{P_k(-\Delta)} i\sigma\cdot \nabla(V\cdot \un{\ph_k}-r_{k,\uparrow})+\alpha y_c\dfrac{c^2(1-\mu_k y_c)}{c P_k(-\Delta)}\big[ V\cdot \un{\chi_k}-r_{k,\downarrow}\big]
\end{array}
\right.
\end{equation}
There holds similar equation for $h_k$ but with additional terms $\tfrac{1}{\alpha c}(\delta_k h)_{\uparrow}$ with $-r_{k,\uparrow}$ and $\tfrac{1}{\alpha c}(\delta_k h)_{\downarrow}$ with $-r_{k,\downarrow}$. 

There holds:
\[
\alpha c(1-\mu_k y_c)=\alpha c(1-\mu_k)+\alpha c\mu_k(1-y_c).
\]
For any $A\ge \G(\rgg)\rgg$, we multiply each term by $|D_0|^{1/2}$ and then by $d_{A,\la}(\cdot)$ defined by $d(\cdot)\xi_A(\cdot)\boldsymbol{\eta}^{\la}_{\rgg}$. 

We take the $L^2$-norm, let us show estimates \emph{independent} of $A$ (but depending on $\xi_1$):
\[
\nlp{2}{d_{A,\la}|D_0|^{1/2}\un{\psi_k}}\le K_\la+\eps_{(\la)} \nlp{2}{E_{A,\la}^{1/2}|D_0|^{1/2}\un{\psi_k}},\text{\ with\ }\eps_{(\la)}<1.
\]
This will end the proof, the family $(K_\la)_\la$ depending on $(\eps_{(\la)})_\la$ and the latter being nonincreasing in $\la\in (\la_0, 2^{-1})$.

We prove the estimation of $\nlp{2}{d^{(2)}_{A,\la} |D_0|^{1/2}\un{\psi_j}}$ with $j=1,2$ by the same method: we need finiteness of $\nlp{2}{d(\cdot)\etabfc{\la/2} |D_0|^{1/2}\un{\psi_k}}$ with $k=1,2$ and of $\ns{2}{|x-y|\g}$. We refer to Appendix \ref{no_bind_elds_wf} for more details.

\medskip

\noindent-- In Appendix \ref{no_bind_elds_wf}, we show:
\begin{equation}\label{no_bind_Eala}
|d_{A,\la}(x)-d_{A,\la}(y)|\apprle |x-y|.
\end{equation}

Let us first multiply \eqref{no_bind_eqdecay} by $|D_0|^{1/2}$: let $\mathcal{F}_{j,k}:=\frac{|D_0|^{1/2} \partial_j}{P_k(-\Delta)}$ and $\mathcal{F}_{0,k}:=\frac{|D_0|^{1/2}}{P_k(-\Delta)}$. It is clear that they are bounded (convolution) operators, we show in Appendix \ref{no_bind_elds_wf} that
\begin{equation}\label{no_bind_Fjk}
\nlp{1}{|\cdot|\mathcal{F}_{j,k}}\apprle 1,\  j\in\{1,2,3\},\,k\in\{1,2\}.
\end{equation}
The function associated to $y_c$ is a Yukawa potential $Y_c$ \cite[Section 6.23]{LL}:
\[
Y_c(x-y)=\sqrt{\frac{\pi}{2}}\frac{(c\La)^2 e^{-c\La|x-y|}}{ |x-y|},
\]
in particular $\nlp{1}{|\cdot| Y_c}\apprle \tfrac{1}{c\La}$.
The idea is to take first the commutator $[d_{A,\la},\mathcal{F}_{j,k}]$ and $[d_{A,\la},y_c]$. Then we study $d_{A,\la}v\un{\varpi_k}$ ($\varpi_k\in\{\ph_k,\chi_k\}$) and $d_{A,\la} r_{\uparrow/\downarrow}$.

\medskip

\paragraph{Estimate of $\alpha c\nlp{2}{V\cdot \un{\ph_k}},\alpha c\nlp{2}{V\cdot \un{\chi_k}}$} We use the same method for both cases. We recall the following:
\[
v_{\un{\g}}=\big(-\un{\check{\Fla}}*\un{n}+(\delta_0-\un{\check{\Fla}})*(\mathfrak{t}_{\un{N}}-\alpha^2\un{\wit{\tau}_2})\big)*\frac{1}{|\cdot|}=-\un{\check{\Fla}}*\un{n}*\frac{1}{|\cdot|}+\un{\rho_{rem}}*\frac{1}{|\cdot|}.
\]
By \eqref{no_bind_Fla2}:
\[
\begin{array}{rl}
%\big|\alpha c \un{\check{\Fla}}*\un{n}*\frac{1}{|\cdot|}\big|(x)&\apprle \frac{\nlp{1}{\Fla}}{\alpha \llo}\un{n}*\frac{1}{|\cdot|}(x)\apprle \un{n}*\frac{1}{|\cdot|}(x)\\
\big|\alpha c \un{\check{\Fla}}*\un{n}*\frac{1}{|\cdot|}\big|(x)&\le \un{n}*\frac{1}{|\cdot|}(x)+\alpha c\big|(\un{\check{\Fla}}-\Fla(0)\delta_0)*v_{\un{n}}(x)\big|\\
          &\le  \un{n}*\frac{1}{|\cdot|}(x)+\mathcal{O}\big(\frac{1}{\sqrt{c}}\big).
\end{array}
\]
We used $\nlp{\infty}{f}\apprle \nlp{1}{\wh{f}}$, split the integral in Fourier space at level $2c$ and used Cauchy-Schwarz inequality. By Appendix \ref{no_bind_ph2} and Proposition \ref{no_bind_RR}:
\[
\begin{array}{rl}
\big|\alpha c \un{\rho_{rem}}*\frac{1}{|\cdot|}\big|(x)&\apprle \alpha c (c^{1/2}\ncc{\rho_{rem}}+c^{3/2}\nlp{2}{\rho_{rem}})\\
      &\apprle \alpha c^{3/2}(\alpha c^{-1}+\alpha^2 c^{-1})+\alpha c^{5/2}(c^{-2}+c^{-1}(\alpha (a_{12}+a_{21}))^{1/2})\\
      &\apprle \frac{\alpha}{\sqrt{\llo}}+\frac{1}{\sqrt{\llo}}+\frac{\alpha^{5/4}}{\sqrt{\llo}}\apprle \frac{1}{\sqrt{\llo}}.
\end{array}
\]
We recall $a_{jk}=\nlp{2}{v_k \psi_k-v_{kj} \psi_k}$ and by Proposition \ref{no_bind_RR} we know it is $\mathcal{O}(c^{-1}\alpha^{3/2})$. We decompose each $\un{\psi_j}$ in sum of $h_1,h_2$: $\un{\psi_k}=c_{k1}h_1+c_{k2}h_2$. Then:
\[
\begin{array}{rl}
v_{\un{\g}} \un{\psi_k}&=v_{\un{\g}}(c_{k1}h_1+c_{k2}h_2)\\
(v_{\un{n}}-R_{\un{N}})\un{\psi_k}&= c_{k1}(v_{|h_2|^2}h_1-v_{h_2^*h_1}h_2)+c_{k2}(v_{|h_1|^2}h_2-v_{h_1^*h_2}h_1).
\end{array}
\]
We write $h_k=\delta_k+\phi_k$ where $\phi_k\in\mathscr{P}$: as in Section \ref{no_bind_ptt} $\nso{1}{\delta_k}^2\apprle \alpha$. By fast decay of the $\phi_k$'s: $(|\phi_k|^2*\tfrac{1}{|\cdot|}(x))^2=\Theta (|\phi_k|^2*\tfrac{1}{|\cdot|^2}(x))$ and for $|x|\apprge 1$ this is $\mathcal{O}(\tfrac{1}{|x-z_k|^2})$. 

In particular for $|x|>\la \rgg$
\[
v\big[|h_k|^2\big](x)\apprle \frac{1+\nlp{2}{\delta_k}}{|x-z_k|}+\psh{|\nabla| \delta_k}{\delta_k}\apprle \frac{1}{\la \rgg}+\alpha,
\]
we choose $C_0>1$ such that $\tfrac{\alpha c}{\la \rgg}<1-\eps_0$ where $0<\eps_0<1$ is fixed (for instance $2^{-1}$).

By Cauchy-Schwarz inequality we have $v\big[h_1^*h_2\big](x),v\big[h_2^*h_1\big](x)=\mathcal{O}(\nlp{2}{\delta})$. It follows that
\[
\alpha c \nlp{2}{d_{A,\la} V\cdot \un{\ph_k}}\apprle \eps'_{(\la)}\nlp{2}{d_{A,\la} \un{\ph_k}},\text{\ with\ }0<\eps'_{(\la)}<1.
\]
\paragraph{Estimate of $\alpha cd_{A,\la} R_{\un{\g}\un{\psi_k}}$}
\[
\begin{array}{rl}
|[d_{A,\la},R_{\un{\g}}](x,y)|&\apprle |\un{\g}(x,y)|\text{\ so:}\\
\alpha c \nlp{2}{[d_{A,\la},R_{\un{\g}]\un{\psi_k}}}&\apprle \alpha c \ns{2}{\un{\g}}\nlp{2}{\un{\psi_k}}\apprle \alpha^2 c^{1/2}=\mathcal{O}(\tfrac{\alpha}{\sqrt{\llo}}).\\
\nlp{2}{R_{\un{\g}}d_{A,\la}\un{\psi_k}}^2&\apprle \ttr(\g R_{\un{\g}})\psh{|\nabla| d_{A,\la}\un{\psi_k}}{d_{A,\la}\un{\psi_k}}\\
           &\apprle c^{-1}\nlp{2}{|D_0|^{1/2} d_{A,\la}\un{\psi_k}}^2.
\end{array}
\]
By Lemma \eqref{no_bind_commd}, $[|D_0|^{1/2},d_{A,\la}]|D_0|^{-1/2}$ is a bounded operator (with norm $\mathcal{O}(\nlp{\infty}{\nabla d_{A,\la}})$) and at last we get:
\[
\alpha c \nlp{2}{d_{A,\la}R_{\un{\g}}\un{\psi_k}}\apprle \alpha c^{1/2}(1+\nlp{2}{d_{A,\la} |D_0|^{1/2}\un{\psi_k}})\text{\ and\ }\alpha c^{1/2}=\mathcal{O}\big(\frac{1}{\sqrt{\llo}}\big).
\]
We know deal with the case of $d^{(2)}_{A,\la}R_{\un{\g}\un{\psi_k}}$, using \eqref{no_bind_taylor}, proved below.

The aim is to prove:
\begin{equation}
\begin{array}{l}
\nlp{2}{d^{(2)}_{A,\la} R_{\un{\g} \un{\psi_k}}}\apprle \ns{2}{|x-y| \un{\g}}+\ns{2}{\un{\g}}\nlp{2}{d(\cdot) \etabfc{\la/2} \un{\psi_k}}\\
\ \ \ \ \ \ \ \ \ +c^{1/2}\nqq{\g}(\nlp{2}{\un{\psi_k}}+\nlp{2}{d(\cdot)\etabfc{\la/2} \un{\psi_k}}).
\end{array}
\end{equation}
First of all we use Taylor's formula \eqref{no_bind_taylor} to get:
\[
\nlp{2}{[d^{(2)}_{1,\la}, R_{\un{\g}] \un{\psi_k}}}\apprle \ns{2}{|x-y| \un{\g}}+\ns{2}{\un{\g}}\nlp{2}{d(\cdot) \etabfc{\la/2} \un{\psi_k}}.
\]
Let us prove at the end $\ns{2}{|x-y| \un{\g}}=c^{-1}\ns{2}{|x-y| \g}\apprle \alpha c^{-1}.$

There remains $\nlp{2}{R_{\un{\g} d^{(2)}_{A,\la} \un{\psi_k}}}\apprle \nlp{2}{|D_0|^{1/2} d^{(2)}_{A,\la} \un{\psi_k}}$. 

We commute: using \eqref{no_bind_sauve}, there holds
\[
\begin{array}{rl}
[|D_0|^{1/2}, d^{(2)}_{A,\la}]&=\dfrac{1}{2^{-1/2}\pi}\dint_0^{+\infty}\frac{s^{1/4} ds}{1-\Delta+s} [-\Delta, d^{(2)}] \frac{1}{1-\Delta+s},\\
\,[-\Delta, d^{(2)}]&=(-\Delta d^{(2)})-2\ssum_{j=1}^3(\partial_j d^{(2)})\partial_j.
\end{array}
\]
First $\nlp{\infty}{\Delta d^{(2)}}\apprle 1$. Then thanks to \eqref{no_bind_taylor}:
\[
\begin{array}{rl}
\nlp{2}{(\partial_j d^{(2)})\tfrac{\partial_j}{1-\Delta+s}\un{\psi_k}}&\apprle \nlp{2}{d(\cdot)\etabfc{\la/2} \tfrac{\partial_j}{1-\Delta+s}\un{\psi_k}}\\
    &\apprle \nlp{1}{|x-y|\mathscr{F}^{-1}(\tfrac{p_j}{1+s+|p|^2})}\nlp{2}{\un{\psi_k}}+\frac{\nlp{2}{d(\cdot)\etabfc{\la/2} \un{\psi_k}}}{1+s}\\
    &\apprle \dfrac{1}{1+s}(\nlp{2}{\un{\psi_k}}+\nlp{2}{d(\cdot)\etabfc{\la/2} \un{\psi_k}}).
\end{array}
\]

To end this section we prove $\ns{2}{|x-y| \g}, \ns{2}{|x-y| |\Dbf|^{1/2}\g}\apprle \alpha$. This is almost trivial: for each $j\in\{1,2,3\}$ we consider $(x_j-y_j)\g(x,y)$ and use the Cauchy expansion of $\g$. For each $Q_{0,k}$, $k\in[|1,5|]$, we replace at least one $P^0_{\eps} v'_\g P^0_{-\eps}$ as in \eqref{no_bind_comm} (\cite{gs}) and write:
\[
x_j-y_j=x_j-\ell_j^{(1)}+\ell_j^{(1)}-\ell_j^{(2)}+\cdots +\ell_j^{(n)}-y_j.
\]
For each convolution operator $\tfrac{|\Dbf|^{1/2}}{\Dbf+i\eta}(x-y), \tfrac{P^0_\eps}{\Dbf+i\eta}(x-y),\tfrac{1}{D_0+i\omega}(x-y)$, multiplying by $(x_j-y_j)$ corresponds to take the derivative $\partial_j$ in Fourier space enabling us to take KSS inequalities \eqref{no_bind_kss} under the integral sign. Indeed we have:
\[
\begin{array}{ll}
|\partial_j \Ebf{p}^{1/2}|\apprle \Ebf{p}^{1/6},&|\partial_j \tfrac{1}{\Ebf(p)+i\eta}|\apprle \frac{1}{|\Ebf{p}+i\eta|^{1+3^{-1}}},\\
|\partial_j\tfrac{1}{E(p)+i\omega}|\apprle \frac{1}{E(\omega_2)+|p|^2},&|\partial_j P^0_\eps(p)|\apprle \frac{1}{E(p)}.
\end{array}
\]
Then operators of type $\rho*\tfrac{1}{|\cdot|}$ or $\alpha_k\partial_k (\rho*\tfrac{1}{|\cdot|})$ remains unchanged while operators of type $(x_j-y_j) R_Q(x,y)$ are trivially Hilbert-Schmidt. This end the proof ; the biggest term comes from $Q_{1,0}((x_j-y_j) \g'(x,y))$.

%We notice $\alpha^2 c=\mathcal{O}(\tfrac{1}{\llo})=o(1)$.
%yyyyyyyyy
%\end{dem}

\subsection{Proof of \eqref{no_bind_Eala} and variation for $d^{(2)}_{A,\la}$}
\noindent 1.\,We recall that $\xi_1$ is a \emph{radial} smooth function with $\xi_1(x)=1$ for $|x|\le 1$ and $\xi_1(x)=0$ for $|x|\ge 2$. We study $d_{A,\la}:=d(\cdot)\xi_{A}(\cdot)\boldsymbol{\eta}^{\la}_{\rgg}(\cdot).$ 

\noindent First remark to be done: $H=\{x: |x-z_1|=|x-z_2| \}$ splits the space into two half-spaces $E_1$ (set of points closest to $z_1$) and $E_2$. Let $s_H$ be the orthogonal symmetry with respect to $H$: $s_H(z_1)=z_2$. If $x\in E_{1}$ and $y\in E_{2}$, then
\[
|d(x)-d(y)|=\big||x-z_1|-|s_H(y)-z_1|\big|\le |x-s_H(y)|\le |x-y|.
\]
Moreover $d_{A,\la}(y)=d_{A,\la}(s_H(y))$ and
\[
\big| d_{A,\la}(x)-d_{A,\la}(y)\big|=\big| d_{A,\la}(x)-d_{A,\la}(s_H(y))\big|.
\]
So we may assume that $d(x)=|x-z_1|$ and $d(y)=|y-z_1|$, and in this case we can write:
\[
d_{A,\la}(x)=F_{\la}(d(x))\xi_A(x):=d(x)\sqrt{1-\xi_{\la \rgg^2(d(x))}}G_A(|x-z_m|)
\]
the same holds for $y$. We will write $F_\la(\cdot )$ for $x\mapsto F_\la(d(x))$ for convenience. There holds
\[
\nabla d_{A,\la}(x)=\big(\nabla F_{\la}(x)\big) \xi_A(x)+F_{\la}(x)(\frac{\nabla \xi_1(x/A)}{A}),
\]
and as we have chosen $A\gg \rgg$ we may assume that if $\nabla \xi_A(x)\neq 0$, then $|x-z_m|=\Theta(d(x))$.
By simple computation:
\begin{equation}\label{no_bind_Ealadeb0}
\begin{array}{rl}
|\nabla d_{A,\la}(x)| &\apprle \big(1+\nlp{\infty}{|\cdot| \nabla \xi_1}+\nlp{\infty}{|\cdot|\nabla \boldsymbol{\eta}^{\la}_1}\big).
\end{array}
\end{equation}

\noindent 2.\,For $x,y\in E_{\eps}$, $\eps=1,2$ (say $E_1$) and $A\gg \rgg$, there holds:
\begin{equation}\label{no_bind_taylor}
\begin{array}{l}
d^{(2)}_{A,\la}(x)-d^{(2)}_{A,\la}(y)=|x-z_1|^2\xi_A(x)\etabfc{\la}(x)-|y-z_1|^2\xi_A(y)\etabfc{\la}(y)\\
    \ \ \ =|y-z_1|^2\big(\tfrac{\etabfc{\la}(y)}{A}\nabla\xi_1(\tfrac{y}{A})+\tfrac{\xi_A(y)}{c\la\rgg}\nabla(\boldsymbol{\eta}_1^{1})(\tfrac{y}{c\la\rgg})\big)\cdot(x-y)\\
    \ \ \  \ \ \  \ \ \  \ \ \ +\xi_A(y)\etabfc{\la}(y)\psh{y-z_1}{x-y} +|y-z_1|^2+\mathcal{O}\big( |x-y|^2\big)\\
       \ \ \ = \mathcal{O}\big(d(y)\etabfc{\la/2}(y)|x-y|+|x-y|^2\big).
\end{array}
\end{equation}
Above we used $\nabla \etabfc{\la}=\etabfc{\la/2}\nabla \etabfc{\la}$ and the $\mathcal{O}(\cdot)$ depends on $\xi_1,\boldsymbol{\eta}_1^1$. This estimate enables us to consider commutators with $\tfrac{|D_0|^{1/2}\sigma\cdot\nabla}{P_k(-\Delta)}$ and $y_c:=\tfrac{(c\La)^2}{(c\La)^2-\Delta}$, as shown in the next section.

\subsection{Proof of \eqref{no_bind_Fjk} and variation for $d^{(2)}_{A,\la}$}
\noindent 1.\,For any borelian function $\mathcal{F}$:
\[
\dint_{\mathbb{R}^3} |x||\mathcal{F}(x)|dx\le \Big\{\dint |x|^4E(x)^2 |\mathcal{F}(x)|^2dx\dint\frac{dx}{|x|^2 E(x)^2}\Big\}^{1/2}.
\]
To prove $|\cdot|\mathcal{F}\in L^1$ it suffices to check all integrals on the right side converge: in Fourier space, we have to prove:
\[
\nlp{2}{\Delta \wh{\mathcal{F}}}^2+\nlp{2}{\nabla \Delta \wh{\mathcal{F}}}^2<+\infty.
\]
Applying this method for $\mathcal{F}_{j,k}(x-y):=\tfrac{|D_0|^{1/2}\partial_j}{P_k(-\Delta)}(x-y)$:
\[
\wh{\mathcal{F}_{j,k}}(p)=\frac{E(p)^{1/2} p_j}{a_k+|p|^2}\Big\{1+\frac{\mu_k^2|p|^2}{\La^2(a_k+|p|^2)}\frac{2b^4+b^2|p|^2}{(b^2+|p|^2)^2} \Big\}^{-1}
\]
where we recall $b=c\La$, $a_k=c^2(1-\mu_k)$. From this expression, it is easy to see that for $\ell=1,2,3$ and $m=1,2$ we have
\[
\nlp{2}{\partial_{\ell}^m \mathcal{F}_{j,k}}^2\apprle 1.
\]
The constant depends on $a_k$ but for sufficiently small $\alpha,L,\La^{-1}$ then $a_k>\eps_0>0$. 

\noindent 2.\,By the same method we can show that: 
\[
\dint_{\mathbb{R}^3} |x|^2|\mathcal{F}(x)|dx\le \Big\{\dint |x|^6E(x)^2 |\mathcal{F}(x)|^2dx\dint\frac{dx}{|x|^2 E(x)^2}\Big\}^{1/2},
\]
enabling us to treat $d^{(2)}_{A,\la}$.

%\subsection{\nlp{2}{d^{(2)}_{A,\la} R_{\un{\g} \un{\psi_k}}}}\label{no_bind_eldsrg}

\section{Localisation in Direct space: $\g$}\label{no_bind_elds}
We recall we explain in Remark \ref{no_bind_imbriquer} how we use the technical results proved here: Propositions \ref{no_bind_estdens}, \ref{no_bind_ns2kin} and \ref{no_bind_estint}.

\subsection{Estimates on the localised density}\label{no_bind_locdens}
Let $Q\in\mathcal{K}$ and $0\le \zeta\le 1$ a smooth function (\emph{e.g.} $\xi_{\la \rgg}$ or $\boldsymbol{\eta}_{\rgg}^{\la}$).  Our aim is to give a semi-quantitative estimate of the localisation of the function $\zeta^2 \rho_Q=\rho_{\zeta Q \zeta}$ around the support of $\zeta$.
\begin{proposition}\label{no_bind_estdens}
Let $Q$ and $\zeta$ be as above, then we have:
\begin{equation}\label{no_bind_zetag}
\lVert\zeta^2 \rho_{Q}-\rho[\zeta^{++} Q \zeta^{++}+\zeta^{--} Q \zeta^{--}]\rVert_{\mathcal{C}}\le F_{\text{est}}[\La,\zeta,Q],
\end{equation}
with
\begin{equation}
\begin{array}{l}
F_{\text{est}}[\La,\zeta,Q]=(\sqrt{\llo}\nlp{3}{\nabla \zeta}+\nlp{\infty}{\nabla \zeta})(\ns{2}{\zeta P^0_{\pm}|D_0|^{\ala}Q}+\nlp{\infty}{\nabla\zeta}\ns{2}{Q})\\
\ +\nlp{6}{\nabla \zeta}^2 \ns{2}{|D_0|^{\ala} Q}+\sqrt{\llo}(\ns{2}{\zeta Q^{\pm\mp}|D_0|^{\ala} \zeta}+\ns{2}{\zeta Q^{\pm\mp}}\nlp{\infty}{\nabla \zeta})\\
\ +\sqrt{\llo}\nlp{\infty}{\nabla \zeta}(\nlp{\infty}{\nabla \zeta}\ns{1}{Q^{\pm \pm}}+\ns{1}{\zeta |D_0|^{\ala} Q^{\pm\pm}})\\
\ +(\llo)^{1/6}\nlp{\infty}{\nabla\zeta}^2\ns{1}{|D_0|^{\ala}Q^{\pm\pm}}.
%\sqrt{\llo}(\ns{2}{\zeta |D_0|^{\ala}Q}+\ns{2}{|D_0|^{\ala} Q}\nlp{\infty}{\nabla \zeta})+\nlp{6}{\nabla \zeta}^2\ns{2}{|D_0|^{\ala}}.
\end{array}
\end{equation}
Moreover there holds for $\eps=\pm$:
\begin{equation}
\begin{array}{rl}
\ncc{\rho[\zeta^{\eps \eps} Q \zeta^{\eps \eps}]}&\le \nb{[\zeta^{\eps \eps},|D_0|^{\ala}]}\ns{1}{Q^{\eps \eps}}+\ns{1}{\zeta^{\eps \eps} Q^{\eps \eps}|D_0|^{\ala} \zeta^{\eps \eps}}\\
    &\apprle \nlp{\infty}{\nabla \zeta}\ns{1}{Q^{\eps \eps}}+\ns{1}{\zeta^{\eps \eps} Q^{\eps \eps}|D_0|^{\ala} \zeta^{\eps \eps}}.
\end{array}
\end{equation}
\end{proposition}

\begin{remark}\label{no_bind_remestdens}
\noindent1.\ In the case $Q=\Pi-P^0_-$ with $\Pi^*=\Pi^2=\Pi$ then (\emph{cf} \cite{ptf}):

$Q^{2}=Q^{++}-Q^{--}\ge Q^{++}$. As shown in \cite{sokd} we can consider an orthonormal family of eigenvectors of $Q^{2}$ that split into those in $\text{Ran}(P^0_+)$ and those in $\text{Ran}(P^0_-)$. It is then clear that:
%\[
%\begin{array}{rl}
%\zeta^{++}|D_0|^{\ala} Q^{++}(\zeta^{++})^2Q^{++}|D_0|^{\ala}\zeta^{++}&\le \zeta^{++}|D_0|^{\ala} (Q^2)^2 |D_0|^{\ala} \zeta^{++}.%\big|\zeta^{++}|D_0|^{\ala}Q^{++}|D_0|^{\ala}\zeta^{++}\big|^2
%\end{array}
%\]
%Thus:
\[
\begin{array}{rl}
\ns{1}{\zeta^{++} Q^{++}|D_0|^{\ala} \zeta^{++}}&\le \ns{1}{\zeta Q^{++} |D_0|^{\ala} \zeta}\\
        &\le \ns{2}{\zeta |D_0|^{\ala} Q}\ns{2}{\zeta Q}\\
         %&\le \ns{2}{Q}\ns{2}{\zeta |D_0|^{\ala}P^0_+ Q} % &\le \ttr\big(\zeta^{++} |D_0|^{\ala} Q^2|D_0|^{\ala} \zeta^{++}\big)\le \ns{2}{\zeta |D_0|^{\ala} Q}^2.
\end{array}
\]

\noindent2.\ There is also an analogous estimate if we choose two different functions $\zeta_1,\zeta_2$, that is with $\zeta_1\zeta_2\rho(Q)=\rho(\zeta_1 Q \zeta_2)$. 
The same proof shows also localisation estimates, but we have to "polarize" the inequalities just like for a quadratic form and its associated bilinear form. 
\end{remark}

\begin{dem}
We prove it by duality. Let $V$ be some Schwartz function: we study $\ttr_0 (\zeta Q \zeta V).$ By symmetry we just treat $(\zeta Q \zeta V)^{++}$.
There holds:
\[
\begin{array}{l}
P^0_+ \zeta Q \zeta VP^0_+=P^0_+ \zeta (P^0_++P^0_-) Q(P^0_++P^0_-) \zeta (P^0_++P^0_-) VP^0_+\\
   \ \       =\zeta^{++}Q^{++}\zeta^{++} V^{++}+\zeta^{++} Q^{++}\zeta^{+-}V^{-+}+\zeta^{++} Q^{+-}\zeta^{-+} V^{++}+\zeta^{++} Q^{+-}\zeta^{--} V^{-+}\\
    \ \      +\zeta^{+-}Q^{-+}\zeta^{++}V^{++}+\zeta^{+-}Q^{-+}\zeta^{+-}V^{-+}+\zeta^{+-}Q^{--}\zeta^{-+}V^{++}+\zeta^{+-}Q^{--}\zeta^{--}V^{-+}.
\end{array}
\]
We first show those operators are trace-class and then prove \eqref{no_bind_zetag}.
\begin{remark}
We recall that by Sobolev inequality: $\nlp{6}{V}\apprle \nlp{2}{\nabla V}$.

Moreover $\ns{2}{|D_0|^{-\ala} V}\apprle \sqrt{\llo}\nlp{2}{V}$.

As shown in Appendix \ref{no_bind_appA}:
\begin{equation}
\zeta^{-+}=\frac{i}{2\pi}\dint_{-\infty}^{+\infty}\frac{1}{D_0+i\eta}\boldsymbol{\alpha}\cdot \nabla \zeta \frac{P^0_+d\eta}{D_0+i\eta}.
\end{equation}
It can be rewritten as:
\begin{equation}\label{no_bind_eqjs}
\zeta^{-+}=\frac{i}{2}\dint_0^{+\infty}e^{-s|D_0|}P^0_-\boldsymbol{\alpha}\cdot\nabla \zeta P^0_+e^{-s|D_0|}ds,
\end{equation}
by writing $\tfrac{1}{E(p)+E(q)}=\int_0^{+\infty}e^{-s(E(p)+E(q))}$ in the kernel of its Fourier transform \emph{cf} Appendix \ref{no_bind_appA}.
\end{remark}

\paragraph{ $\zeta^{++}Q\zeta^{++} V^{++}$:}
\[
\zeta^{++}Q\zeta^{++} V^{++}=\zeta^{++}(Q^{++}\zeta^{++} |D_0|^{\ala})\frac{1}{|D_0|^{\ala}}V^{++}
\]
and $(Q^{++}\zeta^{++} |D_0|^{\ala})\in \mathfrak{S}_1$, $\frac{1}{|D_0|^{\ala}}V^{++}\in\mathfrak{S}_6$ with norm $\mathcal{O}((\llo)^{1/6} \nlp{2}{\nabla V})$ by the KSS inequality \eqref{no_bind_kss}. We write 
\[
\begin{array}{rl}
\ns{1}{\zeta Q^{++} \zeta^{++}|D_0|^{\ala}}&\le \ns{1}{\zeta Q^{++}}\nb{[\zeta^{++}, |D_0|^{\ala}]}+\ns{1}{\zeta Q^{++} |D_0|^{\ala} \zeta}\\
       &\apprle \ns{1}{\zeta Q^{++}}\nlp{\infty}{\nabla\zeta}+\ns{1}{\zeta Q^{++}|D_0|^{\ala}\zeta}.
\end{array}
\]

In general whenever there is  $Q^{++}$ or $Q^{--}$ we can easily estimate. 
\[
\begin{array}{rl}
|\ttr(\zeta^{++} Q^{++}\zeta^{+-}V^{-+})|&=|\ttr(V^{-+}\tfrac{1}{|D_0|^{\ala}} |D_0|^{\ala} \zeta^{++} Q^{++} \zeta^{+-})|\\
                 &\apprle \sqrt{\llo}\nlp{2}{\nabla V}\nlp{\infty}{\nabla \zeta}(\nlp{\infty}{\nabla \zeta}\ns{1}{Q^{++}}+\ns{1}{\zeta |D_0|^{\ala} Q^{++}}),\\
|\ttr(\zeta^{+-} Q^{--} \zeta^{-+} V^{++})|&\le \ns{6}{\tfrac{1}{|D_0|^{\ala}}V}\nb{\zeta^{+-}}\ns{1}{Q^{--} \zeta^{-+} |D_0|^{\ala}}\\
    &\apprle (\llo)^{1/6}\nlp{2}{\nabla V}\nlp{\infty}{\nabla \zeta}^2\ns{1}{Q^{--}|D_0|^{\ala}},\\
|\ttr(\zeta^{+-}Q^{--}\zeta^{--}V^{-+})|&\apprle \sqrt{\llo}\nlp{2}{\nabla V}\nlp{\infty}{\nabla\zeta}(\nlp{\infty}{\nabla\zeta}\ns{1}{Q^{--}}+\ns{1}{\zeta|D_0|^{\ala}Q^{--}}).
\end{array}
\]

\paragraph{The term $\zeta^{+-}Q^{-+}\zeta^{+-}V^{-+}$: }
\[
\begin{array}{rl}
\ns{1}{\zeta^{+-}Q^{-+}\zeta^{+-}V^{-+}}&\le \ns{6}{\zeta^{-+}}\ns{2}{Q^{+-}|D_0|^{\ala}}\ns{3}{\tfrac{1}{|D_0|^{\ala}} \zeta^{-+}V^{++}}\\
\ns{3}{\tfrac{1}{|D_0|^{\ala}} \zeta^{-+}V^{++}}&\apprle \ssum_{j=1}^3\frac{1}{2\pi}\dint_{-\infty}^{+\infty}\ns{3}{\tfrac{1}{|D_0|^{\ala}(D_0+i\eta)} \partial_j \zeta \tfrac{P^0_+}{D_0+i\eta} V}d\eta\\
   &\apprle \ssum_{j=1}^3 \nlp{6}{\partial_j \zeta}\nlp{6}{V}\nlp{6}{\tfrac{1}{E(\cdot)^{5/8}}}^2\dint_{-\infty}^{+\infty}\frac{d\eta}{E(\eta)^{5/4}},\\
\ns{6}{\zeta^{-+}}&\apprle \nlp{6}{\nabla \zeta}.
\end{array}
\]

\paragraph{The term $\zeta^{++} Q^{+-}\zeta^{--} V^{-+}:$}
\[
\begin{array}{rl}
|\ttr(\zeta^{++} Q^{+-}\zeta^{--} V^{-+})|&\apprle \sqrt{\llo}\nlp{2}{\nabla V}(\ns{2}{\zeta Q^{+-}|D_0|^{\ala} \zeta}+\ns{2}{\zeta Q^{+-}}\nlp{\infty}{\nabla \zeta}).
\end{array}
\]
\paragraph{The terms $\zeta^{+-} Q^{-+} \zeta^{++} V^{++}$ and $\zeta^{++} Q^{+-}\zeta^{-+} V^{++}$} 

These operators  are difficult to handle. We use Lemma \ref{no_bind_commd} (Appendix \ref{no_bind_appA}). First:
\[
\zeta^{+-}Q^{-+}\zeta^{++}V^{++}=\big(\zeta^{+-}\frac{1}{|D_0|^{\tfrac{\eps_\La}{4}}}\big)\big(|D_0|^{\tfrac{\eps_\La}{4}}Q^{-+}\zeta^{++}|D_0|^{\tfrac{1}{2}+\tfrac{\eps_\La}{4}}\big)\big(\frac{1}{|D_0|^{\tfrac{1}{2}+\tfrac{\eps_\La}{4}}} V^{++} \big)\in\mathfrak{S}_1,
\]
with norm $\mathcal{O}((\llo)^{3/2} \nlp{3}{\nabla \zeta} \nlp{6}{V} \ns{2}{|D_0|^{\ala} Q})$. We used the KSS inequality and Hölder-type inequality for $\mathfrak{S}_p$. Similarly we can show that $\zeta^{++} Q^{+-}\zeta^{-+} V^{++}\in\mathfrak{S}_1$. Then by density of $\mathfrak{S}_1$ in $\mathfrak{S}_2$, we approximate $\big(|D_0|^{\tfrac{\eps_\La}{4}}Q^{-+}\zeta^{++}|D_0|^{\tfrac{1}{2}+\tfrac{\eps_\La}{4}}\big)$ by trace-class operators enabling us to say that:
\[
\ttr(\zeta^{+-}Q^{-+}\zeta^{++}V^{++})=\ttr\Big(\big(|D_0|^{\tfrac{\eps_\La}{4}}Q^{-+}\zeta^{++}|D_0|^{\tfrac{1}{2}+\tfrac{\eps_\La}{4}}\big) \big(\frac{1}{|D_0|^{\tfrac{1}{2}+\tfrac{\eps_\La}{4}}} V^{++}\big) \big(\zeta^{+-} \frac{1}{|D_0|^{\tfrac{\eps_\La}{4}}} \big)\Big).
\]
Let us show that $Q^{-+}\zeta^{++} V^{++}\zeta^{+-}\in\mathfrak{S}_1$. It suffices to show $\tfrac{1}{|D_0|^{\ala}} V^{++} \eta^{+-}\in\mathfrak{S}_2$. We go in Fourier space and used formula \eqref{no_bind_eqjs} to show $[V,P^0_+ e^{-sE|D_0|}]\in\mathfrak{S}_2$.
\[
\mathscr{F}([V,P^0_+ e^{-sE|D_0|}];p,q)=\frac{1}{(2\pi)^{3/2}}\wh{V}(p-q) \big(P^0_+(q)e^{-sE(q)}-P^0_-(p)e^{-sE(p)} \big);
\]
then (\emph{cf} Appendix \ref{no_bind_appA})
\[
\begin{array}{rl}
P^0_+(q)e^{-sE(q)}-P^0_-(p)e^{-sE(p)} &=(P^0_+(q)-P^0_+(p))e^{-sE(q)}+P^0_+(p)(e^{-sE(q)}-e^{-sE(p)})\\
\big|P^0_+(q)-P^0_+(p) \big|&\apprle \dfrac{|p-q|}{\max(E(p),E(q))}\\
\big|e^{-sE(q)}-e^{-sE(p)} \big|&= s|E(p)-E(q)|\dfrac{|e^{-sE(q)}-e^{-sE(p)}|}{s|E(p)-E(q)|}\\
     &\le s|p-q|\min(e^{-sE(p)},e^{-sE(q)})\\
      &\le s|p-q|(e^{-sE(p)}+e^{-sE(q)}).
\end{array}
\]
By easy computation: $\ns{2}{[V,P^0_+ e^{-sE|D_0|}]}\apprle s^{-1/2}e^{-s/\sqrt{2}}\nlp{2}{\nabla V}$:
\[
\dint_{s=0}^{+\infty}\ns{2}{[V,P^0_+e^{-s|D_0|}] \boldsymbol{\alpha}\cdot \nabla \zeta e^{-s|D_0|}}ds\apprle \nlp{\infty}{\nabla \zeta}\nlp{2}{\nabla V}\dint_0^{+\infty}\frac{e^{-s}ds}{s^{1/2}}.
\]
At last there remains to show:
\[
\mathcal{A}[V,\zeta]=\dint_0^{+\infty}\frac{e^{-s|D_0|}}{|D_0|^{\ala}}P^0_+ \big(V\boldsymbol{\alpha}\cdot \nabla \zeta\big)P^0_-e^{-s|D_0|}ds\in\mathfrak{S}_2,
\]
as in Appendix \ref{no_bind_appA} it suffices to go in Fourier space and remark $\nlp{2}{V \partial_j \zeta}\le \nlp{6}{V}\nlp{3}{\partial_j \zeta}$:
\[
\ns{2}{\mathcal{A}[V,\zeta]}\apprle \sqrt{\llo}\nlp{2}{V \nabla \zeta}\apprle \sqrt{\llo}  \nlp{6}{V}\nlp{3}{\partial_j \zeta}.
\]
The case of $\zeta^{++} Q^{+-}\zeta^{-+} V^{++}$ is similar: first we prove by density that
\[
\ttr(\zeta^{++} Q^{+-}\zeta^{-+} V^{++})=\ttr(\zeta^{-+} V^{++} \zeta^{++} Q^{+-}),
\]
and we get \emph{in fine}
\begin{equation}
\begin{array}{l}
\ncc{\rho[\zeta^{++} Q^{+-}\zeta^{-+} V^{++}]}+\ncc{\rho[\zeta^{-+}Q^{+-}\zeta^{-+}V^{++}]}\\
 \ \ \  \ \ \  \ \ \  \ \ \ \apprle (\sqrt{\llo}\nlp{3}{\nabla \zeta}+\nlp{\infty}{\nabla \zeta})(\ns{2}{\zeta P^0_+|D_0|^{\ala}Q}+\nlp{\infty}{\nabla\zeta}\ns{2}{Q}).
 \end{array}
\end{equation}

\end{dem}

\subsection{Estimates on the localised operator $\g$}%Estimation of $\ns{2}{\zeta|\Dbf|^{1/2} \g}$ and $\ns{2}{\zeta |D_0|^{\ala} \g}$.
Here $\g$ is the vacuum part of a (hypothetical) minimizer of $E^0_{\text{BDF}}(2)$ or a minimizer of $E^0_{\text{BDF}}(1)$. Our aim is to prove:
\begin{proposition}\label{no_bind_ns2kin}
Let $\zeta$ be a smooth function with:
\[
\left\{
\begin{array}{l}
\nlp{\infty}{\nabla \zeta},\nlp{\infty}{\partial_j\partial_k \zeta}<+\infty,\ j,k\in\{1,2,3 \}\\
\nlp{6}{\zeta v'},\nlp{2}{\zeta \nabla v'},\nqq{\zeta \g},\ns{2}{\zeta R_N}<+\infty.
\end{array}
\right.
\]
Then there holds:
\begin{equation}
\begin{array}{l}
\ns{2}{\zeta|\Dbf|^{1/2} \g}\apprle c^{-1/2} \nlp{2}{\zeta \nabla v'}+\alpha (\nqq{\zeta\g}+\ns{2}{\zeta R_N})\\
\ \ \ \ \ \ \ \ \ \ \ \ +\alpha^2(\nlp{2}{\zeta \nabla v'}+\nlp{6}{ \zeta v'}+\nqq{\zeta\g}+\ns{2}{\zeta R_N})^2\\
\ \ \ \ \ \ \ \ \ \ \ \ +\big\{\nlp{\infty}{\nabla\zeta}+\sum_{1\le j,k\le 3}\nlp{\infty}{\partial_j\partial_k \zeta}\big\}\big\{\alpha (\ncc{\rho'_\g}+\ns{2}{|\nabla|^{1/2} \g'})\big\}.
\end{array}
\end{equation}
The same holds for $\ns{2}{\zeta|D_0|^{\wt{a}} \g}$ with $\wt{a}\in\{ \tfrac{1}{2},\ala\}$. 

\noindent We can replace $\nqq{\zeta\g}+\ns{2}{\zeta R_N}$ by $\nqq{\g'}$ and put $P^0_{\pm}\g$ instead of $\g$. 
\end{proposition}

\subsubsection{Idea of the proof}
We will focus on the Cauchy expansion of $\g$: $\g=\ssum_{j=1}^{+\infty}\alpha^jQ_j(\g',\rho'_\g).$

As shown in \cite{gs,sokd,these}, we substitute $P^0_{\pm} (\rho_{\g}'*\tfrac{1}{|\cdot|})P^0_{\mp}$ by its expression \eqref{no_bind_comm} whenever it is necessary (in $Q_{0,1}, Q_{0,3},Q_{0,5}$)

We multiply $\g$ by $|D_0|^{\wt{a}}$ (or $|\Dbf|^{1/2}$) and then by $\zeta$. We consider $\tfrac{|D_0|^{\wt{a}}}{\Dbf+i\eta}$ (or $\tfrac{|\Dbf|^{1/2}}{\Dbf+i\eta}$) as a whole operator and we then commute $\zeta$ with this operator and maybe some $P^0_\eps$ and $\tfrac{1}{D_0+i\omega}$ (if it was necessary to use \eqref{no_bind_comm}) in order to stick $\zeta$ with a $v\rho'_\g*\tfrac{1}{|\cdot|}$, a $R'_\g$ or a $\partial_j \rho'_\g*\tfrac{1}{|\cdot|}$ (if \eqref{no_bind_comm} was used). For instance in the case of $Q_{0,1}$:
\begin{equation}
\begin{array}{rl}
Q_{0,1}^{+-}&=\dint_{-\infty}^{+\infty}\frac{|\Dbf|^{1/2}P^0_+}{\Dbf+i\eta} v'\frac{P^0_-}{\Dbf+i\eta}\\
  &=\dfrac{i}{2\pi}\underset{\mathbb{R}\times \mathbb{R}}{\diint} \dfrac{|\Dbf|^{1/2}}{\Dbf+i\eta}\dfrac{1}{D_0+i\omega}\boldsymbol{\alpha}\cdot \nabla v'\dfrac{P^0_-}{D_0+i\omega}\dfrac{d\eta d\omega}{\Dbf+i\eta}.
\end{array}
\end{equation}
We multiply by $\zeta$ and under the integral sign:
\begin{equation}
\begin{array}{rl}
\zeta \dfrac{|\Dbf|^{1/2}}{\Dbf+i\eta}\dfrac{1}{D_0+i\omega}\boldsymbol{\alpha}\cdot \nabla v'&=\Big[\zeta, \dfrac{|\Dbf|^{1/2}}{\Dbf+i\eta}\Big]\dfrac{1}{D_0+i\omega}\boldsymbol{\alpha}\cdot \nabla v'+\dfrac{|\Dbf|^{1/2}}{\Dbf+i\eta}\Big[\zeta,\dfrac{1}{D_0+i\omega}\Big]\boldsymbol{\alpha}\cdot \nabla v'\\
        &\ \ \ \ \ \ \ \ \ +\dfrac{|\Dbf|^{1/2}}{\Dbf+i\eta}\dfrac{1}{D_0+i\omega}\zeta\boldsymbol{\alpha}\cdot \nabla v'.
\end{array}
\end{equation}
We treat the first two terms in Section \ref{no_bind_comzeta}. For the latter we go in Fourier space and up to a constant the kernel of its Fourier transform is:
\[
\frac{\Ebf{p}^{1/2}}{\Ebf{p}+\Ebf{q}}\frac{P^0_+(p)}{E(p)+E(q)}\big(\mathscr{F}(\zeta \boldsymbol{\alpha}\cdot \nabla v';p-q) \big)P^0_-(q).
\]
In particular its Hilbert-Schmidt norm is $\mathcal{O}(\sqrt{\llo}\nlp{2}{\zeta \nabla v'_{\rho_{\g}}})$.

Doing the same for the other $Q_{k,\ell}$, we get terms with commutators treated in \ref{no_bind_comzeta} and other terms with $\zeta v'_{\rho_\g}$, $\zeta \boldsymbol{\alpha}\cdot \nabla v'$ and $\zeta R_{\g'}=R_{\zeta \g'}$. In particular taking the $\ns{2}{\cdot}$ under the integral sign, we get the following estimates on those terms.

\begin{equation}
\mathcal{O}\Big(c^{-1/2} \nlp{2}{\zeta \nabla v'}+\alpha \nqq{\zeta\g'}+\alpha^2(\nlp{2}{\zeta \nabla v'}+\nlp{6}{ \zeta v'}+\nqq{\zeta\g'})^2\Big).
\end{equation}
\begin{remark}\label{no_bind_addendum}
The term $\nqq{\zeta \g'}$ is due to Ineq. \eqref{no_bind_main_ingred} (l.h.s). Moreover we can deal with $\g$ and $N$ in $\g'$ differently. Indeed as $R_N\in\mathfrak{S}_2$, $\nqq{\zeta \g'}$ can be replaced by $K(\nqq{\zeta \g}+\ns{2}{\zeta R_N})$.%mmmmmmmmmmmm
\end{remark}
\begin{remark}
The term $\mathcal{T}[\zeta,v']:=\zeta \boldsymbol{\alpha}\cdot \nabla v'$ appears in $P^0_{-\eps}v'P^0_{\eps}$, that equals up to a multiplicative constant to% (to deal with $Q_{0,1},Q_{0,3},Q_{0,5}$): 
\[
\dint_{\omega=-\infty}^{+\infty}\frac{d\omega}{D_0+i\omega}\mathcal{T}[\zeta,v']\frac{P^0_\eps}{D_0+i\omega}.
\]
Up to a constant its Fourier transform is
\[
\dfrac{P^0_{-\eps}(p)\wh{\mathcal{T}}(p-q)P^0_{\eps}(q)}{E(p)+E(q)},
\]
and we deal with this term as $\wh{P^0_{-\eps}}(p) \wh{v'}(p-q)\wh{P^0_{\eps}}(q)$ in \cite{ptf,sokd,these}.
\end{remark}

\subsubsection{Commutating $\zeta$}\label{no_bind_comzeta}

We recall here that $[\zeta,P^0_{\eps}]$ is treated in \eqref{no_bind_comm}, Appendix \ref{no_bind_appA}.

In the same spirit of Lemma \ref{no_bind_commd}, we have the following Lemma.
\begin{lemma}\label{no_bind_premier}
Let $\eta\in\mathbb{R}$ and $\zeta$ smooth with 
\[
\nlp{\infty}{\nabla \zeta},\nlp{\infty}{\partial_j\partial_k \zeta}<+\infty,k,j\in\{1,2,3\}.
\]
Then there holds:
\[
\Big\lvert\Big\lvert \Big[\zeta , \frac{|\Dbf|^{1/2}}{\Dbf+i\eta}\Big]|\Dbf+i\eta|^{7/12}\Big\rvert \Big\rvert_{\mathcal{B}}\apprle \nlp{\infty}{\nabla \zeta}+\underset{1\le j,k\le 3}{\ssum}\nlp{\infty}{\partial_j\partial_k \zeta}.
\]
\end{lemma}

\begin{remark}
We can do the same with $|D_0|^{\ala}$ or $|D_0|^{1/2}$ instead of $|\Dbf|^{1/2}$ by using the following formula \cite[p. 87]{stabilitymatter}:
\[
|D_0|^{a}=\frac{\sin(a \pi)}{\pi}\dint_{s=0}^{+\infty}\frac{ds}{s^{1-a}}\frac{|D_0|}{|D_0|+s},\ a=\ala,1/2.
\]
Here we show the proof for $|\Dbf|^{1/2}$ because it enables us to localise the kinetic energy. But we can replace every $|D_0|^{\ala}$ by $|\Dbf|^{1/2}$ and vice-versa.
\end{remark}
There is also:
\begin{lemma}\label{no_bind_second}
There exists $K>0$ such that for any $\eta\in\mathbb{R}$ and any smooth function $\zeta$ with $\nlp{\infty}{\nabla \zeta}<+\infty$:
\begin{equation}
\Big| \Big[\zeta,\frac{1}{D_0+i\omega}\Big](x-y)\Big|\le K\nlp{\infty}{\nabla \zeta}\frac{e^{-E(\eta)/2(x-y)}}{|x-y|}.
\end{equation}
\end{lemma}
\begin{remark}
We recall that up to some constant $\tfrac{1}{a^2-\Delta}(x-y)=\sqrt{\frac{\pi}{2}}\frac{e^{-a|x-y|}}{|x-y|}$ \cite{LL}.
\end{remark}

\noindent-- The interesting fact here is that by taking the commutator of $\zeta$ and some function of $-i\nabla$ we gain some exponent for $\eta$ or $\omega$. Thus by using KSS inequalities under the integral sign we get the following estimates for the term with commutators:
\begin{equation}\label{no_bind_estcomm}
\mathcal{O}\Big(\big(\nlp{\infty}{\nabla\zeta}+\sum_{1\le j,k\le 3}\nlp{\infty}{\partial_j\partial_k \zeta}\big)\big(\alpha (\ncc{\rho'_\g}+\ns{2}{|\nabla|^{1/2} \g}+\ns{2}{\nabla N})\big)\Big)
\end{equation}
\medskip
\paragraph{Proof of Lemma \ref{no_bind_premier}:}
We decompose $\zeta=\zeta^{++}+\zeta^{+-}+\zeta^{-+}+\zeta^{--}$. We write for each term $\zeta^{\eps \eps'}$, $\eps,\eps'\in\{+,-\}$:
\[
\Big[\zeta^{\eps \eps'} , \frac{|\Dbf|^{1/2}}{\Dbf+i\eta}\Big]=[\zeta^{\eps \eps'},|\Dbf|^{1/2}]\frac{1}{\Dbf+i\eta}+|\Dbf|^{1/2}\Big[ \zeta^{\eps \eps'},\frac{1}{\Dbf+i\eta}\Big].
\]
It follows that:
\begin{equation}\label{no_bind_dundemi}
|\Dbf|^{1/2}\Big[ \zeta^{\eps \eps'},\frac{1}{\Dbf+i\eta}\Big]=\frac{|\Dbf|^{1/2}P^0_{\eps}}{\Dbf+i\eta}[\Dbf,\zeta]\frac{P^0_{\eps'}}{\Dbf+i\eta}.
\end{equation}
\subparagraph{The term $|\Dbf|^{1/2}\Big[ \zeta^{\eps \eps'},\frac{1}{\Dbf+i\eta}\Big]$} By simple computation we have:
\begin{equation}\label{no_bind_subst}
\begin{array}{rl}
[\Dbf,\zeta]&=\Big(1-\dfrac{\Delta}{\La^2}\Big)(-i\boldsymbol{\alpha}\cdot \nabla \zeta)+\dfrac{(-\Delta \zeta)}{\La^2}D_0+2\nabla \zeta\cdot\dfrac{\nabla D_0}{\La^2}\\
      &=(-i\boldsymbol{\alpha}\cdot \nabla \zeta)-\ssum_{j=1}^3 \Big( \frac{\partial_j}{\La^2} (-i\boldsymbol{\alpha}\cdot \nabla \partial_j \zeta)-2(\partial_j^2\zeta)\frac{D_0}{\La^2}\Big)\\
      &+(-\Delta \zeta)\dfrac{D_0}{\La^2}-\ssum_{j=1}^3\frac{\partial_j}{\La}\Big( (-i\boldsymbol{\alpha}\cdot \nabla  \zeta)\frac{\partial_j}{\La}-(\partial_j \zeta)\frac{D_0}{\La}\Big).
\end{array}
\end{equation}
Then there holds:
\begin{equation}
\nb{\frac{|D_0|}{\La |\Dbf|^{1/3}}}\apprle 1.
\end{equation}
Thus substituting in \eqref{no_bind_dundemi}, on the right of derivatives of $\zeta$, there is still an operator $\frac{1}{|\Dbf+i\eta|^{2/3}}$ available for some KSS inequality. The $\nb{\cdot}-$norm of the operator on their left is $\mathcal{O}(\Ebf{\eta}^{-1/6})$. The $\nb{\cdot}-$norm of derivatives of $\zeta$ are $\mathcal{O}(\nlp{\infty}{\nabla \zeta}+\nlp{\infty}{\Delta \zeta})$.
%There remains to treat the case of $[\zeta^{\eps \eps'},|\Dbf|^{1/2}]\frac{1}{\Dbf+i\eta}$
\subparagraph{The term $[\zeta^{\eps \eps'},|\Dbf|^{1/2}]\frac{1}{\Dbf+i\eta}$} By symmetry it suffices to study $\zeta^{++}$ and $\zeta^{+-}$. First:
\[
\begin{array}{rl}
[\zeta^{++},|\Dbf|^{1/2}]\frac{1}{\Dbf+i\eta}&=\frac{1}{\pi}\dint_0^{+\infty}\sqrt{s}ds\frac{P^0_+}{\Dbf+s}[\Dbf, \zeta]\frac{P^0_+}{\Dbf+s}\frac{1}{\Dbf+i\eta}.
\end{array}
\]
Once again, if we replace $[\Dbf,\zeta]$ by its expression in \eqref{no_bind_subst}, we see that taking $|\Dbf+i\eta|^{-1/4}$ from $\tfrac{1}{\Dbf+i\eta}$, there remains $\frac{|\Dbf+i\eta|^{1/4}}{\Dbf+i\eta}$ for some KSS inequality. 

This enables us to get a finite integral over the $s$ variable:
\[
\dint_0^{+\infty} \frac{\sqrt{s}ds}{(1+s)^{2/3}}\frac{1}{(1+s)^{11/12}}<+\infty.
\]
At last:
\[
\begin{array}{rl}
[\zeta^{+-},|\Dbf|^{1/2}]\dfrac{1}{\Dbf+i\eta}&=-\dfrac{1}{\pi}\dint_0^{+\infty}\sqrt{s}ds\frac{P^0_+}{|\Dbf|+s}(\zeta \Dbf+\Dbf \zeta)\frac{P^0_-}{|\Dbf|+s}\frac{1}{\Dbf+i\eta}\\
         &=-\dfrac{1}{\pi}\dint_0^{+\infty}\sqrt{s}ds\frac{P^0_+}{|\Dbf|+s}(2\zeta \Dbf+[\Dbf,\zeta])\frac{P^0_-}{|\Dbf|+s}\frac{1}{\Dbf+i\eta}.
\end{array}
\]
The term with $[\Dbf,\zeta]$ is dealt with as before. There remains:
\begin{equation}\label{no_bind_forms}
\begin{array}{rl}
\dint_0^{+\infty} \frac{\sqrt{s}ds}{|\Dbf|+s}\zeta^{+-}\frac{\Dbf}{|\Dbf|+s}\frac{1}{\Dbf+i\eta}.
\end{array}
\end{equation}
We write (\emph{cf} \eqref{no_bind_comm}):
\begin{equation}
\zeta^{+-}=P^0_+[\zeta,P^0_-]=\frac{P^0_+}{2\pi}\dint_{-\infty}^{+\infty}\frac{d\omega}{\Dbf+i\omega}[\Dbf,\zeta]\frac{1}{\Dbf+i\omega},
\end{equation}
and substitute $\zeta^{+-}$ by this expression in \eqref{no_bind_forms}. We must compensate $\frac{|D_0|}{\La}$ on the left side of $\zeta$ and $\frac{|D_0|\Dbf}{\La}$ on its right side: we use $\frac{1}{|\Dbf+i\omega|^{1/3}}$ on the left side and $\{|\Dbf+i\omega|^{1/2}|\Dbf+i\eta|^{5/12}(|\Dbf|+s)^{5/12}\}^{-1}$ on the right side: there remains $\tfrac{1}{|\Dbf+i\eta|^{7/12}}$ for some KSS inequality and:
\[
\dint_{s=0}^{+\infty}\dint_{\omega=-\infty}^{+\infty}\frac{\sqrt{s}dsd\omega}{(1+s)^{19/12}E(\omega)^{7/6}}<+\infty.
\]

\paragraph{Proof of lemma \ref{no_bind_second}:}
This is straightforward because everything is computable:
\[
\frac{1}{D_0+i\eta}=\frac{D_0-i\eta}{E(\eta)^2-\Delta}.
\]
However $\dfrac{1}{E(\eta)^2-\Delta}(x-y)=\dfrac{e^{-E(\eta)|x-y|}}{4\pi |x-y|}$ so it is clear that:
\[
\Big|\frac{1}{D_0+i\eta}(x-y)\Big|\apprle \frac{e^{-E(\eta)|x-y|/2}}{|x-y|^2}.
\]
In Direct space we use $|\zeta(x)-\zeta(y)|\le \nlp{\infty}{\nabla \zeta}|x-y|$ and
\[
\Big| \Big[\zeta,\frac{1}{D_0+i\omega}\Big](x-y)\Big|\apprle \nlp{\infty}{\nabla \zeta}\frac{e^{-E(\eta)/2(x-y)}}{|x-y|}
\]

\subsubsection{Localisation of $\nabla v_{\rho_\g'}$ and $R_N$}\label{no_bind_estmoche}
%\subsubsection{Estimates of $\nlp{2}{\etabfc{\la} \partial_j v},\nlp{6}{\etabfc{\la} v},\nqq{\etabfc{\la} \g},\ns{2}{\etabfc{\la} R_N}$}\label{no_bind_estmoche}

We recall that $\etabfc{\la}$ is the following function:
\[
\etabfc{\la}(x):=\big\{1-\xi_{c\la\rgg}^2(x-cz_1)-\xi_{c\la\rgg}^2(x-cz_2) \big\}^{-1/2},\ \la_0<\la<2^{-1}.
\]
We will take $\la_0\le \la\le 3^{-1}$ ($\la_0(L,\rgg)$ is defined in \eqref{no_bind_def_la_0}). More generally except for $\nlp{2}{\etabfc{\la} \partial v},\nlp{6}{\etabfc{\la} v}$, the estimates are true with $\zeta$ instead of $\etabfc{\la}$ in the case where $\zeta$ is $\zeta(x)=\zeta_0(x/A)$ with $0\le \zeta_0\le 1$ fixed . This part gives estimates with respect to $\zeta_0$ and $A$.

\begin{notation}
We write $\theta_1^{1}(x):=\sqrt{1-\xi_1^2(x)}$, it is clear that
\[
\nlp{\infty}{\nabla \etabfc{\la}}\le \dfrac{\nlp{\infty}{\nabla \theta_1^{1}}}{c\la\rgg}\ \text{and\ so\ on}.
\]
\end{notation}

\begin{proposition}\label{no_bind_estint}
Let $\g+N$ be a minimizer for $E^0(2)$ (or $E^0(1)$), $\rho\in L^1\cap L^2$ (\emph{e.g.} $\rho=\rho_\g,\rho_N$) and $\la_0\le \la<2^{-1}$. With the previous notations, there holds:
\begin{equation}
\left\{
\begin{array}{rl}
\ns{2}{\etabfc{\la}R[N_j]}^2&\apprle \nlp{2}{\nabla \psi_j}^2\dint_x (\etabfc{\la})^2(x)|\psi_j(x)|^2dx\apprle \big\{(\la \rgg)^2 c^2\big\}^{-1},\\
\nqq{\etabfc{\la} \g}&\apprle\nlp{\infty}{\nabla \theta_1^1}(c\la\rgg)^{-1}\ns{2}{|D_0|^{1/2} \g}+\ns{2}{\etabfc{\la} |D_0|^{1/2} \g},\\
\nlp{6}{\etabfc{\la} v_\rho}&\apprle \nlp{2}{(\nabla \etabfc{\la}) v_\rho}+\nlp{2}{\etabfc{\la} \nabla v_\rho},\\
\nlp{2}{(\nabla \etabfc{\la}) v_\rho}&\apprle \nlp{1}{\rho}\nlp{2}{\nabla |\nabla| \boldsymbol{\theta}_1^{1}}(c\la\rgg)^{-1/2},\\
\nlp{2}{\etabfc{\la} \partial_j v_\rho}&\apprle \ncc{\etabfc{\la}\rho\,\etabfc{\la/2}}+\nlp{1}{\rho}\Big(\dfrac{\nlp{\infty}{\nabla \theta_1^1}^{1/4}}{(c\la \rgg)^{1/2}}+\dfrac{\nlp{\infty}{\nabla \theta_1^1}}{(c\la\rgg)^{3/4}}\Big)\\
   &\ \ \ +\nlp{2}{\rho}^{1/6}\nlp{1}{\rho}^{5/6}\dfrac{\nlp{\infty}{\nabla \theta_1^1}^{3/4}}{(c\la \rgg)^{1/2}}+\nlp{1}{\rho}\Big(\dfrac{1+\nlp{\infty}{\nabla \theta_1^1}}{(c\la\rgg)^{1/2}}\Big).
\end{array}
\right.
\end{equation}
Moreover if we write $\g=\alpha Q_{0,1}+\alpha Q_{1,0}+\alpha^2\wit{Q}_2,\ \rho_N=n$ we also have:
\begin{equation}\label{no_bind_why}
\left\{
\begin{array}{rl}
\ncc{\etabfc{\la} \rho_\g}&\apprle \frac{\alpha}{c\la\rgg} \nlp{\infty}{\nabla \theta_1^1}(\ncc{n}+\nlp{6/5}{\alpha \rho_{1,0}+\alpha^2\wit{\rho}_2})\\
     &+L\ncc{\etabfc{\la} n}+\ncc{\etabfc{\la} (\alpha \rho_{1,0}+\alpha^2\wit{\rho}_2)}.
\end{array}\right.
\end{equation}
\end{proposition}

We recall that $\nlp{6/5}{\rho}\apprle \nlp{2}{\rho}^{1/3}\nlp{1}{\rho}^{2/3}$.

\begin{dem}
We will write $v_\rho=v$ for convenience.

\subparagraph{The term $\ns{2}{\etabfc{\la} R_N}$}
\[
\begin{array}{rl}
\ns{2}{\etabfc{\la}N_j}^2&=\diint \frac{(\etabfc{\la})^2(x)|\psi_j(x)|^2|\psi_j(y)|^2}{|x-y|^2}dxdy\\
          &=\dint_x dx(\etabfc{\la})^2(x)|\psi_j(x)|^2 \dint_y\frac{|\psi_j(y)|^2}{|x-y|^2}dy\\
          &\le 4\nlp{2}{\nabla \psi_j}^2\dint_x (\etabfc{\la})^2(x)|\psi_j(x)|^2dx\apprle \dfrac{1}{(\la \rgg)^2}\dfrac{1}{c^2}
\end{array}
\]
where we have used Lemma \ref{no_bind_decaylemma}. %The same Lemma enables us to say:
%\[
%\begin{array}{rl}
%\nlp{2}{\etabfc{\la} \partial_j v_n}^2&\le 
%\end{array}
%\]

\subparagraph{The term $\nqq{\etabfc{\la} \g}$}
\[
\begin{array}{rl}
\nqq{\etabfc{\la} \g}&\le \sqrt{\frac{\pi}{2}}\ns{2}{|D_0|^{1/2}\etabfc{\la} \g}\\
      &\le \sqrt{\frac{\pi}{2}}\big( \ns{2}{[|D_0|^{1/2},\etabfc{\la}] \g}+\ns{2}{\etabfc{\la} |D_0|^{1/2} \g}\big)\\
      &\apprle \nlp{\infty}{\nabla \etabfc{\la}}\ns{2}{|D_0|^{1/2} \g}+\ns{2}{\etabfc{\la} |D_0|^{1/2} \g},
\end{array}
\]
and we can treat $\ns{2}{\etabfc{\la} |D_0|^{1/2} \g}$ as $\ns{2}{\etabfc{\la} |D_0|^{\ala} \g}$.

\subparagraph{The term $\nlp{6}{\etabfc{\la} v}$} We use the Sobolev inequality:
\[\begin{array}{rl}
\nlp{6}{\etabfc{\la} v}&\apprle \nlp{2}{(\nabla \etabfc{\la}) v}+\nlp{2}{\etabfc{\la} \nabla v}.
\end{array}\]

We get a term $\nlp{2}{\etabfc{\la} \nabla v}$ we will treat later.

\noindent -- For the term $\nlp{2}{(\nabla \etabfc{\la}) v}$, we use the fact that $\rho*\tfrac{1}{|\cdot|}$ is $L^3_{w}$ with weak norm of order $\nlp{1}{\rho}$ \cite{stein} and we use rearrangement inequalities \cite{LL}: $\dint |fg|\le \dint |f|_*|g|_*$ and $\nlp{2}{\nabla |f|_*}\le \nlp{2}{\nabla |f|}$.
\[
\begin{array}{rl}
\nlp{2}{(\nabla \etabfc{\la}) v}^2&=\dint |\nabla \etabfc{\la}|^2 |v|^2\le \dint (|\nabla \etabfc{\la}|^2)_* (|v|^2)_*\\
        &\apprle \dint (|\nabla \etabfc{\la}|^2)_*(x)\frac{\nlp{1}{\rho}^2}{|x|^2}dx\\
        &\apprle \nlp{1}{\rho}^2\nlp{2}{\nabla \sqrt{(|\nabla \etabfc{\la}|^2)_*}}^2=\nlp{1}{\rho}^2\nlp{2}{\nabla (\sqrt{|\nabla \etabfc{\la}|^2})_*}^2\\
        &\apprle \nlp{1}{\rho}^2\nlp{2}{\nabla |\nabla \etabfc{\la}|}^2\apprle \nlp{1}{\rho}^2\frac{\nlp{2}{\nabla |\nabla| \theta_1^{1}}^2}{c\la\rgg}.
\end{array}
\]

\noindent -- For the term $\nlp{2}{\etabfc{\la} \partial_j v}$, we write: 
\begin{equation}\label{no_bind_chiant}
\etabfc{\la} \partial_j v(x)=\dint\frac{(y_j-x_j)}{|x-y|^3}(\etabfc{\la}(x)-\etabfc{y})\rho(y)dy+(\etabfc{\la} \rho)*\Big(\partial_j\frac{1}{|\cdot|}\Big).
\end{equation}
The last term will give $\ncc{\etabfc{\la} \rho}$. From this point, due to the particular form of $\etabfc{\la}$ there holds:
\begin{equation}
\etabfc{\la}=\etabfc{\la}\etabfc{\la/2}\text{\ so\ }\ncc{\etabfc{\la} \rho}=\ncc{\etabfc{\la} \rho\,\etabfc{\la/2}}.
\end{equation}

Let us treat the first term of \eqref{no_bind_chiant}. More generally we take $\zeta(x)=\zeta_0(x/A)$ and we use the properties of $\etabfc{\la}$ at the very end.

Taking the squared norm we have:
\[
\begin{array}{l}
\diint \rho(x)\rho(x)dxdy\dint \frac{(\zeta(t)-\zeta(x))(\zeta(t)-\zeta(y))(t_j-x_j)(t_j-y_j)}{|t-x|^3|t-y|^3}dt.
\end{array}
\]
We split at level $|x-y|= \sqrt{A}$: first if $|x-y|\ge \sqrt{A}$, then
\[
\begin{array}{l}
\underset{|x-y|\ge \sqrt{A}}{\diint}\dfrac{|\rho(x)||\rho(y)|}{|x-y|^{1/2}}\dint\frac{\nlp{\infty}{\nabla \zeta}^{1/2}dt}{|t|^{7/4}|t-\mathbf{e}|^{7/4}}\\
\ \ \ \ \ \ \ \ \ \le \dfrac{\nlp{\infty}{\nabla \zeta}^{1/2}}{\sqrt{A}}\nlp{1}{\rho}^2\apprle \dfrac{L^2\nlp{\infty}{\nabla\zeta_0}^{1/2}}{A}.
\end{array}
\]
If $|x-y|\le \sqrt{A}$ then there holds $|x-y|\nlp{\infty}{\nabla \zeta}\le \nlp{\infty}{\nabla \zeta_0}\tfrac{1}{\sqrt{A}}$, thus $\zeta(x)=\zeta(y)+\zeta(x)-\zeta(y)$ and we substitute in the integral over $t$.
We split $\mathbb{R}^3$ in three: $|t-x|< |x-y|/2$, $|t-y|<|x-y|/2$ and the remainder domain. 

\noindent a. For the first ball $B(x,|x-y|/2)=B_x$:
\[
\begin{array}{rl}
\dint_{B_x}\frac{|\zeta(x)-\zeta(t)||\zeta(y)-\zeta(t)|}{|t-x|^2|t-y|^2}dt\le \dfrac{\nlp{\infty}{\nabla \zeta_0}^2}{A^{3/2}}\dint_{B_x}\frac{dt}{|t-x|^2|t-y|}
   &\apprle \dfrac{\nlp{\infty}{\nabla \zeta_0}^2}{A^{3/2}}.
\end{array}
\]
\noindent b. The same holds for the ball $B_y$. c. For the remainder domain $C_{xy}$:

\noindent c.1. we first deal with the term $(\zeta(y)-\zeta(x)(\zeta(t)-\zeta(y))$:

\[
\begin{array}{rl}
\underset{t\in C_{xy}}{\dint} dt\dfrac{|(\zeta(x)-\zeta(y))(\zeta(t)-\zeta(y))|}{|x-t|^2|y-t|^2}&\le \dfrac{(\nlp{\infty}{\nabla \zeta_0})^{3/2}}{A}\dint\frac{dt}{|x-t|^2|y-t|^{3/2}}\\
     &\apprle \dfrac{(\nlp{\infty}{\nabla \zeta_0})^{3/2}}{A}\dfrac{1}{|x-y|^{1/2}}
\end{array}
\]
\[
\text{and:}\diint\frac{|\rho(x)||\rho(y)|}{|x-y|^{1/2}}dxdy\apprle \nlp{2}{\rho}^{1/3}\nlp{1}{\rho}^{5/3}\apprle L^2c^{-1/2}.
\]
We used above the Hardy-Littlewood-Sobolev inequality \cite[Theorem 4.3]{LL}.

\noindent c.2. At last we must handle the term $(\zeta(t)-\zeta(x))^2$:
\[
\underset{|x-y|\le \sqrt{A}}{\diint}\rho(x)\rho(y)dxdy\underset{t\in C_{xy}}{\dint}\frac{(\zeta(t)-\zeta(y))^2(t_j-y_j)(t_j-x_j)}{|x-t|^3|y-t|^3}dt.
\]
%We suppose $\sqrt{A}\le \tfrac{A}{2}$ that is $A\ge 4$. We note $\delta=y-x$
As $t\in C_{xy}$ we can replace $|x-t|^{-2}$ by $K|y-t|^{-2}$.
\[
\begin{array}{rl}
\underset{t\in C_{xy}}{\dint} \dfrac{(\zeta(t)-\zeta(y))^2}{|t-y|^4}dt&\le \dint_t \frac{(\zeta(t)-\zeta(y))^2}{|t-y|^4}dt. 
\end{array}
\]
We use now the properties of the function $\etabfc{\la}$. It is easy to see that no matter where $y\in\mathbb{R}^3$ is, this last integral is $\mathcal{O}((c\la\rgg)^{-1}K(\theta_1^{1}))$. Indeed let $\text{Ext}$ be the domain defined by $\text{Ext}=\{y\in\mathbb{R}^3: f(y):=\text{dist}( y,\{\etabfc{\la}\neq 1 \})>2c\la \rgg \}$. 

\noindent c.2.1. If $y\in \text{Ext}$, then it is clear that the previous integral is an
\[
\mathcal{O}\Big(\frac{(c\la \rgg)^3}{f(y)^4} \Big)=\mathcal{O}\big(\frac{1}{c\la \rgg} \big).
\]
\noindent c.2.2. Else we split $\RR$ at level $|t-y|= 2c\la \rgg$:
\[
\underset{|t-y|\le 2c\la\rgg}{\dint}dt\frac{(\etabfc{\la}(t)-\etabfc{\la}(y))^2}{|t-y|^4}\apprle \frac{\nlp{\infty}{\nabla \theta_1^{1}}^2}{(c\la \rgg)^2} (c\la \rgg)=\mathcal{O}\big(\frac{\nlp{\infty}{\nabla \theta_1^{1}}^2}{ c\la \rgg}\big).
\]
\[
\underset{|t-y|> 2c\la\rgg}{\dint}dt\frac{(\etabfc{\la}(t)-\etabfc{\la}(y))^2}{|t-y|^4}\le \underset{|t-y|> 2c\la\rgg}{\dint}\frac{dt}{|t-y|^4}=\mathcal{O}\Big(\frac{1}{c\la\rgg}\Big).
\]

\subparagraph{Proof of \eqref{no_bind_why}} To begin with we remark that by the Hardy-Littlewood-Sobolev inequality \cite{LL}: $\ncc{\rho}\apprle \nlp{6/5}{\rho}.$ Then we use formula \eqref{no_bind_form_dens} of $\rho_\g$. 
%\[
%\left\{
%\begin{array}{rl}
%\rho_\g&=-\check{b}_\La*n+(\delta_0-\check{b}_\La)*(\alpha \rho_{1,0}+\alpha^2\wit{\rho}_2),\\
%\check{b}_\La&=\mathscr{F}^{-1}\left( \frac{\alpha B_\La}{1+\alpha B_\La}\right),
%\end{array}
%\right.
%\]
%where
%\[
%\left\{
%\begin{array}{l}
%B_\La(k)=\dint_0^1\frac{8\La^2 dt}{\pi t^3}\dint_0^{\sqrt{1-t^2}}\frac{(1-u^2)du}{1+3u^2}\frac{1}{\mu_1(t)^2+|k|^2}\times \frac{1}{\mu_2(t,u)^2+|k|^2},\\
%\mu_1(t)=\dfrac{2}{t}\text{\ and\ }\mu_2(t,u)=2\La(1-\La^{-2}+t^{-2})^{1/2}(1+3u^2)^{-1/2}.
%\end{array}\right.
%\]
We write 
\[
\etabfc{\la}(x)\check{F}_\La*\rho(x)=\dint_y (\etabfc{\la}(x)-\etabfc{\la}(y))\check{F}_\La(x-y)\rho(y)dy+\check{F}_\La*(\etabfc{\la}\rho)(x).
\]
So it suffices to show $\nlp{1}{|\cdot| \check{F}_\La}\apprle \alpha$ to end the proof: this is precisely \eqref{no_bind_est_fla_no}-\eqref{no_bind_est_fla_no_1}, applied with $\ell=1$ to $\check{F}_\La$ (true if $\alpha$ is less than some $K(\ell=1)$).
%\[
%\check{b}_\La=\ssum_{j=1}^{+\infty}(-1)^{j+1}\{\alpha \check{B}_\La\}^{*(j)}\text{\ and\ }\nlp{1}{\alpha \check{B}_\La}=\alpha B_\La(0)\apprle \alpha \llo=\Theta(L),
%\]
%it suffices to prove $\nlp{1}{|\cdot| \check{B}_\La}\apprle \alpha$, up to take sufficiently small $L$. It can be shown $\check{B}_\La$ has exponential falloff but we will prove it in a simpler way. It is known $\mathscr{F}^{-1}(\tfrac{1}{a^2+|\cdot|^2})$ is the Yukawa potential $\sqrt{\tfrac{2}{\pi}}\tfrac{e^{-a|\cdot|}}{|\cdot|}>0$. By direct computation there also holds:
%\[
%\mathscr{F}^{-1}\Big( \frac{a}{(a^2+|\cdot|^2)^2}\Big)=\sqrt{\frac{\pi}{2}}e^{-a|\cdot|}>0.
%\]
%It is then straightforward to see that for $\La>1$ we have:
%\[
%\dint_0^1\frac{8\La^2 dt}{\pi t^3}\dint_0^{\sqrt{1-t^2}}du\frac{(1-u^2)}{1+3u^2}\frac{\mu_1(t)^{-1}+\mu_2(t,u)^{-1}}{\mu_1(t)^2 \mu_2(t,u)^2}\apprle 1.
%\]
\end{dem}

\subsubsection{Proof of Lemma \ref{no_bind_ultimchiant}}

We write $\xi$ instead of $\xi_j^{(\tfrac{1}{3})}$ and $Q$ instead of $\g'$ for convenience.

First remark: for any $\eps,\eps'\in\{+,-\}$:
\begin{equation}\label{no_bind_fiant}
P^0_{\eps}\xi P^0_\eps Q P^0_{\eps'}  \xi P^0_{\eps'}=[P^0_{\eps},\xi]Q^{\eps\, \eps'}[\xi, P^0_{\eps'}]+[P^0_{\eps},\xi]Q^{\eps\, \eps'}\xi+\xi Q^{\eps\, \eps'}[\xi, P^0_{\eps'}]+\xi Q^{\eps\,\eps'}\xi.
\end{equation}
This gives the error term between $\xi Q \xi$ and $\xi[Q]$.
We estimate their density as in Section \ref{no_bind_locdens}, that is by duality. 

Second remark: $\partial_j \xi^{(\la)}=(\partial_j \xi)\etabfc{\la/2}$.

As in this section, by using \eqref{no_bind_comm}, it is clear that
\[
\ncc{[P^0_{\eps},\xi]Q^{\eps\, \eps'}[\xi, P^0_{\eps'}]}\apprle \ns{2}{|D_0|^{\ala} Q}\nlp{6}{\xi}^2\apprle \dfrac{\ns{2}{|D_0|^{\ala} Q}}{(c\rgg)^2}.
\]

We can drop terms involving the density of these operators.

We write:
\[
\xi^{+-}=\dfrac{i}{2\pi}\dint_{-\infty}^{+\infty} \dfrac{1}{D_0+i\omega} (\etabfc{\la/2})^2(\boldsymbol{\alpha}\cdot \nabla\xi)\dfrac{d\omega}{D_0+i\omega}.
\]
We commute $\etabfc{\la/2}$ with $(D_0+i\omega)^{-1}$ on the right and on the left. As shown before there holds:
\[
\Big|(\etabfc{\la/2}(x)-\etabfc{\la/2}(y))\frac{1}{D_0+i\omega}(x-y) \Big|\apprle \dfrac{e^{-E(\omega)|x-y|/2}}{|x-y|}\nlp{\infty}{\nabla \etabfc{\la/2}}.
\]
So taking KSS inequalities under the integral sign we obtain for instance:
\[
\begin{array}{l}
\ttr\big( P^0_-\xi Q \xi^{+-} VP^0_-\Big)=\ttr\Big( P^0_- \xi Q \etabfc{\la/2} \xi^{+-} \etabfc{\la/2} V P^0_-\Big)\\
+\mathcal{O}\Big( \nlp{6}{V}\ns{2}{Q|D_0|^{1/2}}\nlp{3}{\nabla \xi}\nlp{\infty}{\nabla \etabfc{\la/2}}\dint_{\mathbb{R}}\frac{d\omega}{E(\omega/2)^{5/4}}\Big)\\
+\mathcal{O}\Big( \nlp{\infty}{\nabla \etabfc{\la/2}}^2\nlp{6}{V}\nlp{3}{\nabla \xi}\ns{2}{Q}\dint_{\mathbb{R}}\dfrac{d\omega}{E(\omega/2)^2}\Big).
\end{array}
\]
There remains the first trace. First of all, for any $V$ Schwartz, we can show as in Section \ref{no_bind_locdens} that the operator is trace-class with norm controlled by
\[
\sqrt{\llo}\nlp{2}{\nabla (\xi\etabfc{\la/2}V)}\ns{2}{|D_0|^{1/2} \etabfc{\la/2} Q^{+-}}+\nlp{\infty}{\nabla \xi}\nlp{2}{\nabla (\etabfc{\la/2}V)}\ns{2}{Q P^0_+\etabfc{\la/2}}.
\]

We have \emph{a priori} 
$\begin{array}{l}
\nlp{2}{\nabla (\xi\etabfc{\la/2}V)}\apprle \nlp{2}{\etabfc{\la/2} \nabla V}+\nlp{3}{\nabla (\xi\etabfc{\la/2})}\nlp{6}{V}.
\end{array}$

In particular:
\[
\ncc{[P^0_\eps,\xi] Q^{\eps\,\eps'} \xi}\apprle \sqrt{\llo}\ns{2}{|D_0|^{1/2} \etabfc{\la/2} Q^{\eps'\,\eps}}\apprle \dfrac{L}{\sqrt{c\la\rgg}}.
\]

We use now the fact that we want the trace \emph{for a particular $V$}, namely $\rho[\xi Q \xi]*\tfrac{1}{|\cdot|}$. So as in Proposition \ref{no_bind_estint}, the function $(\xi^2 \rho_\g')*\tfrac{1}{|\cdot|}$ is in $L^3_w$ and
\[
\nlp{2}{(\nabla \etabfc{\la/2}) \big[(\xi^2 \rho_\g')*\tfrac{1}{|\cdot|}\big]}\apprle \frac{\sqrt{2}}{\sqrt{\la c\rgg}}\nlp{1}{\rho'_\g}\apprle \frac{\sqrt{2}}{\sqrt{\la c\rgg}}.
\]
Then we write $(\xi^2 \rho_\g')*\tfrac{1}{|\cdot|}=\rho'_\g *\tfrac{1}{|\cdot|}-((\etabfc{\la})^2\rho'_\g)*\tfrac{1}{|\cdot|}$ and
\[
\begin{array}{rl}
\nlp{2}{\etabfc{\la/2} \nabla \big[ (\xi^2 \rho_\g')*\tfrac{1}{|\cdot|}\big]}&\apprle \nlp{2}{\nabla ( (\etabfc{\la})^2\rho'_\g)*\tfrac{1}{|\cdot|}}+\nlp{2}{\etabfc{\la/2} \nabla (\rho_\g')*\tfrac{1}{|\cdot|}}\\
 &\apprle \ncc{(\etabfc{\la})^2 \rho'_\g}+\ssum_{j=1}^3\nlp{2}{\etabfc{\la/2} \partial_j v'_{\rho_\g}},
\end{array}
\]
and those terms are dealt with Propositions \ref{no_bind_estint} and \ref{no_bind_titim}.

Putting everything together, we get an error term of order:
\[
\sqrt{\llo}\times \frac{1}{c\sqrt{\rgg}}\times\dfrac{1}{\sqrt{c\rgg}}=\mathcal{O}\Big( \dfrac{L}{c\rgg}\Big).
\]

\end{appendix}

\bibliographystyle{plain}
{\small\bibliography{bibliothese.bib}}

\end{document}